\documentclass[12pt]{article}
\usepackage{lscape}
\usepackage{geometry}
\usepackage[onehalfspacing]{setspace}
\usepackage{amssymb}
\usepackage{amsfonts}
\usepackage{graphicx}
\usepackage{amsmath}
\usepackage{bbm}
\usepackage{epstopdf}
\usepackage{natbib}
\usepackage{bibunits}
\usepackage{morefloats}
\usepackage{float}
\usepackage{comment}
\usepackage{pdflscape}
\usepackage{adjustbox}
\usepackage{subcaption}
\usepackage{longtable}
\usepackage{apalike}
\usepackage{transparent}
\usepackage{xfrac}
\usepackage[dvipsnames,table]{xcolor}
\usepackage{etoolbox,siunitx}
\robustify\bfseries

\usepackage{hhline,multirow}
\usepackage{yfonts}

\usepackage{longtable}
\usepackage{threeparttablex}

\newcolumntype{d}[1]{D{.}{.}{#1}}
\usepackage{qtree}
\usepackage{multicol}
\usepackage{booktabs}
\usepackage{threeparttable}
\usepackage{tabularx}
\usepackage{dcolumn}
\newcolumntype{d}[1]{D..{#1}} 

\usepackage{siunitx}
\usepackage{mathpazo} 
\usepackage[scaled]{helvet} 
\usepackage{courier} 
\normalfont
\usepackage[T1]{fontenc}
\usepackage{microtype}
\usepackage{listings}
\usepackage{epigraph}
\def\sym#1{\ifmmode^{#1}\else\(^{#1}\)\fi}

\usepackage[pdftex,bookmarks,colorlinks]{hyperref}
\hypersetup{
  colorlinks,
  citecolor=darkpowderblue,
  linkcolor=red,
  urlcolor=blue,
  pdftitle={Building Macroeconomically Relevant Climate Indices via the Assemblage VAR},
  pdfauthor={Christophe Barrette, Philippe Goulet Coulombe, Tim Reinicke}}
\usepackage{cleveref}

\usepackage[defaultlines=2,all]{nowidow}

\DeclareMathOperator*{\argmin}{arg\,min}

\usepackage{etoolbox}

\definecolor{dukeblue}{rgb}{0.0, 0.0, 0.61}
\definecolor{darkred}{rgb}{0.8,0,0}
\definecolor{darkgreen}{rgb}{0,0.5,0}

\makeatletter
\patchcmd{\epigraph}{\@epitext{#1}}{\itshape\@epitext{#1}}{}{}
\makeatother

\makeatletter
\def\munderbar#1{\underline{\sbox\tw@{$#1$}\dp\tw@\z@\box\tw@}}
\makeatother

\usepackage{graphicx}
\usepackage{wrapfig}

\usepackage{algorithm,algorithmic}

\usepackage{xspace}

\definecolor{bred}{RGB}{122, 0, 0}
\definecolor{darkpowderblue}{rgb}{0.0, 0.05, 0.5}

\definecolor{dpd}{rgb}{0.0, 0.05, 0.5}
\definecolor{dpd2}{rgb}{0.0, 0.035, 0.38}
{}

\definecolor{reviewerorange}{RGB}{255, 140, 0}
\definecolor{responseblue}{RGB}{0, 100, 200}

\newcommand{\ascn}{MRCI$_\text{\scriptsize comps}^\text{\scriptsize Nation}$\xspace}
\newcommand{\asn}{MRCI$^\text{\scriptsize Nation}$\xspace}
\newcommand{\asrn}{MRCI$_\text{\scriptsize ranks}^\text{\scriptsize Nation}$\xspace}
\newcommand{\ascr}{MRCI$_\text{\scriptsize comps}^\text{\scriptsize Region}$\xspace}
\newcommand{\asr}{MRCI$^\text{\scriptsize Region}$\xspace}
\newcommand{\asrr}{MRCI$_\text{\scriptsize ranks}^\text{\scriptsize Region}$\xspace}
\newcommand{\asco}{MRCI$_\text{\scriptsize comps}^\text{\scriptsize NOAA}$\xspace}
\newcommand{\aso}{MRCI$^\text{\scriptsize NOAA}$\xspace}
\newcommand{\asro}{MRCI$_\text{\scriptsize ranks}^\text{\scriptsize NOAA}$\xspace}

\newcommand{\mn}[1]{{\fontfamily{phv}\selectfont #1}}

\usepackage{soul}

\makeatletter
\newcommand\halftiny{\@setfontsize\halftiny\@vipt\@viipt}
\newcommand\notsotiny{\fontfamily{phv}\selectfont\@setfontsize\notsotiny{7.5}{9.5}}
\newcommand\superlarge{\@setfontsize\superlarge{24}{27}}
\renewcommand\paragraph{\@startsection{paragraph}{4}{\z@}%
  {1.75ex \@plus .5ex \@minus .2ex}%
  {-1em}%
  {\normalfont\normalsize\bfseries}}
\makeatother

\usepackage{ulem}

\renewenvironment{abstract}
 {\small
  \begin{center}
  \bfseries \abstractname\vspace{-.5em}\vspace{0pt}
  \end{center}
  \list{}{
        \setlength{\leftmargin}{1.4cm}    \setlength{\rightmargin}{\leftmargin}  }  \item\relax}
 {\endlist}
\begin{document}
\sloppy
\title{\vspace*{0.5cm} \superlarge {\textbf{\color{dpd} \textls[-15]{Building Macroeconomically Relevant Climate Indices via the Assemblage VAR}}} \vspace{0.5cm}}
\small
\small
\author{ \hspace{-1.15em} Christophe Barrette \\
	\textbf{\color{black} \texttt{\fontfamily{phv}\selectfont \footnotesize\normalsize \quad \quad \quad Bocconi University} \quad \quad \quad}  \newline \smallskip
	\and Philippe Goulet
	Coulombe\thanks{%
		Contact: \href{mailto:p.gouletcoulombe@gmail.com}{\textcolor{dpd}{\fontfamily{phv}\selectfont p.gouletcoulombe@gmail.com}}. We thank Karin Klieber for helpful discussions. We also thank Mohamed L. A. Guirmeye for research assistance. Replication codes are available \href{https://github.com/timmy-o-toole/Assemblage_VAR}{\color{dpd}here}. This draft: 09-22-26.} \\
	\textbf{\color{black} \texttt{\fontfamily{phv}\selectfont \footnotesize \normalsize \quad \quad \quad \quad \quad Universit\'{e} du Qu\'{e}bec \`{a} Montr\'{e}al} \quad \quad \quad \quad}  \newline   \smallskip
	\and Tim Reinicke \\
	\textbf{\color{black} \texttt{\fontfamily{phv}\selectfont \footnotesize \normalsize \quad \quad \quad \quad \quad \quad \quad \enskip \phantom{..}ETH Zürich} \quad \quad \quad \quad \quad \quad}
}

\date{}

\setstretch{0.98}

\maketitle
\vspace{0.4cm}
\setstretch{1.2}

\begin{abstract}
What should a macroeconomically relevant climate index contain? The composition is inherently ambiguous, and aggregation choices affect structural inference. We introduce the Assemblage VAR, which jointly estimates nonnegative aggregation weights and VAR parameters by maximizing the system likelihood-gain criterion, effectively outsourcing aggregation to observed macroeconomic dynamics. Two variants operate in component-space and rank-space, reweighting named subcomponents or emphasizing regions of the cross-sectional distribution. \textcolor{black}{Applied to disaggregated U.S.\ climate data from the Actuaries Climate Index and NOAA, VARs using the assembled climate measures yield contractionary impulse responses that are substantially larger than those estimated using fixed-weight benchmarks. Weights emphasize high-wind variables and distributional tails over slow-moving components such as sea level.}
\end{abstract}

\thispagestyle{empty}

\clearpage


\clearpage
\setcounter{page}{1}

\newgeometry{left=2 cm, right= 2 cm, top=2.3 cm, bottom=2.3 cm}
\setstretch{1.27} 

\section{Introduction}\label{sec:intro}
What is climate from a macroeconomic perspective? Climate is a high-dimensional phenomenon: temperatures, precipitation, winds, drought indices, sea levels, all tracked across many locations and time scales. When studying the effects of climate on the economy, empirical work typically represents climate with one or a small number of scalar measures (composite indices, temperature or precipitation anomalies, or shock series) and estimates their dynamic effects in VARs, local projections, or related macroeconometric frameworks \citep{deschenes07,DellJonesOlken2012,Kahn2021,Faccia2021,Huber2023ClimatePVAR,ciccarelli2024demand,KimMatthesPhan2025,BilalKaenzig2026Temperature}. This creates a measurement problem, not one of noise, but of relevance. Which dimensions of climate actually matter for macroeconomic fluctuations is far from obvious.

Consider the Actuaries Climate Index (ACI), a composite of six standardized climate variables developed by North American actuarial organizations \citep{ACIAPrimer, ACIHome}; it is increasingly used in climate-macro research \citep{sheng2024time, liao2024extreme, KimMatthesPhan2025}. Sea level rise is first-order for coastal economies but largely irrelevant for the U.S.\ Midwest. Different types of extremes propagate through different channels and at different speeds. Given the input-output linkages and geographic complexity of the economy, it is far from obvious how one would aggregate these disparate phenomena into a single number that captures what matters for the business cycle. What we would like is a \emph{maximally macro-relevant} climate index.

\paragraph{The Assemblage VAR.}
We outsource the aggregation problem to the macroeconomic data. Rather than building a climate index in a first step and then estimating a VAR in a second, we do both at once. 
The Assemblage VAR endogenizes variable measurement within the VAR itself: aggregation weights on disaggregated climate components and VAR parameters are optimized jointly to maximize the system-level likelihood gain over a benchmark. The resulting index has a precise interpretation: it is, by construction, the climate series that is maximally coherent with a specified VAR of the economy. This goes both ways: the assembled variable improves the predictability of macroeconomic outcomes from climate, but also improves the predictability of the climate variable itself from the macro system.

The approach comes in two variants, both building on the assemblage regression framework of \citet{GouletCoulombeEtAl2024}. Component-space assemblage learns how the aggregation weights are allocated across named constituents (e.g., winds versus sea level). Rank-space assemblage learns how the weights are allocated across the cross-sectional distribution, producing a supervised trimming rule analogous to the trimmed-mean inflation \citep{bryan1994measuring}. The key departure from \citet{GouletCoulombeEtAl2024}, who learn core inflation weights by maximizing univariate predictability of future headline inflation, is that the objective here is system-wide: the assembled variable must improve the coherence of the entire VAR, not just forecast a single target.

\paragraph{Empirical results.}
We assemble a Macro-Relevant Climate Index (MRCI) at three levels of disaggregation: national (reweighting five of the six ACI components), regional (34 constituents across seven U.S.\ regions), and a NOAA-based specification (82 climate and atmospheric variables across ten zones).\footnote{Consecutive dry days (\texttt{CDD}) is excluded; see Section~\ref{sec:climate_data}.} Under the ACI benchmark, impulse responses of industrial production to an innovation in the benchmark climate measure are barely distinguishable from zero. In each specification the object of interest is the recursively identified innovation to that specification's own climate measure, which is ordered first in the VAR. The benchmark specification uses the transformed equal-weight analogue of the ACI, while the assembled specifications replace that benchmark with a learned climate aggregate. The picture changes markedly once the climate measure is assembled rather than fixed. The national specification, which reweights the five components, yields a response close to the benchmark's; it is the regional and NOAA specifications, with their richer panels, \textcolor{black}{that yield industrial-production contractions up to twice the response estimated using the benchmark measure and render them statistically significant}. The responses of the other variables are more nuanced, as detailed below. Spatial detail matters: the estimator exploits geographic heterogeneity to assign greater weight to regions and climate types that are more strongly associated with the macro system according to the learning criterion.

The assembled index tracks the ACI through the 1960s--2000s but the two series separate in level afterward. The ACI trends persistently upward, reflecting components that may be less relevant at business-cycle frequencies. The index was designed by actuaries to price insurance risk, so it naturally loads on sea level, relevant for insured coastal property but not for business-cycle transmission. Component-space weights confirm this pattern in the regional specification: high-wind extremes receive about \textcolor{black}{40\%} of the weight while sea level drops to \textcolor{black}{single} digits, consistent with episodic, high-damage events rather than gradual trends. Rank-space weights concentrate on the tails of the cross-sectional distribution, locating the macro-relevant variation in the extremes.

\paragraph{Related literature.}
Several approaches address the tension between high-dimensional disaggregate information and low-dimensional VARs; we position the Assemblage VAR relative to each.

\vspace{0.3em}
\noindent \textit{Climate and macroeconomics.}
A rapidly growing literature studies the macroeconomic effects of climate by embedding low-dimensional \emph{scalar representations} of climate in VARs, local projections, and related macroeconometric frameworks. In practice, these scalars take several forms: selected physical variables and shock series (temperature/weather anomalies and identified innovations) \citep{Faccia2021,LucidiPisaTancioni2024,ciccarelli2024demand_24_EER,BilalKaenzig2026Temperature}, and the shocks built from richer climate information (e.g., drought-exposure measures) \citep{Huber2023ClimatePVAR}. A related strand uses such scalar measures to separate the demand and supply channels through which climate change reaches the macroeconomy \citep{ciccarelli2024demand}.

These scalar objects are typically constructed from expert judgment or equal weighting, unsupervised dimension reduction (such as the cross-country temperature factor \citep{MarottaMumtaz2023} or the climate risk factors \citep{ByrneVitenuSackey2024MacroClimateRisk}) or text- and market-based indices \citep{EngleEtAl2020HedgingClimateNews,Gavriilidis2021CPU,ArdiaEtAl2023MCCC}. The resulting object is not necessarily the one most relevant for macroeconomic dynamics: a latent factor that loads heavily on sea level or slow-moving temperature trends may explain most of the cross-section but carry little information about business-cycle transmission. The Assemblage VAR takes the opposite approach, learning the aggregation that maximizes coherence with a macroeconomic system rather than fidelity to the climate panel itself, applying the train-for-what-you-aim principle of \citet{GouletCoulombeEtAl2024}.

\vspace{0.3em}
\noindent \textit{Disaggregate information in VARs.}
One option, following \citet{BanburaGiannoneReichlin2010}, is to include all disaggregate series directly in a large Bayesian VAR and rely on shrinkage for tractability. This yields impulse responses for every constituent, but not a single coherent index, and averaging those impulse responses does not recover the effect of a shock to a composite variable, because the shock to a portfolio is not a linear combination of isolated shocks. This connects to \citet{BilalKaenzig2026Temperature}, who argue that aggregate climate shocks are more informative for macroeconomic outcomes than granular local ones: what matters is the joint event, not the sum of its parts. A more structured approach, taken by \citet{ChangChenSchorfheide2024} and \citet{MarcellioRenzettiTornese2024}, embeds the entire cross-sectional distribution as a functional object in the VAR; the Assemblage VAR takes a more parsimonious route, reducing the panel to a single scalar but learning that reduction from the data. Conceptually, this is also related to the MIDAS approach \citep{GhyselsSantaClaraValkanov2004}, which parameterizes and learns supervised temporal aggregation to condense high-frequency data at a lower frequency; we address the analogous problem in the cross-sectional dimension. Finally, \citet{HuberMarcellinoScheckel2025} propose coarsened Bayesian VARs that replace the exact likelihood with a robust alternative to correct for model misspecification, addressing the downstream consequences of incorrect aggregation, whereas the Assemblage VAR corrects the aggregation itself.

\vspace{0.3em}
\noindent \textit{Factor-augmented VARs.}
The FAVAR of \citet{BernankeBoivinEliasz2005} is an economical way to compress a large panel and bring its information into a VAR. Its factors, though, are principal components with unrestricted loadings, so the leading factor need not correspond to the combination most relevant for the macroeconomic system, especially when the panel is drawn from an outside source, as our climate data are. One naturally expects the dominant factor to matter, but the two need not coincide. The Assemblage VAR instead builds a single, interpretable index, a non-negative weighted average of named constituents, endogenously within the VAR, so that the measurement step is aligned with the system by construction rather than carried out in a separate first stage as in principal components.

\vspace{0.3em}
\noindent \textit{Reduced-rank VARs.}
The method is also related to the literature on reduced-rank and index-structured VARs \citep{Reinsel1983, VeluReinselWichern1986, Cubadda2025Survey}. When the macroeconomic variables depend on the lagged climate panel only through a single linear combination, the panel enters the macro block at each lag through one shared cross-sectional direction, a per-lag rank-one restriction. The Assemblage VAR induces such a structure, imposing the same index direction $\boldsymbol{w}$ at every lag, while further restricting the loading vector to a non-negative weighted average of named constituents, which makes it closely related to index-augmented autoregressions \citep{CubaddaGuardabascio2019}. Unlike an unconstrained reduced-rank direction, which may combine constituents with both positive and negative weights, the simplex restriction prevents the index from being formed as a contrast of oppositely signed constituents and preserves a transparent interpretation of it as an aggregate measure of climate stress. It also avoids parameterizing the potentially high-dimensional dynamics of the constituent panel itself, whose dimension can become substantial when the panel contains dozens or hundreds of climate indicators. The rank-space variant departs further from this linear literature, replacing the index $\boldsymbol{w}'\mathbf{X}_t$ with a weighted average of order statistics, a nonlinear, threshold-emphasizing measurement.

\vspace{0.3em}
\noindent \textit{Supervised dimension reduction.}
The Assemblage VAR also builds on a methodological tradition of supervised aggregation. A large literature constructs core inflation measures by stripping out volatile components \citep{bryan1994measuring,DolmasKoenig2019}, and \citet{GouletCoulombeEtAl2024} formalize this as a supervised aggregation problem, learning trimming weights that maximize predictability of future headline inflation. The broader supervised dimension reduction literature typically optimizes a single equation: the partial least squares \citep{KellyPruitt2015}, the targeted predictors \citep{BaiNg2008TargetedPredictors}, or the pre-screened factors \citep{BoivinNg2006}. Recent approaches use flexible function approximation to learn nonlinear mappings from disaggregate inputs to macro outcomes \citep{ClarkHuberKoopMarcellino2024,GouletCoulombe2025NPC}. The Assemblage VAR optimizes a \emph{multivariate} objective: the assembled variable must not only predict other variables in the system but must itself be predictable by them, so that the entire VAR is dynamically coherent.

\paragraph{Outline.}
Section \ref{sec:model} formalizes the Assemblage VAR and introduces the system-level objective under interpretability constraints and regularization. Section \ref{sec:app_climate} assembles a macro-relevant climate index from U.S.\ climate data. Section \ref{sec:conclusion} concludes.

\section{Methodology}\label{sec:model}
\setcounter{equation}{0}

This section formalizes the Assemblage VAR in its component- and rank-space variants.

\subsection{Setup and Learning Criterion}

Let $\{\mathbf{y}_t\}_{t=1}^T$ be a $k$-dimensional time series containing the benchmark variable of interest, $x_t$, alongside other macroeconomic variables. The benchmark variable could be a climate index, a price index, or any similar aggregate constructed from a panel of disaggregated constituents. Let $\mathbf{X}_t\in\mathbb{R}^d$ collect the disaggregated constituents of the benchmark variable $x_t$; in our climate application, $\mathbf{X}_t$ contains regional or component-specific climate measures. For weights $\boldsymbol{w}\in\mathbb{R}^d$, the assembled series is
\begin{equation}\label{eq:xstar}
	x_t^*(\boldsymbol{w}) \equiv \boldsymbol{w}'\mathbf{X}_t.
\end{equation}
We form the VAR state by replacing $x_t$ with $x_t^*(\boldsymbol{w})$:
\begin{equation}\label{eq:yomega}
	\mathbf{y}_t(\boldsymbol{w}) \equiv \big(x_t^*(\boldsymbol{w}), \mathbf{y}_t^{(-x)}\,\big)\in\mathbb{R}^{k},
\end{equation}
where $\mathbf{y}_t^{(-x)}$ denotes all variables except $x_t$. 
For each fixed $\boldsymbol{w}$, consider the reduced-form VAR($p$) 
projection of $\mathbf{y}_t(\boldsymbol{w})$:
\begin{equation}\label{eq:var_lag}
    \mathbf{y}_t(\boldsymbol{w}) = \boldsymbol{\nu}(\boldsymbol{w}) + \sum_{j=1}^{p} \mathbf{A}_j(\boldsymbol{w})\,\mathbf{y}_{t-j}(\boldsymbol{w}) + \boldsymbol{u}_t(\boldsymbol{w}),
\end{equation}
where $\boldsymbol{\nu}(\boldsymbol{w})$ is a $k\times1$ intercept, $\mathbf{A}_j(\boldsymbol{w})$ are the $k\times k$ lag-coefficient matrices, and $\boldsymbol{u}_t(\boldsymbol{w})$ is the innovation with covariance $\boldsymbol{\Sigma}(\boldsymbol{w})$. Equation~(\ref{eq:var_lag}) is a reduced-form description, not a structural one: it holds for each $\boldsymbol{w}$ as the projection of $\mathbf{y}_t(\boldsymbol{w})$ on its own past. Changing $\boldsymbol{w}$ changes the assembled series, and with it every moment of $\mathbf{y}_t(\boldsymbol{w})$, so the intercept, the coefficients, the innovation, and its covariance all depend on $\boldsymbol{w}$. The model does not by itself single out a $\boldsymbol{w}$; the objective below does.

For estimation, stack observations $t=p+1,\ldots,T$ into matrices. Let $n=T-p$ denote the effective sample size. Define $\mathbf{Y}(\boldsymbol{w})\in\mathbb{R}^{n\times k}$ as the matrix of dependent observations, with row $t$ equal to $\mathbf{y}_t(\boldsymbol{w})'$. Define the regressor matrix $\mathbf{Z}(\boldsymbol{w})\in\mathbb{R}^{n\times (kp+1)}$, with row $t$ equal to $(1,\,\mathbf{y}_{t-1}(\boldsymbol{w})',\ldots,\mathbf{y}_{t-p}(\boldsymbol{w})')$. Stacking the intercept and lag coefficients as $\mathbf{B}(\boldsymbol{w})=(\boldsymbol{\nu}(\boldsymbol{w})',\,\mathbf{A}_1(\boldsymbol{w})',\ldots,\mathbf{A}_p(\boldsymbol{w})')'\in\mathbb{R}^{(kp+1)\times k}$, the system is
\begin{equation}\label{eq:var1}
	\mathbf{Y}(\boldsymbol{w}) = \mathbf{Z}(\boldsymbol{w})\,\mathbf{B}(\boldsymbol{w}) + \mathbf{U}(\boldsymbol{w}),
\end{equation}
where $\mathbf{U}(\boldsymbol{w})\in\mathbb{R}^{n\times k}$ is the matrix of residuals. For any fixed $\boldsymbol{w}$, OLS applied equation by equation yields $\widehat{\mathbf{B}}_T(\boldsymbol{w})=\big[\mathbf{Z}(\boldsymbol{w})'\mathbf{Z}(\boldsymbol{w})\big]^{-1} \mathbf{Z}(\boldsymbol{w})'\mathbf{Y}(\boldsymbol{w})$, residuals $\widehat{\mathbf{U}}_T(\boldsymbol{w})=\mathbf{Y}(\boldsymbol{w})-\mathbf{Z}(\boldsymbol{w})\widehat{\mathbf{B}}_T(\boldsymbol{w})$, and residual covariance $\widehat{\boldsymbol{\Sigma}}_{1,T}(\boldsymbol{w})=n^{-1}\widehat{\mathbf{U}}_T(\boldsymbol{w})'\widehat{\mathbf{U}}_T(\boldsymbol{w})$.

\vspace{0.2em}
\noindent \textit{Learning criterion.} We choose $\boldsymbol{w}$ to maximize the VAR's log-likelihood gain over a mean-only benchmark, namely an intercept-only model in which each component of $\mathbf{y}_t(\boldsymbol{w})$ is predicted by its sample mean. Maximizing the VAR likelihood alone would reward weights that compress the assembled variable's unconditional variance rather than weights that improve its dynamic fit. Differencing against a mean-only benchmark neutralizes this, isolating the incremental fit from dynamics. The resulting criterion is a multivariate generalization of the $R^2$ improvement used in univariate supervised aggregation.

In the spirit of quasi-maximum likelihood, define for a fixed $\boldsymbol{w}$, the Gaussian quasi-log-likelihood of the VAR projection as
\begin{equation}\label{eq:loglik}
	\ell_T(\mathbf{B},\boldsymbol{\Sigma};\boldsymbol{w}) =
	-\frac{nk}{2}\log(2\pi)
	-\frac{n}{2}\log|\boldsymbol{\Sigma}|
	-\frac{1}{2}
	\sum_{t=p+1}^{T}
	\boldsymbol{u}_t(\mathbf{B};\boldsymbol{w})'
	\boldsymbol{\Sigma}^{-1}
	\boldsymbol{u}_t(\mathbf{B};\boldsymbol{w}),
\end{equation}
where $\boldsymbol{u}_t(\mathbf{B};\boldsymbol{w}) = \mathbf{y}_t(\boldsymbol{w}) - \boldsymbol{\nu}- \sum_{j=1}^{p}\mathbf{A}_j\mathbf{y}_{t-j}(\boldsymbol{w})$. The Gaussian likelihood is used as a scoring rule for the reduced-form projection; the construction does not require the VAR innovations to explicitly be Gaussian. For each fixed $\boldsymbol{w}$, all equations share the same regressor matrix $\mathbf{Z}(\boldsymbol{w})$, so equation-by-equation OLS coincides with the maximizer of \eqref{eq:loglik} with respect to $\mathbf{B}$.

Let $\hat{\ell}_{1,T}(\boldsymbol{w})$ denote the maximized VAR quasi-log-likelihood evaluated at $\widehat{\mathbf{B}}_T(\boldsymbol{w})$ and $\widehat{\boldsymbol{\Sigma}}_{1,T}(\boldsymbol{w})$, and let $\hat{\ell}_{0,T}(\boldsymbol{w})$ denote the maximized quasi-log-likelihood of the mean-only benchmark $\mathbf{y}_t(\boldsymbol{w}) = \boldsymbol{\mu}(\boldsymbol{w}) + \boldsymbol{u}^{(0)}_t(\boldsymbol{w})$ with covariance estimator $\widehat{\boldsymbol{\Sigma}}_{0,T}(\boldsymbol{w})$. The feasible sample learning criterion is
\begin{equation}\label{eq:Score_sample}
	S_T(\boldsymbol{w})
	\equiv
	\hat{\ell}_{1,T}(\boldsymbol{w})
	-
	\hat{\ell}_{0,T}(\boldsymbol{w}).
\end{equation}
With matching deterministic terms, $S_T(\boldsymbol{w})$ is the in-sample likelihood gain of the VAR over the nested mean-only benchmark, so weights are rewarded only insofar as the assembled series improves dynamic fit relative to its own mean-only representation. The criterion is a multivariate counterpart to the predictability criteria of \citet{LoMacKinlay1997} and \citet{GouletCoulombeGobel2023}, extended from a single forecasting equation to the full system.

The population counterpart of the average sample criterion is
$ S(\boldsymbol w) \equiv \ell_1^\star(\boldsymbol w)-\ell_0^\star(\boldsymbol w),$ where \(\ell_1^\star(\boldsymbol w)\) and \(\ell_0^\star(\boldsymbol w)\) are the per-observation Gaussian quasi-log-likelihoods of the VAR and mean-only projections, respectively, evaluated at their population projection coefficients and covariance matrices. Under the relevant law of large numbers, $\frac{1}{n}S_T(\boldsymbol w) \xrightarrow{p} S(\boldsymbol w)$. With matching deterministic terms, $S(\boldsymbol w) = \frac{1}{2} \log\!\left( \frac{|\boldsymbol\Sigma_0^\star(\boldsymbol w)|} {|\boldsymbol\Sigma_1^\star(\boldsymbol w)|} \right), $ so \(S(\boldsymbol w)\) measures the population likelihood gain of the VAR over the mean-only benchmark, equivalently one half of the log generalized-variance ratio. \textcolor{black}{The unpenalized population target is the aggregation that makes the system most predictable, $\boldsymbol{w}^* = \argmin_{\boldsymbol{w}\in\mathcal{W}} -S(\boldsymbol{w})$, where $\mathcal{W}\equiv\{\boldsymbol{w}\in\mathbb{R}^d:\boldsymbol{w}\ge\mathbf{0},\ \mathbf{1}'\boldsymbol{w}=1\}$. It is defined by this predictability property and is not a physically ``true'' climate index.} The feasible estimator regularizes the sample criterion,
\begin{equation}
	\widehat{\boldsymbol{w}}(\lambda) = 
	\argmin_{\boldsymbol{w}\in\mathcal{W}}
	\big\{ -S_T(\boldsymbol{w}) + \lambda\,\operatorname{pen}(\boldsymbol{w}) \big\}, 
\end{equation}
\textcolor{black}{where the form of the penalty reflects the parameterization. In component-space, it shrinks named-component weights towards the equal-weight reference; in rank-space, it smooths the profile across adjacent ranks. The precise forms are given in Equations~\eqref{eq:Qcomp} and~\eqref{eq:Qrank}. Because $S_T(\boldsymbol w)$ is a sum over $n=T-p$ observations, the average sample objective has effective penalty $\eta=\lambda/n$. For any fixed $\eta\geq0$, define the penalized population target
\begin{equation}
    \boldsymbol w^*(\eta)=\argmin_{\boldsymbol w\in\mathcal W}\big\{-S(\boldsymbol w)+\eta\,\operatorname{pen}(\boldsymbol w)\big\}.
\end{equation}
The unpenalized target is $\boldsymbol w^*(0)=\boldsymbol w^*$. For each specification, $\widehat\lambda$ is selected by block cross-validation and the reported estimate uses $\widehat\eta=\widehat\lambda/n$. Conditional on that selected value, $\widehat{\boldsymbol w}(\widehat\lambda)$ is interpreted relative to $\boldsymbol w^*(\widehat\eta)$. We do not assume that the penalty vanishes.}

\textcolor{black}{Because $\widehat{\boldsymbol w}(\widehat\lambda)$ is estimated from a finite sample, it carries genuine sampling uncertainty:} it is governed by the number of time-series observations $T$ at fixed cross-sectional dimension $d$, and is not driven to zero by letting $d\to\infty$ as in factor analysis. Section~\ref{down} and Appendix~\ref{sec:emp_climate_boot} take this uncertainty into account.

\subsection{Component-Space Assemblage}

The first variant answers a natural question: which components are relevant? For a climate index, it is far from obvious whether winds, precipitation, temperature extremes, or sea level matter most for macroeconomic transmission, and this likely depends on the country's geography, industrial structure, and exposure to specific hazards. Component-space assemblage lets the VAR itself resolve this ambiguity by learning basket weights over the original components.

When constituents are numerous, overlapping, or weakly informative, the map $\boldsymbol{w}\mapsto S_T(\boldsymbol{w})$ can be flat or locally irregular, producing unstable finite-sample solutions. Following \citet{GouletCoulombeEtAl2024}, we impose two forms of discipline. First, simplex and nonnegativity constraints keep $x_t^*(\boldsymbol{w})$ in the benchmark variable units, ensuring the assembled variable remains a proper weighted average of its constituents. Second, ridge-type regularization shrinks towards a reference basket $\boldsymbol{w}^{\text{ref}}$, preventing the optimizer from concentrating weight on one or two idiosyncratic series and forcing the solution to remain a reasonably dense mixture.

Component-space assemblage learns basket weights directly over the disaggregated components $\mathbf{X}_t$:
\begin{equation}\label{eq:Qcomp}
	\widehat{\boldsymbol{w}}_c(\lambda) = \argmin_{\boldsymbol{w}} \; -S_T(\boldsymbol{w}) +  \lambda\|\boldsymbol{w} - \boldsymbol{w}^{\text{ref}}\|_2^2
	\quad\text{s.t.}\quad \boldsymbol{w} \ge \mathbf{0},\; \mathbf{1}' \boldsymbol{w} = 1,
\end{equation}
where the ridge penalty shrinks towards reference weights $\boldsymbol{w}^{\text{ref}}$. In our climate application, $\boldsymbol{w}^{\text{ref}}$ corresponds to the equal weights of the established ACI benchmark. As $\lambda\to\infty$, the solution approaches $\boldsymbol{w}^{\text{ref}}$ and the standard VAR with official measurement. As $\lambda\to 0$, weights maximize system coherence subject only to simplex restrictions.

\textcolor{black}{Equal weights therefore serve as the shrinkage reference in component-space and the empirical comparator in the figures. They do not serve as the second term in the learning criterion, because both likelihoods must be evaluated on the same candidate index $x_t^*(\boldsymbol w)$. Replacing the mean-only likelihood with one evaluated on the equal-weight index would make the benchmark independent of $\boldsymbol w$ and leave the variance-compression incentive intact.}

\subsection{Rank-Space Assemblage}
The second variant addresses a different concern: all constituents may be relevant, but what matters is \emph{when}, specifically, where their realizations fall in the cross-sectional distribution at a given date. Rank-space assemblage reweights the cross-sectional \emph{distribution} of the panel rather than its named components, with the weights chosen to maximize coherence with the VAR's joint dynamics rather than fixed in advance. The construction is deliberately flexible: depending on what the data favor, it can recover a tail-risk measure, an outlier-robust measure of the kind used in trimmed-mean inflation \citep{bryan1994measuring}, or emphasis on a single tail, with the scheme learned rather than imposed. For climate the logic points to the tails, since moderate deviations are absorbed by the economy while extreme realizations carry threshold damages; and because weights attach to ranks rather than names, the composition of the assembled index varies over time, loading on whichever constituent occupies the relevant region of the distribution.

Formally, rank-space assemblage operates on order statistics rather than individual components \citep{GouletCoulombeEtAl2024}. At each date $t$, we sort the $d$ constituent values from lowest to highest and collect them as $\mathbf{O}_t \equiv \text{sort}(\mathbf{X}_t) \in \mathbb{R}^d$. The assembled variable is then a weighted average of these order statistics,
\begin{equation}\label{eq:xstar_rank}
	x_t^*(\boldsymbol{w}) = \boldsymbol{w}' \mathbf{O}_t,
\end{equation}
and the VAR state $\mathbf{y}_t(\boldsymbol{w})$ is formed as before by replacing $x_t$ with $x_t^*(\boldsymbol{w})$. The key difference from the component-space formulation is that the $j$-th weight now attaches to the $j$-th \emph{rank} in the cross-sectional distribution, not to a fixed named constituent. This makes the assemblage invariant to component relabeling and instead learns which regions of the cross-sectional distribution carry the most macro-relevant information.

The sample criterion $S_T(\boldsymbol{w})$ in \eqref{eq:Score_sample} is evaluated on this rank-space assembled variable, $x_t^*(\boldsymbol{w}) = \boldsymbol{w}' \mathbf{O}_t$. We solve
\begin{equation}\label{eq:Qrank}
	\widehat{\boldsymbol{w}}_r(\lambda) = \argmin_{\boldsymbol{w}} \; -S_T(\boldsymbol{w}) +  \lambda\|\mathbf{D}\boldsymbol{w}\|_2^2
	\quad\text{s.t.}\quad \boldsymbol{w} \ge \mathbf{0},\; \mathbf{1}' \boldsymbol{w} = 1,
\end{equation}
where $\mathbf{D}$ is the first-difference operator. In general one would anchor the assembled variable's training-sample mean to that of the benchmark, $n^{-1}\sum_{t=p+1}^{T}\boldsymbol{w}'\mathbf{O}_t =n^{-1}$ $\sum_{t=p+1}^{T}x_t$; because the constituents are standardized to zero mean in our application, that restriction holds for every $\boldsymbol{w}$ on the simplex, and the normalization $\mathbf{1}'\boldsymbol{w}=1$ anchors the weights instead. Both restrictions are linear in $\boldsymbol{w}$, since $\mathbf{O}_t$ does not depend on the weights. The fused ridge penalty, analogous in spirit to \citet{TibshiraniEtAl2005}, enforces smooth weight profiles across adjacent ranks, producing interpretable trimming schemes. A flat weight profile recovers the cross-sectional mean, which coincides exactly with the benchmark series when the latter uses equal weights. Concentrated weights at the tails then recover a tail-risk measure, and weights near the median a robust, median-based one. The non-negativity constraint $\boldsymbol{w}\ge\mathbf{0}$ keeps the object an index in every case: a weighted average of order statistics is a level in the units of the benchmark, whereas signed weights could place positive mass on high ranks and negative mass on low ones, yielding a dispersion or volatility measure rather than a level of climate stress. The sort-and-weight construction makes $x_t^*(\boldsymbol{w}) = \boldsymbol{w}'\mathbf{O}_t$ a nonlinear, threshold-emphasizing function of the panel $\mathbf{X}_t$, even though the VAR that follows is linear in the constructed index; the claim that extremes matter is thus about which realizations enter the index, not that the system responds convexly to it, a distinction we revisit when reading the impulse responses (Section~\ref{sec:climate_results}).

\subsection{Optimization, Stability, and Computation}

The constrained optimization in \eqref{eq:Qcomp}--\eqref{eq:Qrank} is solved \textit{via} Sequential Quadratic Programming (SQP) using the \texttt{solnl} function from the \textbf{NlcOptim} R package \citep{chen2019nlcoptim}. Nonnegativity is imposed \textit{via} lower-bound arguments, the simplex constraint through a linear equality, and, where the mean-targeting restriction binds, through a second linear equality. Benchmark weights serve as initial values (equal weights otherwise), providing a warm start near the feasible region.

The regularization parameter $\lambda$ is selected by non-overlapping block cross-validation to respect temporal dependence. The sample is divided into 5 contiguous blocks; at each fold $m$, one block is held out, the weights $\widehat{\boldsymbol{w}}^{(-m)}(\lambda)$ are estimated on the remaining 4 blocks, and a fold-specific held-out criterion $S_{T_m}^{(m)}(\widehat{\boldsymbol{w}}^{(-m)}(\lambda))$ is computed using a VAR estimated on the held-out block alone. The selected value, $\widehat\lambda$, maximizes the average held-out criterion, ensuring that the tuning criterion is evaluated on data not used to estimate the weights. \textcolor{black}{\label{sec:penalty_selection}The specification-specific value $\widehat\lambda$ is then held fixed in all bootstrap replications. Throughout, inference ``conditional on $\widehat\lambda$'' refers to this fixed-tuning convention: cross-validation is not repeated, so the reported intervals exclude penalty-selection uncertainty.}

Because $S_T(\boldsymbol{w})$ is non-convex, SQP may in principle converge to a local optimum. We verified stability by initializing from random starting points drawn from the simplex. Across all runs, the recovered weight solutions were nearly identical, suggesting that the compactness of the feasible set, the ridge penalty, and the warm start at economically motivated benchmark weights produce a well-behaved objective landscape. The current implementation is well-suited for the application considered here ($d$ on the order of tens to low hundreds of constituents); scaling to higher-dimensional panels would benefit from gradient-based optimization with automatic differentiation.

\subsection{Downstream Use and Inference}\label{down}

\textcolor{black}{Conditional on the cross-validated penalty $\widehat\lambda$, the reported estimator is $\widehat{\boldsymbol w}(\widehat\lambda)$, the solution to the regularized constrained M-estimation problem defined over the unit simplex.} Because the parameter space is compact and the regularized sample objective is continuous in $\boldsymbol{w}$ whenever the relevant VAR covariance matrices are nonsingular, a solution exists. The $\ell_2$ (or fused $\ell_2$) regularization discourages extreme allocations and improves numerical stability, though the objective remains non-convex; the simplex constraint and penalization together bound the estimator and mitigate boundary degeneracy in finite samples.

\textcolor{black}{Given $\widehat{\boldsymbol{w}}(\widehat\lambda)$ from either approach, we construct $x_t^* = x_t^*(\widehat{\boldsymbol{w}}(\widehat\lambda))$ and estimate the VAR on $\{\mathbf{y}_t(\widehat{\boldsymbol{w}}(\widehat\lambda))\}_{t=1}^T$.} Standard impulse responses and forecast error variance decompositions follow under conventional identification schemes.

\noindent\textit{Interpreting a shock to the index.} A one-standard-deviation innovation to $x_t^*$ is a shock to the assembled \emph{index}, not to any single constituent, so its content is composite. \textcolor{black}{Holding $\boldsymbol{w}$ fixed gives the index a stable definition, a fixed mapping from the component panel to the scalar measure, but does not imply a stable physical composition of each realized innovation. The same is true of a shock to headline inflation or to industrial production, each of which is itself an aggregate. In rank-space, the weights are fixed over rank positions, while the named component occupying a given rank may change over time.} But that composite is the point: rather than presuming that a single physical variable carries the macro-relevant climate impulse, the Assemblage VAR estimates the combination of climate phenomena with which the macroeconomy is most coherent, and the estimated weight profiles make that combination legible. Working with such an index, rather than tracing each constituent separately, is an advantage: it pools common variation and averages out idiosyncratic noise in the individual series.

The index is also not redundant with entering the constituents one by one. The orthogonalized innovation to a composite variable is not generally the same object as a weighted sum of orthogonalized innovations to its constituents, so the response to the assembled-index shock need \emph{not} equal a weighted average of component-specific responses. A researcher who wants the response to one specific phenomenon can include that constituent directly; the assembled index is complementary, characterizing the macro-relevant combination rather than any single channel.

Appendix~\ref{sec:emp_climate_boot} reports a complementary \textcolor{black}{weight-reestimating pairs bootstrap} that re-estimates the weights and VAR parameters at the selected penalty. Its conclusions are broadly similar to those obtained from the main-text fixed-weights residual bootstrap. Because the two procedures resample different objects and use different interval constructions, we interpret this similarity as a robustness finding rather than as a clean decomposition of the contribution of weight-estimation uncertainty. Because $\widehat{\boldsymbol{w}}$ is a constrained system $M$-estimator at fixed, moderate $d$, its sampling uncertainty does not vanish in the cross-section, so it must be quantified rather than assumed away.

\section{What is Climate from a Macroeconomic Perspective?}\label{sec:app_climate}
\setcounter{equation}{0}

We assemble a Macro-Relevant Climate Index (MRCI) from U.S.\ climate data and embed it in a structural VAR alongside standard macroeconomic variables. The goal is to construct a climate measure tailored to U.S.\ business-cycle transmission rather than designed for global monitoring.

\subsection{Data and VAR Setup}\label{sec:climate_data}

We describe the constituent panel from which the assembled climate index is constructed, then specify the VAR system that embeds it.

\paragraph{The constituent panel $\mathbf{X}_t$.}
We use the U.S.\ portion of the Actuaries Climate Index (ACI) \citep{ACIHome,ACIAPrimer}, which aggregates six standardized components: high-temperature extremes (\texttt{T90}), low-temperature extremes (\texttt{T10}), heavy precipitation (\texttt{Rx5Day}), consecutive dry days (\texttt{CDD}), high winds (\texttt{WP90}), and sea level. The published ACI applies five-year smoothing to track slow-moving trends, well suited for monitoring but potentially misaligned with business-cycle transmission analysis. Moreover, the index is standardized relative to 1961--1990, thus inheriting pronounced drift that dominates VAR dynamics and contaminates the structural innovation.\footnote{The 30-year normalization window follows standard WMO convention for defining climate normals. The components are also transformed so that higher values correspond to more adverse climate stress. Even the unsmoothed ACI yields rising industrial production after a climate shock, consistent with low-frequency drift contaminating the structural innovation; results available upon request.}

We construct an event-oriented analogue: for each component $X_{j,t}$, we compute month-on-month changes and standardize using the 1961--1990 moments. This preserves the ACI's normalization logic while aligning time-series properties with stationary SVAR analysis. The baseline SVAR uses the same transformed components to construct the benchmark series $x_t$ using equal weights.\footnote{We exclude consecutive dry days (\texttt{CDD}), which is reported annually and interpolated to monthly frequency, and would tilt the likelihood-based optimization towards exploiting this mechanical predictability.} The constituent panel $\mathbf{X}_t$, from which the assembled variable $x_t^*(\boldsymbol{w}) = \boldsymbol{w}'\mathbf{X}_t$ is constructed, is stratified into three specifications of increasing granularity:

\vspace{0.2em}
\noindent\textit{\textsf{National}} (\asn) uses 5 subcomponents corresponding directly to the ACI's national-level climate variables, so that the assemblage simply reweights the five ACI components. 

\vspace{0.2em}
\noindent\textit{\textsf{Regional}} (\asr) uses 34 subcomponents obtained from the ACI's regional disaggregation, allowing the assemblage to exploit spatial heterogeneity by placing more weight on regions whose climate measures are more strongly associated with the macro system.\footnote{The Midwest region does not have a sea level variable.}

\vspace{0.2em}
\noindent\textit{\textsf{NOAA}} (\aso) uses 82 subcomponents drawn from the NOAA dataset containing weather and atmospheric measurements across ten U.S.\ regional zones, including Temperature, Precipitation, Heating Degree Days, Cooling Degree Days, and four Palmer drought indices (PDSI, PHDI, PMDI, and Palmer Z-Index), as well as global atmospheric CO$_2$ concentrations recorded at the Mauna Loa observatory in Hawaii; this is the most granular specification, providing the assemblage estimator with a rich panel of climate series.

\paragraph{The remaining VAR variables $\mathbf{y}_t^{(-x)}$.} The assembled climate variable enters a VAR alongside six macroeconomic series, forming the state vector $\mathbf{y}_t(\boldsymbol{w})$. The variables, in Cholesky ordering, are: the climate index (ACI or its assembled counterpart), industrial production (\mn{INDPRO}), PCE inflation (\mn{PCEPI}), unemployment rate (\mn{UNRATE}), the shadow rate of \citet{WuXia2016} (\mn{SHADOWRATE}), the U.S.\ dollar/pound sterling exchange rate (\mn{EXUSUKx}), and housing starts (\mn{HOUST}). The macroeconomic series are taken from the FRED-MD database, except the shadow rate \citep{WuXia2016}. \mn{INDPRO}, \mn{PCEPI}, \mn{EXUSUKx}, and \mn{HOUST} are transformed by taking log-differences; \mn{UNRATE} is first-differenced, and \mn{SHADOWRATE} is kept in levels.

The climate index is ordered first, reflecting contemporaneous exogeneity of weather to the macroeconomy, the same Cholesky ordering used by \citet{KimMatthesPhan2025} in their ACI-based VAR. The lag length is set to $p=12$. \textcolor{black}{Structural shocks are identified recursively under this ordering. Across specifications, the recursive ordering is unchanged, but the climate variable entering the VAR differs. A one-standard-deviation innovation is therefore defined separately for each climate measure. The impulse responses compare responses to innovations in alternative climate measures, not responses to a common climate shock. They are reported in cumulative form, except for the interest rate.}

Confidence bands are reported at the 68\% level, following the convention adopted by \citet{Faccia2021} and \citet{KimMatthesPhan2025} in climate--macroeconomics.\footnote{\textcolor{black}{The main-text bands are pointwise equal-tailed percentile-$t$ intervals obtained from a fixed-weights residual bootstrap \citep{Kunsch1989,Hall1992,BrueggemannJentschTrenkler2016}. Under the fixed-tuning convention of Section~\ref{sec:penalty_selection}, both the selected penalty $\widehat\lambda$ and the resulting weights $\widehat{\boldsymbol w}(\widehat\lambda)$ are held fixed across bootstrap replications. Appendix~\ref{sec:fixed_weights_boot} details this procedure. Appendix~\ref{sec:emp_climate_boot} reports complementary bands from a weight-reestimating pairs bootstrap that holds $\widehat\lambda$ fixed but re-estimates the weights and VAR parameters jointly in each replication.}}

\subsection{Results}\label{sec:climate_results}

We first examine what the optimized measurement implies for structural impulse responses, then inspect the resulting time series and weight profiles.

\begin{figure}[t!]
	\caption{{\textcolor{black}{Component-Space Climate Measure Shocks}}} \label{fig:Caci_irf}
	\begin{center}
		
		\begin{subfigure}[t]{.344\textwidth}
			\vspace*{-0.4cm}
			\captionsetup{margin={.40cm,0cm}} 
			\centering\caption{National}
			\vspace*{-0.18cm}
			\hspace*{-1mm}\includegraphics[width=1.01\textwidth,trim = 17mm 166mm 11.5mm 47mm, clip]{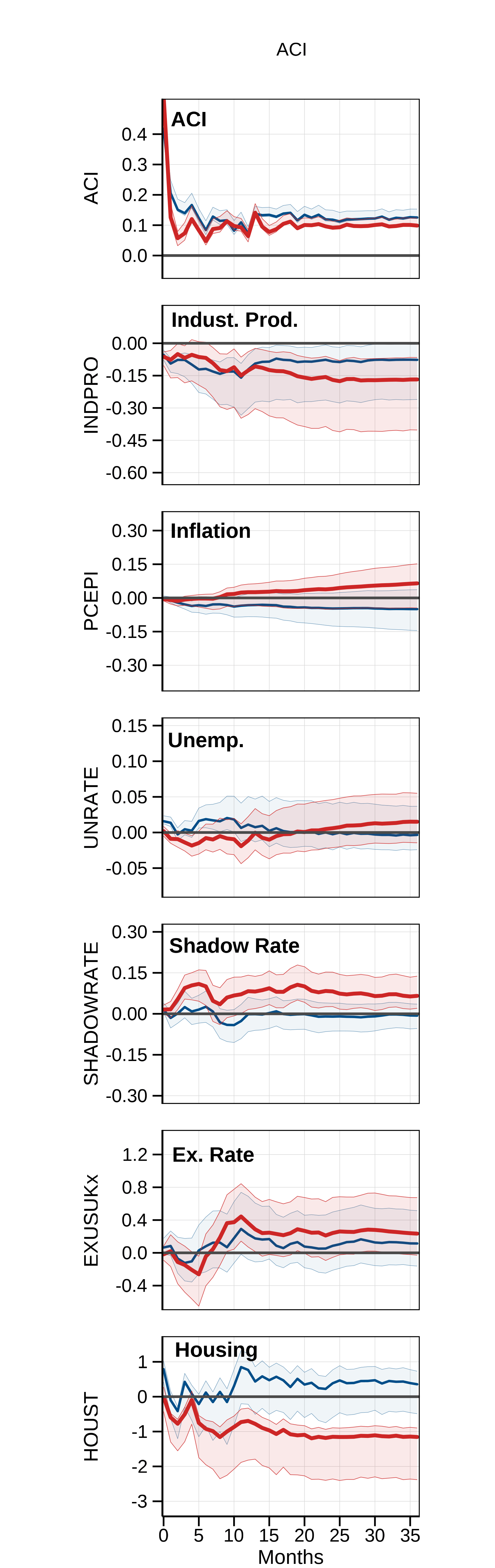}
		\end{subfigure}
		\hspace*{-4.25mm}
		\begin{subfigure}[t]{.344\textwidth}
			\vspace*{-0.4cm}
			\captionsetup{margin={-0.3cm,0cm}} 
			\centering\caption{Regional}
			\vspace*{-0.18cm}
			\hspace*{2.8mm}\includegraphics[width=.89\textwidth, trim = 24.5mm 166mm 9.5mm 47mm, clip]{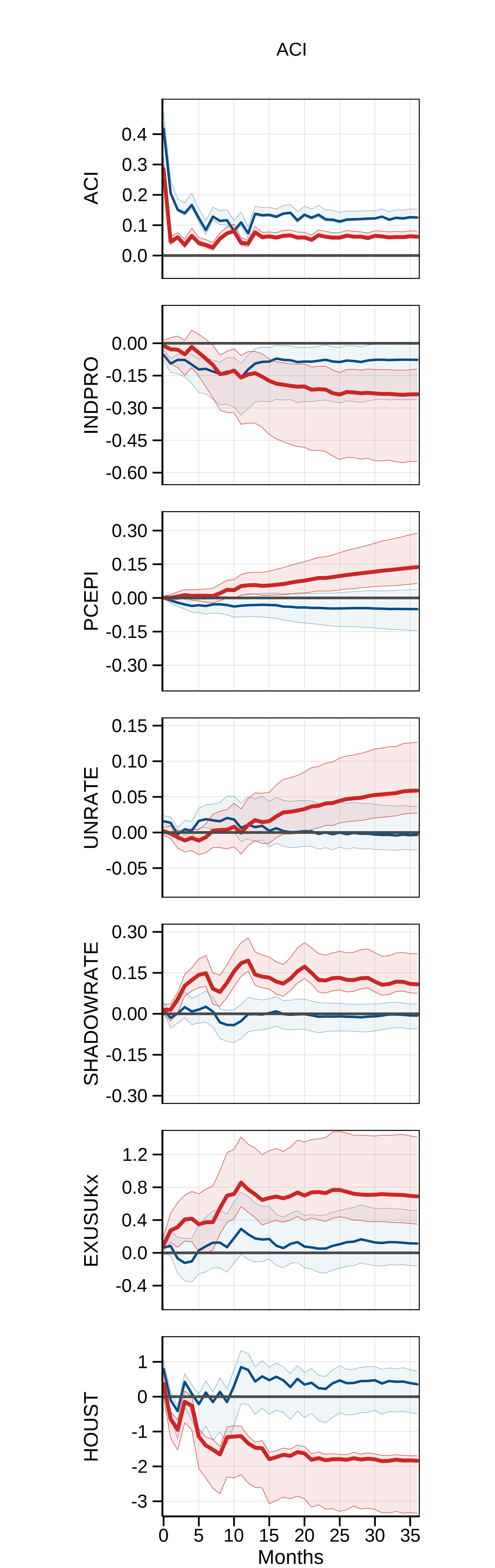}
		\end{subfigure}%
		\hspace*{-7mm}
		\begin{subfigure}[t]{.344\textwidth}
			\vspace*{-0.4cm}
			\captionsetup{margin={-0.3cm,0cm}} 
			\centering\caption{NOAA}
			\vspace*{-0.18cm}
			\hspace*{4.25mm}\includegraphics[width=.89\textwidth, trim = 24.5mm 166mm 9.5mm 47mm, clip]{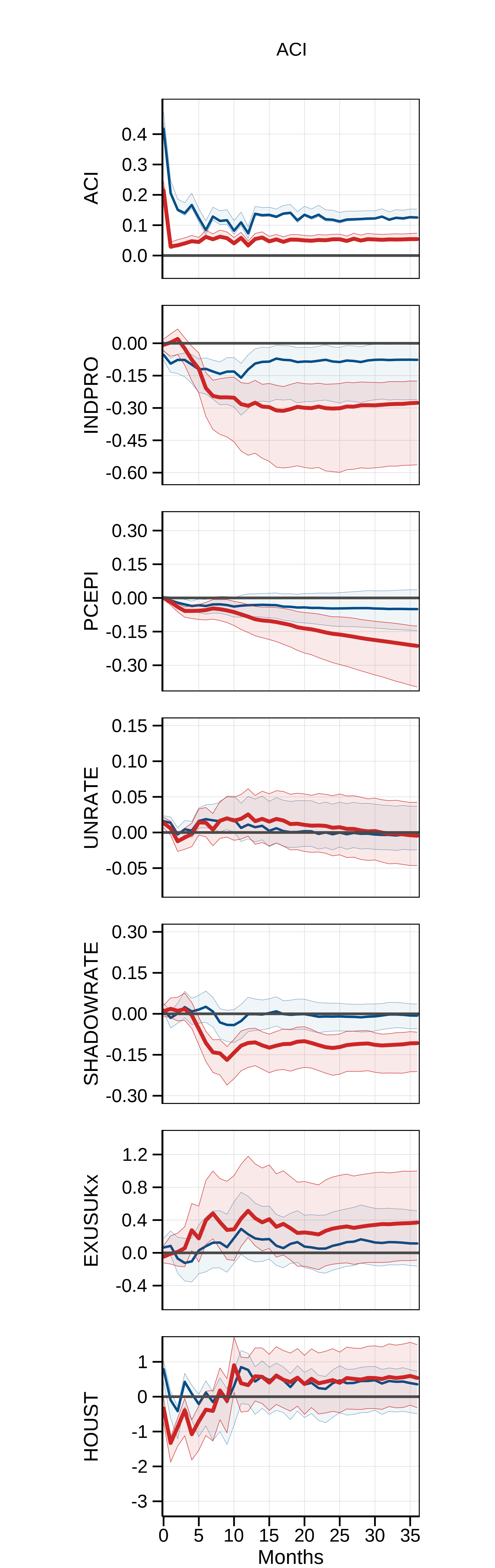}
		\end{subfigure}%
		
		\begin{subfigure}[t]{.344\textwidth}
			\vspace*{0.18cm}
			\hspace*{-1.5mm}\includegraphics[width=1.02\textwidth,trim = 16.5mm 102.9mm 11.5mm 110mm, clip]{figure/ACI_Comp_lvl_1_shockACI_Sfull_RUN0814.1049_v2026-08-14.png}
		\end{subfigure}
		\begin{subfigure}[t]{.344\textwidth}
			\vspace*{.16cm}
			\hspace*{0.6mm}\includegraphics[width=.89\textwidth, trim = 24.5mm 103mm 9.5mm 110mm, clip]{figure/ACI_Comp_lvl_2_shockACI_Sfull_RUN0814.0906_v2026-08-14.png}
		\end{subfigure}%
		\hspace*{-10.5mm}
		\begin{subfigure}[t]{.344\textwidth}
			\vspace*{.16cm}
			\hspace*{5.1mm}\includegraphics[width=.89\textwidth, trim = 24.5mm 103mm 9.5mm 110mm, clip]{figure/ACI_Comp_lvl_3_shockACI_Sfull_RUN0814.0903_v2026-08-14.png}
		\end{subfigure}%
		
		\begin{subfigure}[t]{.344\textwidth}
			\vspace*{-0.075cm}
			\hspace*{-.8mm}\includegraphics[width=1.01\textwidth,trim = 17mm 3mm 11.5mm 202.3mm, clip]{figure/ACI_Comp_lvl_1_shockACI_Sfull_RUN0814.1049_v2026-08-14.png}
		\end{subfigure}
		\begin{subfigure}[t]{.344\textwidth}
			\vspace*{-.09cm}
			\hspace*{.6mm}\includegraphics[width=.89\textwidth, trim = 24.5mm 3mm 9.5mm 202mm, clip]{figure/ACI_Comp_lvl_2_shockACI_Sfull_RUN0814.0906_v2026-08-14.png}
		\end{subfigure}%
		\hspace*{-10.5mm}
		\begin{subfigure}[t]{.344\textwidth}
			\vspace*{-.09cm}
			\hspace*{5.1mm}\includegraphics[width=.89\textwidth, trim = 24.5mm 3mm 9.5mm 202mm, clip]{figure/ACI_Comp_lvl_3_shockACI_Sfull_RUN0814.0903_v2026-08-14.png}
		\end{subfigure}%
		
		\begin{subfigure}[t]{.38\textwidth}
			\vspace*{-0.08cm}
			\hspace*{-1mm}\includegraphics[width=1.05\textwidth, trim = 13mm 3.8mm 5.5mm 235mm, clip]{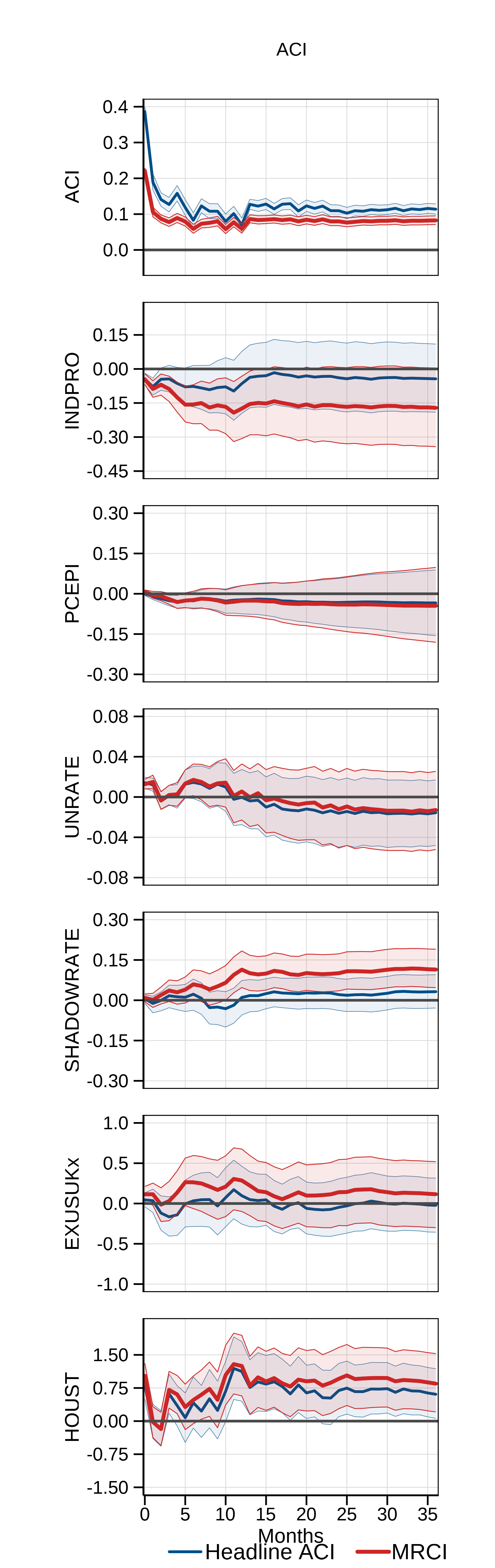}
		\end{subfigure}%
		
		\begin{subfigure}[t]{.38\textwidth}
			\vspace*{-0.07cm}
			\hspace*{-4mm}\includegraphics[width=1.05\textwidth, trim = 13mm 0mm 5.5mm 237.5mm, clip]{figure/Naming_ACI.png}
		\end{subfigure}%
	\end{center}
	
	\begin{threeparttable}
		\centering \vspace*{-0.6cm}
		\begin{minipage}{\textwidth}
			\begin{tablenotes}[para,flushleft]
				\setlength{\lineskip}{0.2ex}
				\notsotiny 
				{\textit{Notes}: Results are based on VAR models estimated using data from 1975:M1 to 2025:M4, excluding the COVID-19 period (2020:M1--2021:M6). All specifications include 12 lags and maintain a consistent variable ordering: ACI/MRCI, \mn{INDPRO}, \mn{PCEPI}, \mn{UNRATE}, \mn{SHADOWRATE}, \mn{EXUSUKx}, and \mn{HOUST}. Confidence intervals at the 68\% level, conditional on $\widehat\lambda$ and $\widehat{\boldsymbol w}(\widehat\lambda)$. For brevity, we display impulse responses for selected variables only; complete results are available in Appendix~\ref{annexe_climate} (Figure~\ref{fig:FULL_Caci_irf}). For the \textcolor{black}{weight-reestimating pairs bootstrap version}, with the aggregation weights re-estimated in every replication, see Figure~\ref{fig:Caci_boot_irf} in Appendix~\ref{sec:weight_uncertainty}.}\vspace{-.3cm}
			\end{tablenotes}
		\end{minipage}
	\end{threeparttable}
\end{figure}

\begin{figure}[t!]
	\caption{{\textcolor{black}{Rank-Space Climate Measure Shocks}}} \label{fig:Raci_irf}
	\begin{center}
		
		\begin{subfigure}[t]{.344\textwidth}
			\vspace*{-0.4cm}
			\captionsetup{margin={.40cm,0cm}} 
			\centering\caption{National}
			\vspace*{-0.18cm}
			\hspace*{-1mm}\includegraphics[width=1.01\textwidth,trim = 17mm 166mm 11.5mm 47mm, clip]{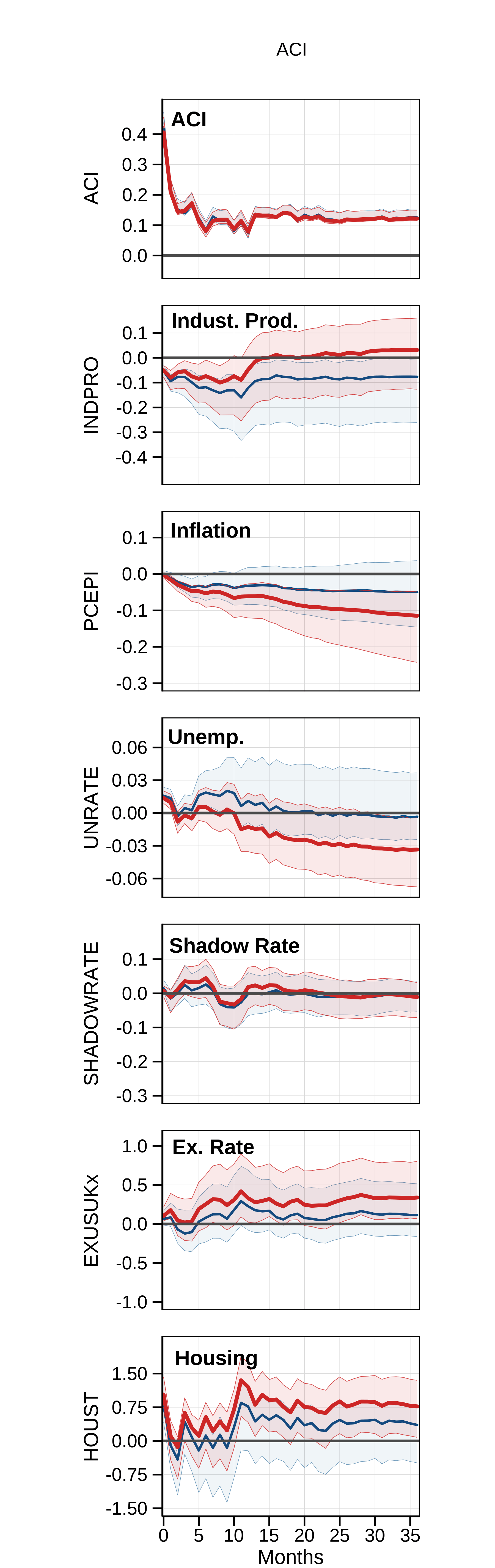}
		\end{subfigure}
		\hspace*{-4.25mm}
		\begin{subfigure}[t]{.344\textwidth}
			\vspace*{-0.4cm}
			\captionsetup{margin={-0.3cm,0cm}} 
			\centering\caption{Regional}
			\vspace*{-0.18cm}
			\hspace*{2.8mm}\includegraphics[width=.89\textwidth, trim = 24.5mm 166mm 9.5mm 47mm, clip]{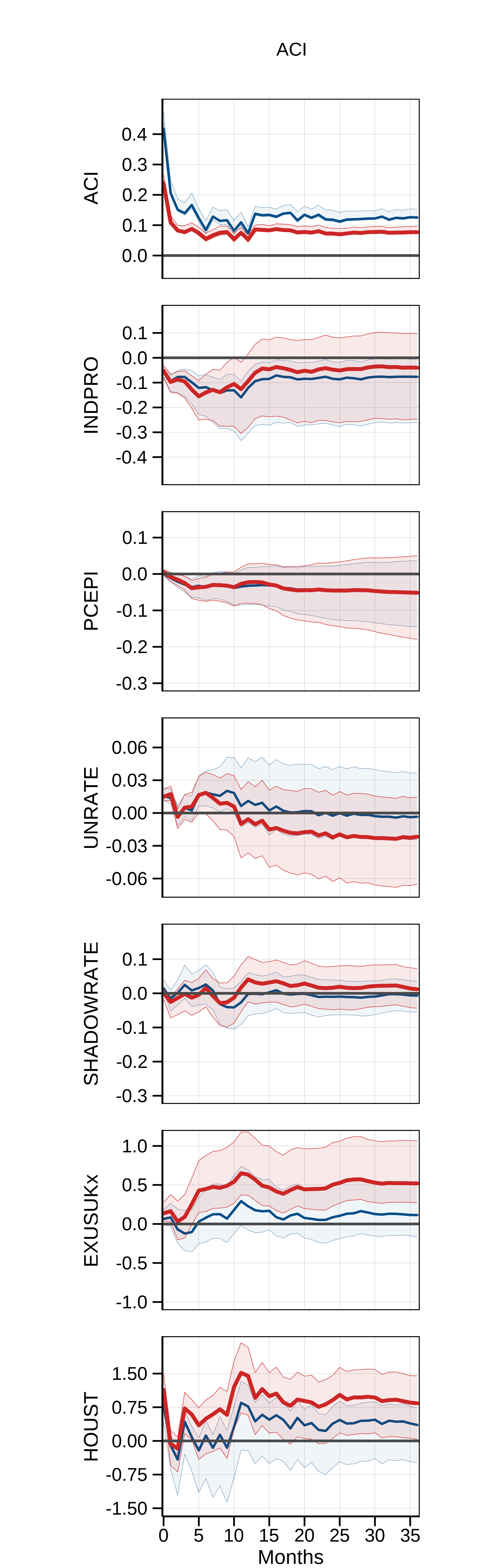}
		\end{subfigure}%
		\hspace*{-7mm}
		\begin{subfigure}[t]{.344\textwidth}
			\vspace*{-0.4cm}
			\captionsetup{margin={-0.3cm,0cm}} 
			\centering\caption{NOAA}
			\vspace*{-0.18cm}
			\hspace*{4.25mm}\includegraphics[width=.89\textwidth, trim = 24.5mm 166mm 9.5mm 47mm, clip]{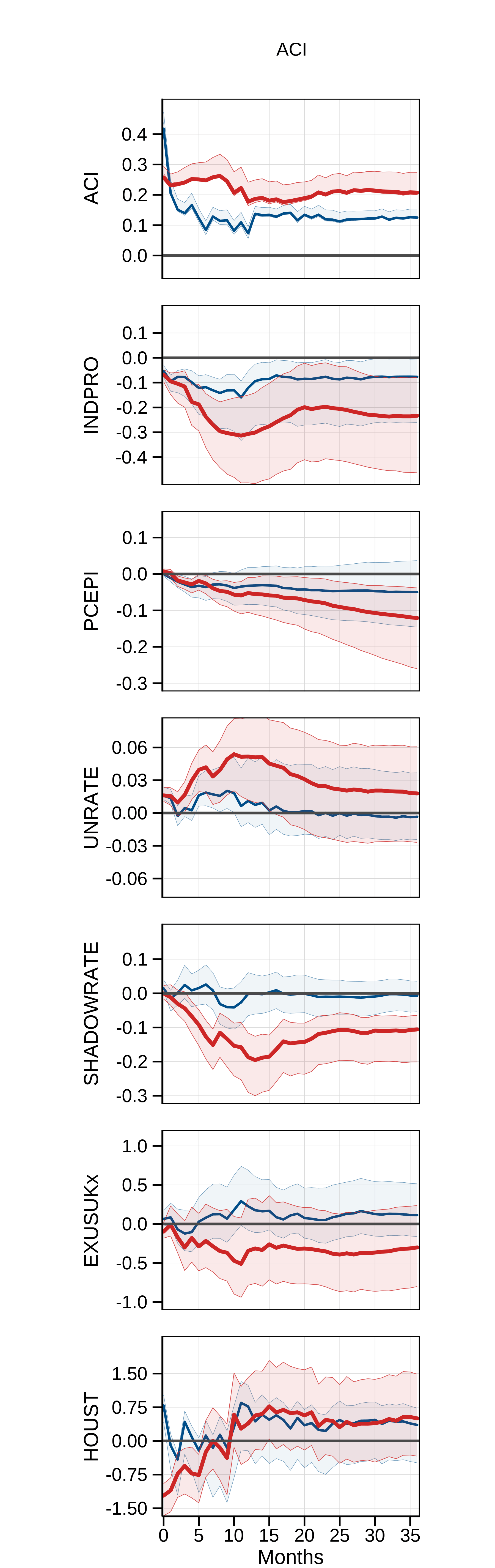}
		\end{subfigure}%
		
		\begin{subfigure}[t]{.344\textwidth}
			\vspace*{0.18cm}
			\hspace*{-1.5mm}\includegraphics[width=1.02\textwidth,trim = 16.5mm 102.9mm 11.5mm 110mm, clip]{figure/ACI_Rank_lvl_1_shockACI_Sfull_RUN0814.1043_v2026-08-19.png}
		\end{subfigure}
		\begin{subfigure}[t]{.344\textwidth}
			\vspace*{.16cm}
			\hspace*{0.6mm}\includegraphics[width=.89\textwidth, trim = 24.5mm 103mm 9.5mm 110mm, clip]{figure/ACI_Rank_lvl_2_shockACI_Sfull_RUN0814.0959_v2026-08-19.png}
		\end{subfigure}%
		\hspace*{-10.5mm}
		\begin{subfigure}[t]{.344\textwidth}
			\vspace*{.16cm}
			\hspace*{5.1mm}\includegraphics[width=.89\textwidth, trim = 24.5mm 103mm 9.5mm 110mm, clip]{figure/ACI_Rank_lvl_3_shockACI_Sfull_RUN0814.1010_v2026-08-19.png}
		\end{subfigure}%
		
		\begin{subfigure}[t]{.344\textwidth}
			\vspace*{-0.075cm}
			\hspace*{-.8mm}\includegraphics[width=1.01\textwidth,trim = 17mm 3mm 11.5mm 202.3mm, clip]{figure/ACI_Rank_lvl_1_shockACI_Sfull_RUN0814.1043_v2026-08-19.png}
		\end{subfigure}
		\begin{subfigure}[t]{.344\textwidth}
			\vspace*{-.09cm}
			\hspace*{.6mm}\includegraphics[width=.89\textwidth, trim = 24.5mm 3mm 9.5mm 202mm, clip]{figure/ACI_Rank_lvl_2_shockACI_Sfull_RUN0814.0959_v2026-08-19.png}
		\end{subfigure}%
		\hspace*{-10.5mm}
		\begin{subfigure}[t]{.344\textwidth}
			\vspace*{-.09cm}
			\hspace*{5.1mm}\includegraphics[width=.89\textwidth, trim = 24.5mm 3mm 9.5mm 202mm, clip]{figure/ACI_Rank_lvl_3_shockACI_Sfull_RUN0814.1010_v2026-08-19.png}
		\end{subfigure}%
		
		\begin{subfigure}[t]{.38\textwidth}
			\vspace*{-0.08cm}
			\hspace*{-1mm}\includegraphics[width=1.05\textwidth, trim = 13mm 3.8mm 5.5mm 235mm, clip]{figure/Naming_ACI.png}
		\end{subfigure}%
		
		\begin{subfigure}[t]{.38\textwidth}
			\vspace*{-0.07cm}
			\hspace*{-4mm}\includegraphics[width=1.05\textwidth, trim = 13mm 0mm 5.5mm 237.5mm, clip]{figure/Naming_ACI.png}
		\end{subfigure}%
	\end{center}
	
	\begin{threeparttable}
		\centering \vspace*{-0.6cm}
		\begin{minipage}{\textwidth}
			\begin{tablenotes}[para,flushleft]
				\setlength{\lineskip}{0.2ex}
				\notsotiny 
				{\textit{Notes}: Results are based on VAR models estimated using data from 1975:M1 to 2025:M4, excluding the COVID-19 period (2020:M1--2021:M6). All specifications include 12 lags and maintain a consistent variable ordering: ACI/MRCI, \mn{INDPRO}, \mn{PCEPI}, \mn{UNRATE}, \mn{SHADOWRATE}, \mn{EXUSUKx}, and \mn{HOUST}. Confidence intervals at the 68\% level, conditional on $\widehat\lambda$ and $\widehat{\boldsymbol w}(\widehat\lambda)$. For brevity, we display impulse responses for selected variables only; complete results are available in Appendix~\ref{annexe_climate} (Figure~\ref{fig:FULL_Raci_irf}). For the \textcolor{black}{weight-reestimating pairs bootstrap version}, with the aggregation weights re-estimated in every replication, see Figure~\ref{fig:Raci_boot_irf} in Appendix~\ref{sec:weight_uncertainty}.}\vspace{-.3cm}
			\end{tablenotes}
		\end{minipage}
	\end{threeparttable}
\end{figure}

\subsubsection{Impulse Responses}
\noindent \textcolor{black}{Figures~\ref{fig:Caci_irf} and~\ref{fig:Raci_irf} present responses to positive one-standard-deviation innovations in the component-space and rank-space climate measures. We compare these responses with the response estimated using the equal-weight benchmark measure. The evidence is strongest for industrial production and more mixed for the remaining variables.}

\paragraph{Industrial production.} Under the fixed-weight benchmark, a one-standard-deviation climate measure innovation reduces industrial production by \textcolor{black}{0.16\%} at peak (month 12), while \citet{KimMatthesPhan2025} reported a 0.15 percentage-point decline using the same ACI in a nonlinear VAR.
\textcolor{black}{Component-space specifications yield larger estimated responses, though unevenly: the national specification \ascn, which reweights only five constituents, differs least from the benchmark, while the richer assembled measures yield larger contractions.}
\textcolor{black}{For innovations in the \ascr and \asco measures, the estimated contraction ranges from \textcolor{black}{0.24\%} to \textcolor{black}{0.32\%} and is statistically significant, roughly one and a half to two times the peak response estimated using the benchmark measure.}
Rank-space assemblage is more selective: \asrn adds little over the benchmark, with only five components to rank; the regional \asrr stays close to the benchmark as well, and \textcolor{black}{only the \asro specification yields an economically meaningful estimated contraction, alongside its tail-concentrated rank-weight profile.} This pattern is consistent with recent climate--macro evidence: \citet{KimMatthesPhan2025} find that severe weather shocks generate meaningful contractions in industrial production, and \citet{ciccarelli2024demand_24_EER} document asymmetric temperature effects on euro-area inflation, both in VAR/LP-type frameworks. \textcolor{black}{Our contribution is to show that the estimated industrial-production response varies materially across alternative climate measures. With greater flexibility over regional components and their aggregation, the assembled specifications yield contractions up to twice the response estimated using the ACI benchmark. This suggests that alternative measures may be worth tracking when assessing climate--macro relationships.}

\paragraph{Unemployment.} The unemployment response is ambiguous across specifications. The \asn and \asr indices point towards an increase, with confidence bands excluding zero at longer horizons in component-space. At the highest level of disaggregation, the confidence interval includes zero in component-space, while exhibiting a brief but significant rise in the short term in rank-space. Unemployment thus yields mixed results depending on the aggregation scheme: rank-space assemblage suggests that tail-events trigger a short-lived spike in unemployment, consistent with the immediate labor market disruptions \citep{KimMatthesPhan2025}, while component-space assemblage suggests persistent, medium-run deterioration.

\paragraph{Housing starts.} 
Housing starts are the outcome for which the benchmark is least informative. Under the ACI benchmark, the point estimate is positive, which is counterintuitive for an adverse climate shock, but never statistically distinguishable from zero. Assemblage yields a clearer contractionary response: in component-space, all three panels imply a significant short-run decline, while in rank-space only the NOAA specification remains negative and significant. NOAA is therefore the only case in which the housing contraction is supported in both spaces. \textcolor{black}{The housing-start response also varies with the climate measure. The response estimated using the ACI benchmark is imprecise, whereas several assembled specifications yield a significant contraction.}

\subsubsection{Time Series and Weights}
\begin{figure}[t!]
	\caption{{Assembled Climate Index (ACI Specification)}} \label{fig:assemblage_aci}
	\begin{center}

		\begin{subfigure}[t]{\textwidth}
			\vspace*{-0.7cm}
			\centering
			\hspace*{-.5cm}\includegraphics[width=1.015\textwidth, trim = -3.2mm -10mm 0mm 5mm, clip]{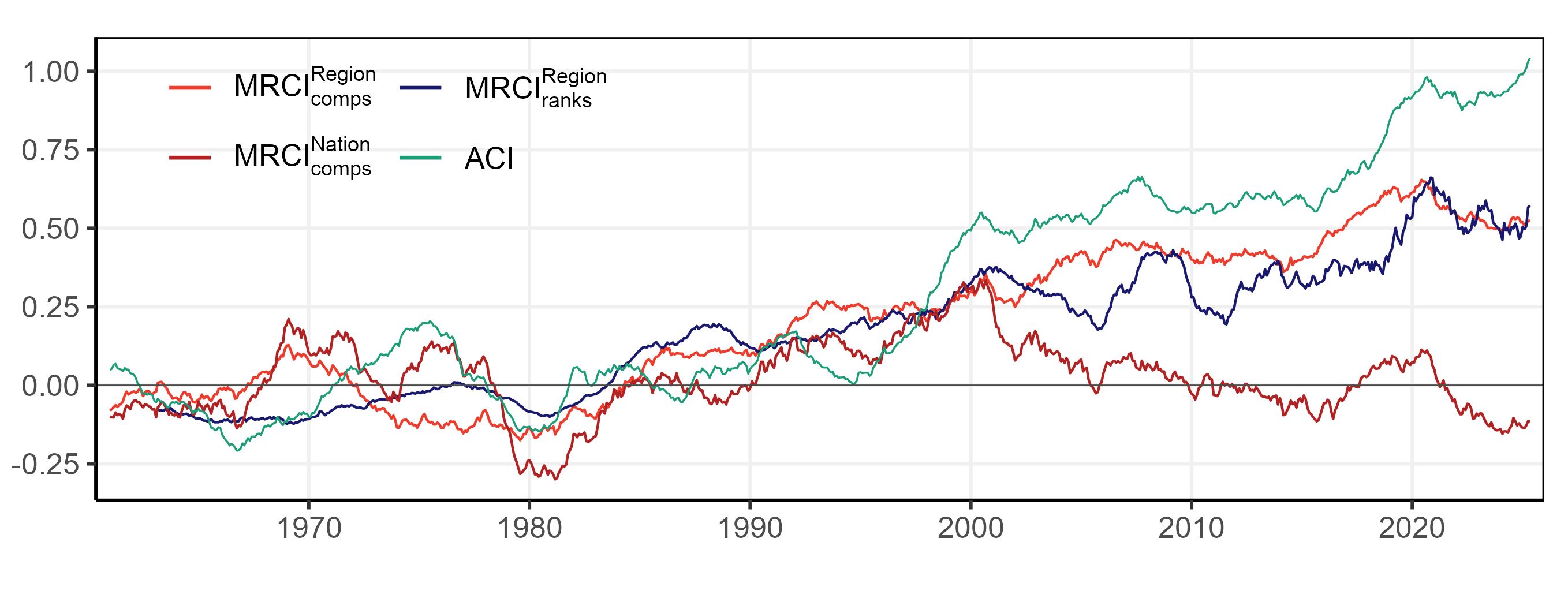}
		\end{subfigure}
        
		\begin{minipage}[t]{0.53\textwidth}
			\vspace*{-0.75cm}
			\centering
			\hspace*{-.21cm}\includegraphics[width=1.07\textwidth, trim = 0mm -20mm -8mm 27mm, clip]{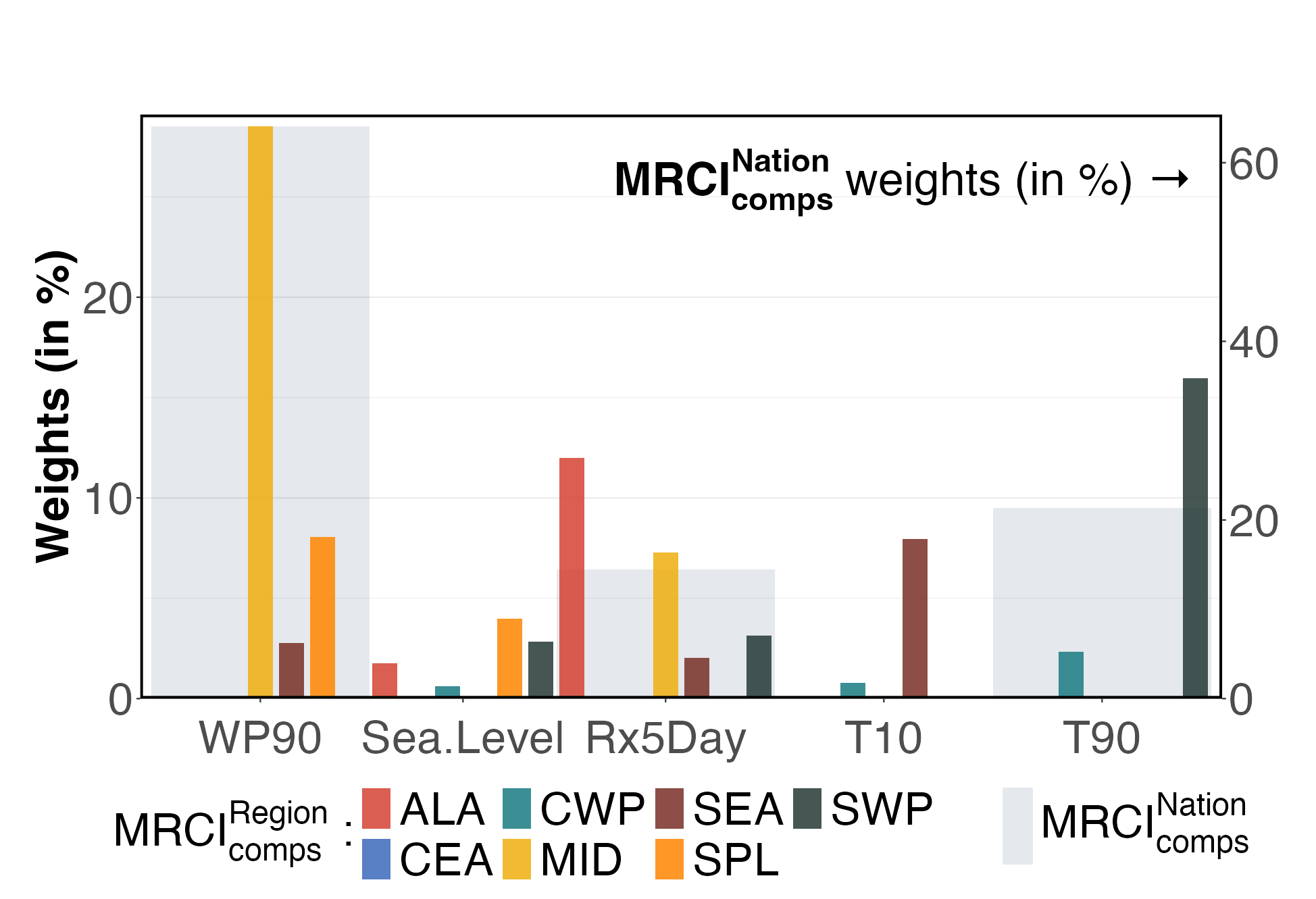}
		\end{minipage}%
		\hspace*{0.3cm}
		\begin{minipage}[t]{0.52\textwidth}
			\vspace*{-1.0cm}
			\centering
			\hspace*{-0.78cm}\includegraphics[width=.96\textwidth, trim = -2mm -33mm 0mm -5mm, clip]{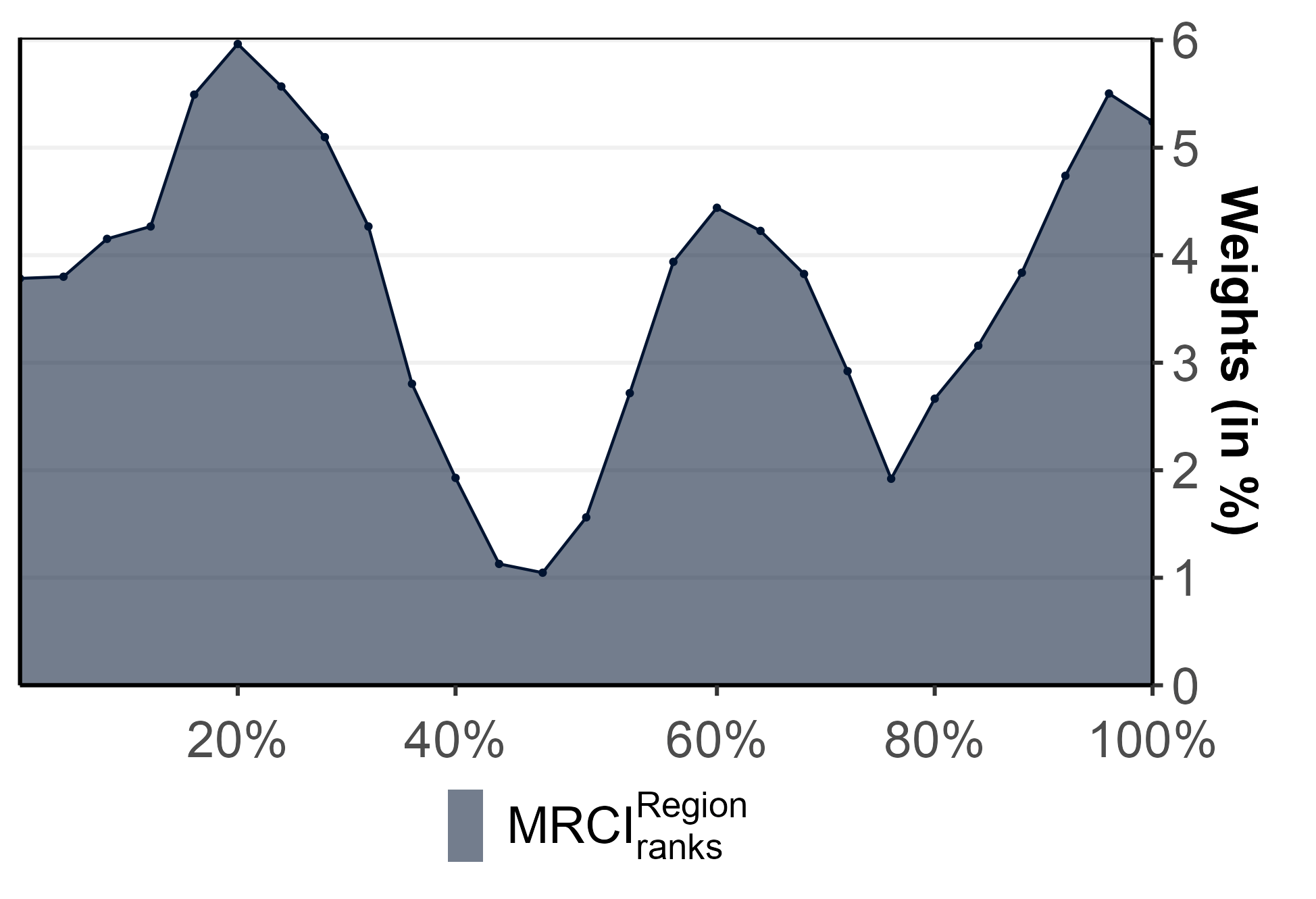}
		\end{minipage}

		\vspace*{-0.7cm}
		\begin{minipage}[t]{0.53\textwidth}\centering (a)~Component Weights\end{minipage}%
		\hspace*{0.3cm}
		\begin{minipage}[t]{0.52\textwidth}\centering\hspace*{-1.2cm}(b)~Rank Weights\end{minipage}
	\end{center}

	\begin{threeparttable}
		\centering\vspace{-0.7cm}
		\begin{minipage}{\textwidth}
			\begin{tablenotes}[para,flushleft]
				\setlength{\lineskip}{0.2ex}
				\notsotiny 
				{\textit{Notes}: The \textbf{upper panel} shows MRCI (National and Regional) and the Headline ACI/MRCI, all shown as five-year moving averages of the standardized components. All series are using the weights estimated from 1975-2025. Moreover, we show a sliding average over two years for the MRCI$_\text{ranks}$.  The \textbf{lower left panel} presents the weights in percent for MRCI$_\text{comps}$ National and Regional. In the \textbf{lower right panel}, we present the resulting weights for MRCI$_{\text{ranks}}^{\text{Region}}$. To ensure an adequate representation, we smooth the displayed weights in rank-space using a trapezoidal kernel spanning eight ranks.}
			\end{tablenotes}
		\end{minipage}
	\end{threeparttable}
\end{figure}

\begin{figure}[t!]
		\caption{{Assembled Climate Index (NOAA Specification)}} \label{fig:assemblage_noaa}
	\begin{center}

		\begin{subfigure}[t]{\textwidth}
			\vspace*{-0.7cm}
			\centering
			\hspace*{-.5cm}\includegraphics[width=1.015\textwidth, trim = -3.2mm -10mm 0mm 5mm, clip]{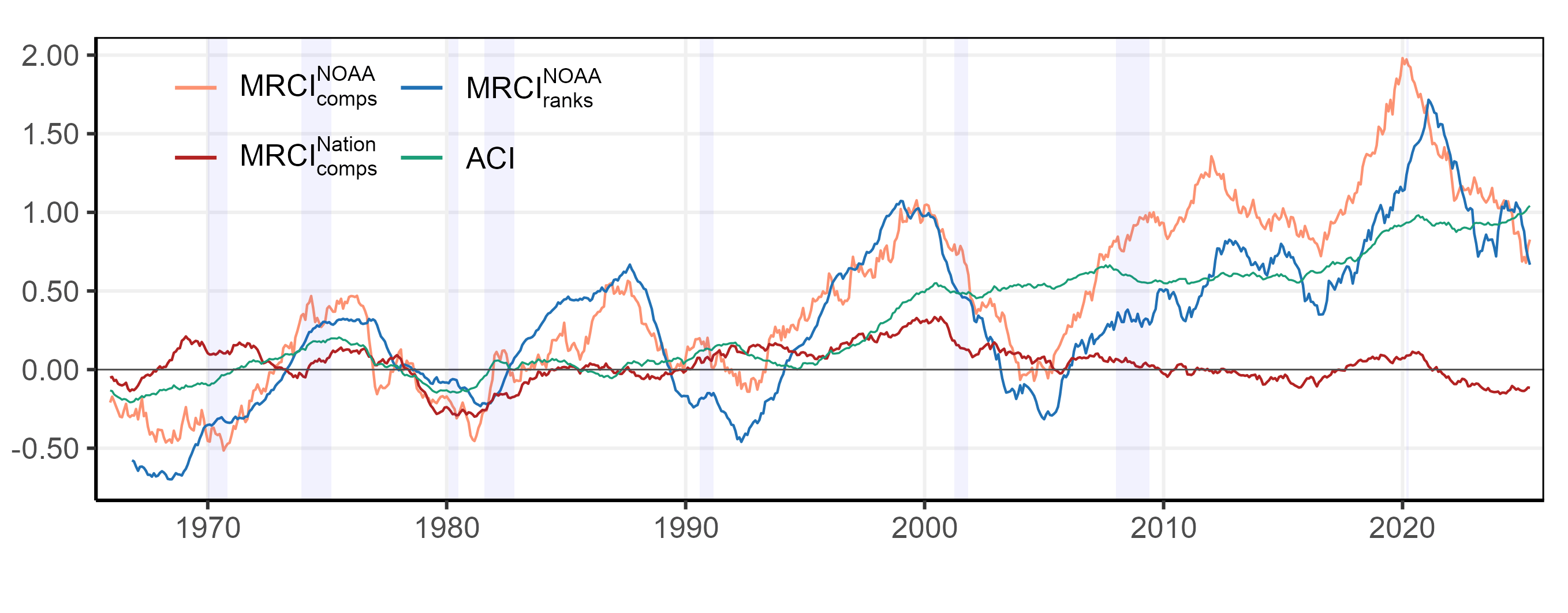}
		\end{subfigure}%

		\begin{minipage}[t]{0.52\textwidth}
			\vspace*{-0.69cm}
			\centering
			\hspace*{-.28cm}\includegraphics[width=1.0712\textwidth, trim = 0mm -20mm -8mm 25mm, clip]{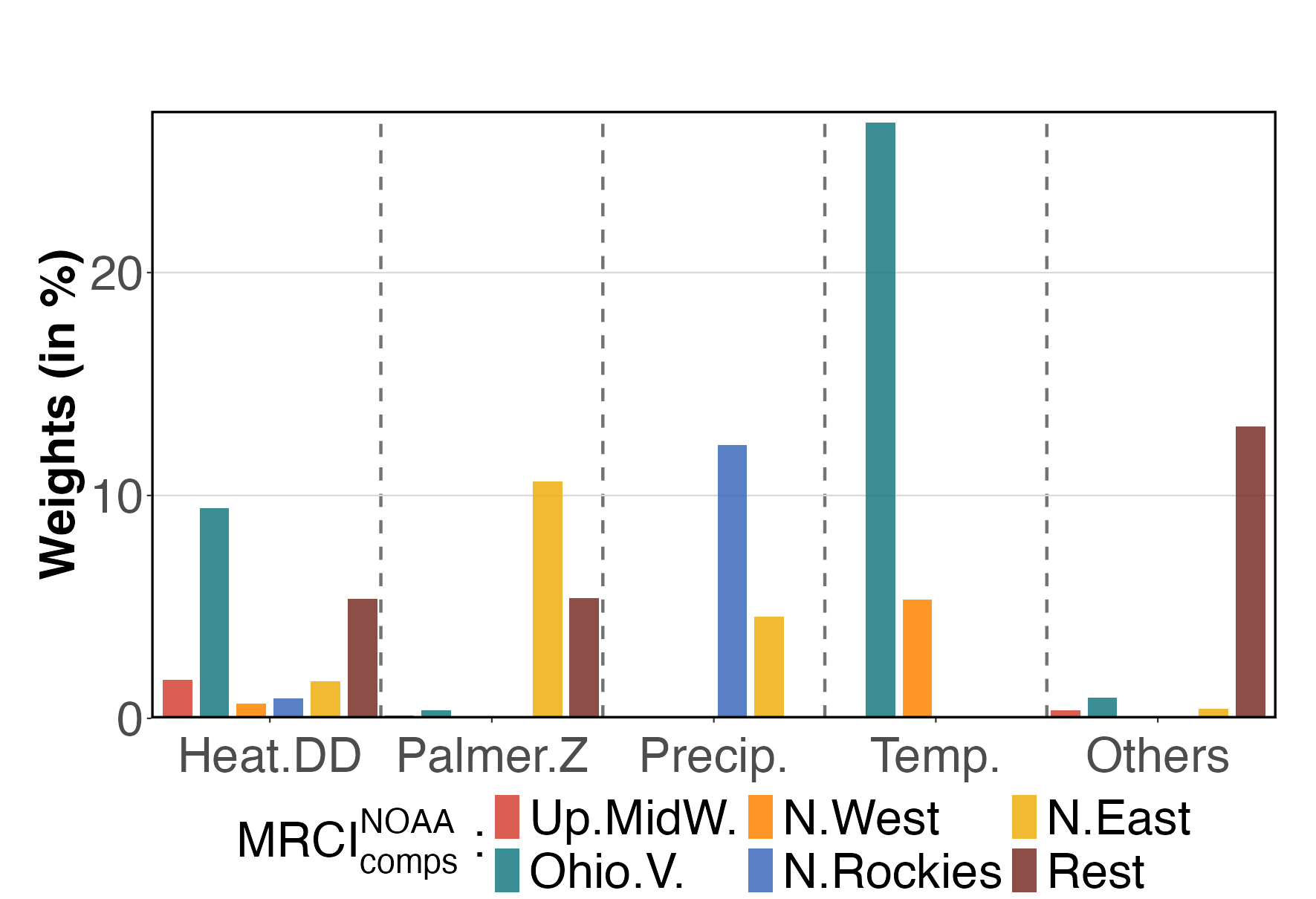}
		\end{minipage}%
		\hspace*{0.3cm}
		\begin{minipage}[t]{0.51\textwidth}
			\vspace*{-1.05cm}
			\centering
			\hspace*{-.25cm}\includegraphics[width=1.03\textwidth, trim = -2mm -33mm 0mm -5mm, clip]{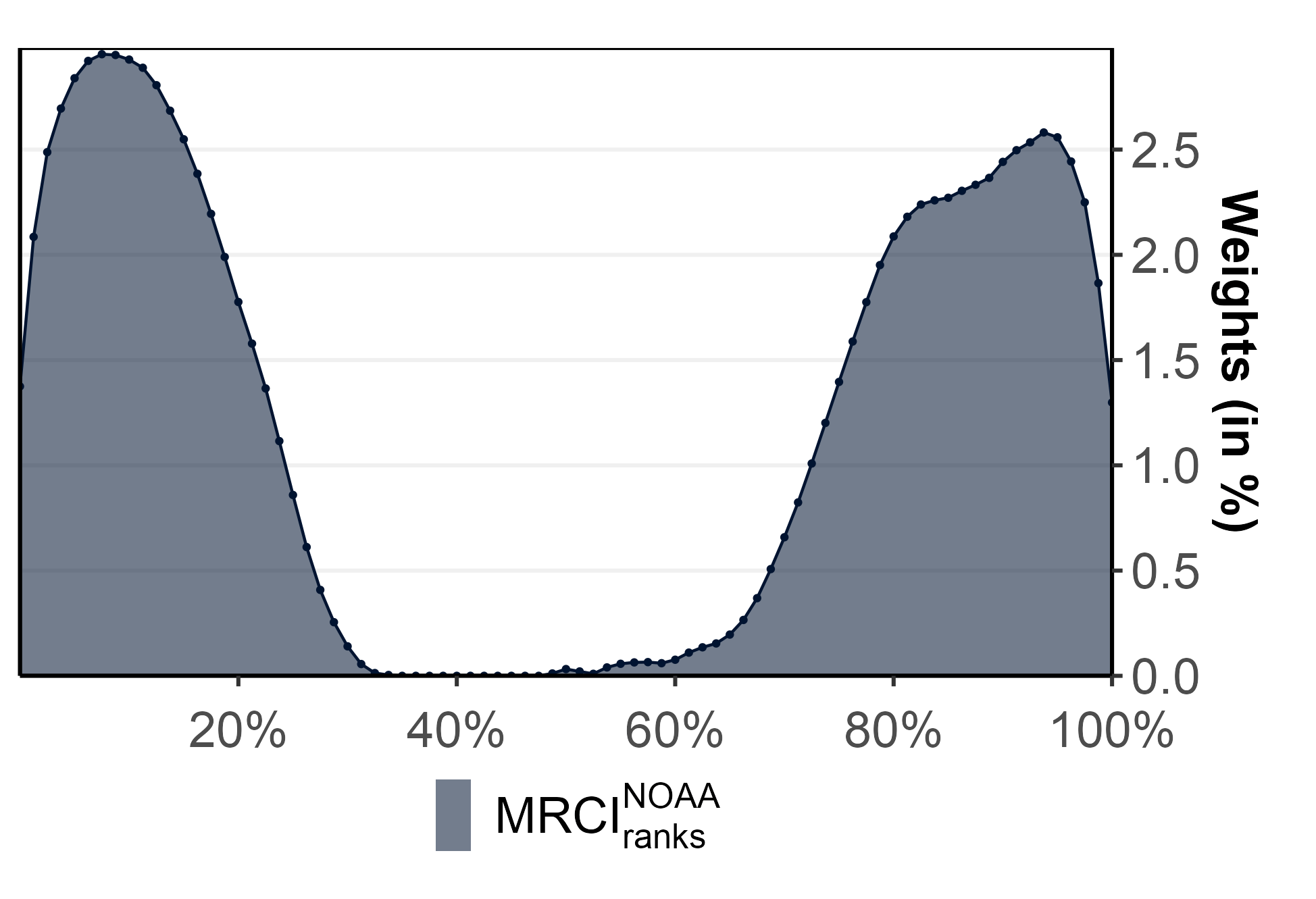}
		\end{minipage}

		\vspace*{-0.7cm}
		\begin{minipage}[t]{0.52\textwidth}\centering (a)~Component Weights\end{minipage}%
		\hspace*{0.3cm}
		\begin{minipage}[t]{0.51\textwidth}\centering\hspace*{-1.2cm}(b)~Rank Weights\end{minipage}
	\end{center}

	\begin{threeparttable}
		\centering\vspace{-0.7cm}
		\begin{minipage}{\textwidth}
			\begin{tablenotes}[para,flushleft]
				\setlength{\lineskip}{0.2ex}
				\notsotiny 
				{\textit{Notes}: The \textbf{top panel} shows MRCI (National and NOAA) and the Headline ACI/MRCI, all shown as five-year moving averages of the standardized components. All series are using the weights estimated from 1975-2025. Moreover, we show a sliding average over a year for the MRCI$_\text{ranks}$. The \textbf{lower left panel} presents the weights in percent for MRCI$_{\text{comps}}^{\text{NOAA}}$. In the \textbf{lower right panel}, we present the resulting weights for MRCI$_{\text{ranks}}^{\text{NOAA}}$. To ensure an adequate representation, we show a sliding average over two ranks to smooth the weights in rank-space. See Appendix~\ref{sec:aggregation} for the graphical aggregation of the NOAA components.}
			\end{tablenotes}
		\end{minipage}
	\end{threeparttable}
\end{figure}

\noindent The assembled indices produce different levels of system coherence and different macroeconomic responses. The natural question is what drives these differences, and whether the composition implied by $\boldsymbol{w}$ makes economic sense. Figures~\ref{fig:assemblage_aci} and~\ref{fig:assemblage_noaa} display the assembled series (top panels) alongside the component-space weights (lower left) and rank-space weights (lower right) for the ACI and NOAA specifications, respectively.

Two broad patterns emerge in every component-space specification, and both cut the same way. The components that gain weight are more closely associated with higher-frequency climate variation, such as wind and heat extremes, while those that lose weight are slower-moving or less cyclical, such as sea level and, to some extent, cold extremes. And the assembled series all bend the same way: they flatten after 2000 rather than tracking the ACI's upward drift, suggesting that the persistent drift in the benchmark reflects components that are less influential for business-cycle dynamics. We discuss each specification in order of increasing complexity.

\paragraph{\ascn (National, component-space).}
\ascn aggregates the five national ACI components: high-temperature extremes (\texttt{T90}), low-temperature extremes (\texttt{T10}), heavy precipitation (\texttt{Rx5Day}), high winds (\texttt{WP90}), and sea level, at the aggregate U.S.\ level, with no spatial disaggregation. Because it has only four degrees of freedom, this is the most constrained specification and provides a clean test of whether composition alone, holding the geographic coverage fixed, can move structural inference; at this most constrained level it does so only modestly. In Figure~\ref{fig:assemblage_aci} (top panel), \ascn tracks the headline ACI through the 1960s--90s; both series register the 1980 heat--drought event, the 1988 shocks, the 1993 Midwest floods, and the early-2000s hot-drought regime. The two diverge after ${\sim}$2000: whereas the ACI continues trending upward, \ascn slows down and undershoots the benchmark because sea level, a strongly trending component, now receives zero weight.

This is consistent with the weight profile in Figure~\ref{fig:assemblage_aci} (lower left, light gray bars): low-temperature extremes (\texttt{T10}) and sea level receive minimal weight (0\% each). The weight released from these two components is absorbed almost entirely by high winds (\texttt{WP90}), whose share rises from the equal-weight benchmark of 20\% to around \textcolor{black}{65\%}. This is consistent with episodic wind events (hurricanes, severe storms, and derechos) serving as the main source of macro-predictive climate variation at business-cycle frequencies. Sea level exhibits minimal month-to-month variation; its relevance is concentrated at lower (long-run) frequencies rather than business cycles. \texttt{T10} contributes little to the nation-wide acute climate stress captured at macro frequencies.
Even this minimal reweighting reshapes the index towards the most informative components, indicating that equal weighting in the standard ACI aggregates variables with varying relevance for business-cycle dynamics.

\paragraph{\ascr (Regional, component-space).}
The regional specification uses the regionally disaggregated ACI data, taking the same five ACI components across seven U.S.\ regions and yielding 34 region-component constituents.
The time series in Figure~\ref{fig:assemblage_aci} (top panel) follows a similar trajectory to \ascn but with lower amplitude post-2000, consistent with spatial averaging diluting localized extremes. The added granularity also makes the regional allocation more credible than the national one. Restricted to only five aggregate components, \ascn can re-express measurement only through a few large shifts, concentrating weight aggressively on the single most predictively relevant component; \ascr can instead distribute the same macro-informative signal across thirty-four region-specific constituents, accommodating genuine spatial heterogeneity rather than forcing it onto a handful of national aggregates. The national index is thus the most constrained object in the menu, which is why it moves the response the least; it is the heterogeneity of the richer regional and NOAA panels that gives assemblage its room to work.

The weight profile in Figure~\ref{fig:assemblage_aci} (lower left) is broadly aligned with the national pattern (\texttt{WP90} dominates and sea level receives minimal weight) but now reveals regional variation. The contrast with the ACI's equal 20\% allocation per component is immediate: high winds receive \textcolor{black}{40\%} aggregated weight, reaching \textcolor{black}{29\%} of the global index in the Midwest region alone. Heat extremes (\texttt{T90}) enter with pronounced regional heterogeneity: \textcolor{black}{16\%} in the Southwest Pacific and around \textcolor{black}{2\%} in the Central West Pacific. Precipitation (\texttt{Rx5Day}) receives up to \textcolor{black}{25\%}, concentrated in flood-prone areas such as Alaska, \textcolor{black}{the Midwest} and the Southwest Pacific.

The regional pattern aligns with known structural exposures across sectors: the Southern Plains, Southwest Pacific and Midwest, regions with substantial exposure through manufacturing, energy infrastructure, and agriculture, allocate more than \textcolor{black}{35\%} of their own regional weight to wind extremes, consistent with episodic, high-damage events (hurricanes, severe storms) that account for substantial macro-level risk \citep{Noy2009,CavalloEtAl2013}. Heat extremes receive their largest weights in the Southwest Pacific, consistent with the evidence that temperature affects economic activity through productivity, energy demand, and sectoral exposure (e.g., agriculture and construction) \citep{BurkeHsiangMiguel2015,ColacitoHoffmannPhan2019}.

\paragraph{\asco (NOAA, component-space).}
The NOAA specification draws on an entirely different data source: 82 subcomponents from NOAA weather and atmospheric measurements across ten U.S.\ regional zones, providing a rich and alternative panel. Figure~\ref{fig:assemblage_noaa} (top panel) shows the time evolution of \asco. Its shape reflects dominant loadings on two regional climate blocks (Figure~\ref{fig:assemblage_noaa}, lower left): an Ohio Valley thermal block (temperature and heating degree days, approximately \textcolor{black}{36\%} of the index weight) and a Northern/Northeast hydroclimate block (precipitation and Palmer Z index, approximately \textcolor{black}{27\%}). The series displays pronounced decadal cyclicality ($\approx$10--15 years), consistent with known low-frequency ocean--atmosphere modes such as the Pacific Decadal Oscillation and the Atlantic Multidecadal Oscillation \citep{Mantua1997,McCabePaleckiBetancourt2004,NewmanEtAl2016}. 

In line with \citet{ColacitoHoffmannPhan2019}, the thermal block captures macro-relevant temperature variability that aligns with energy demand and industrial fluctuations located in the manufacturing corridor, while the hydroclimate block, in the spirit of \citet{SchlenkerRoberts2009} and \citet{BurkeHsiangMiguel2015}, proxies for flood and drought conditions across climate-sensitive sectors. Despite originating from an entirely different data environment, the NOAA component-space weights are coherent with the ACI findings: both specifications upweight episodic, high-damage phenomena (wind, precipitation extremes) and downweight slow-moving or low-variance components. 

Two independent data environments converge on one geographic corridor: the regional ACI loads on the central U.S.\ (Southern Plains and Midwest), and the NOAA panel on the Ohio Valley. Both identify the central U.S.\ region, home to major manufacturing and agricultural assets, as carrying the highest statistical weight for macro-relevant climate variation.

\paragraph{Rank-space specifications.}
The rank-space estimator operates on cross-sectional order statistics rather than named components: it reweights only the shape of the cross-sectional distribution, so it has less room to depart from the equal-weight benchmark than component-space, which can single out specific series. How far it departs depends on how much distributional structure the panel offers, and the three specifications differ sharply on that score (the rank profiles are in Figures~\ref{fig:assemblage_aci} and~\ref{fig:assemblage_noaa}, lower-right panels).

At the national level \asrn adds little: with only five components there is too little cross-sectional variation for the rank representation to exploit. The regional \asrr does lift the tails, near the \textcolor{black}{20th and 95--100th} percentiles, but still places substantial weight across the interior, including near the median, so the assembled index stays close to the cross-sectional average, the equal-weight ACI, and its response is correspondingly mild, much like the benchmark's, though somewhat stronger in the first 12 months; the panel does not offer enough distributional signal for the rank representation to express what the component-space weights can. Only the NOAA \asro carries a sharp distributional signal: about \textcolor{black}{three-quarters} of its mass sits in the two tails, with the single most heavily weighted ranks receiving \textcolor{black}{$2.95\%$} on the left and \textcolor{black}{$2.58\%$} on the right against \textcolor{black}{$1.2\%$} under flat weighting, and only about \textcolor{black}{three} percent falling between the 30th and 55th percentiles. It departs markedly from the flat benchmark and produces a response clearly different from it. The two representations moreover agree: the rank- and component-space NOAA indices both locate the macro-relevant variation in the extremes and read as temperature-led (Appendix~\ref{sec:rankstocomps}).

In the time series (Figure~\ref{fig:assemblage_aci}, top panel), \asrr nonetheless rises more strongly than its component-space counterpart post-2010, accentuating episodes such as the 2012 drought, Hurricane Harvey (2017), the 2020 Midwest derecho (\texttt{WP90}), and the 2021 Pacific Northwest heat dome (\texttt{T90}) \citep{NOAACEIDefinition}.

\paragraph{Mapping ranks back to components.} Because the rank-space weights attach to distributional positions rather than to named series, it is natural to ask which constituents actually occupy the weighted ranks. Appendix~\ref{sec:rankstocomps} pushes the rank weights back through the period-by-period sort to recover a time-varying loading on each named constituent and summarizes it as a size-neutral usage measure. The picture is interpretable: the NOAA panel concentrates its weight on thermal variables, while the ACI panel spreads its weight fairly evenly across names, so the tail-loading documented above does not conceal a single dominant named series.

\paragraph{Robustness and weight uncertainty.} Because the aggregation weights are themselves estimated, Appendix~\ref{sec:weight_uncertainty} asks how far they, and the index they produce, can be trusted, and does three things. First, we trace the weights through time on an expanding window (Appendix~\ref{sec:rolling}, Figure~\ref{fig:rolling_weights}) and find that the dominant components are in place early and that the dominant variable families retain their relative weights over the sample. Second, we bootstrap the weights jointly with the VAR dynamics (weight-reestimating pairs bootstrap), re-estimating them in every replication (Appendix~\ref{sec:emp_climate_boot}, Figure~\ref{fig:boot_weights}): the individual loadings are noisy, but aggregated to variable families they are stable, and the assembled index itself is tightly determined (Figure~\ref{fig:boot_index}). Third, we carry this weight uncertainty into the impulse responses (Figures~\ref{fig:Caci_boot_irf} and~\ref{fig:Raci_boot_irf}) and find that the \textcolor{black}{weight-reestimating pairs bootstrap} responses remain close to their fixed-weights residual-bootstrap counterparts. \textcolor{black}{Because the procedures differ in both their resampling schemes and interval constructions, this similarity is evidence of robustness across procedures; it does not isolate the contribution of weight-estimation uncertainty}.

\section{Conclusion}\label{sec:conclusion}
\setcounter{equation}{0}

The Assemblage VAR endogenizes variable measurement within structural VARs by jointly estimating aggregation weights and dynamics, offering component-space and rank-space variants each regularized towards conventional benchmarks. \textcolor{black}{Applied to U.S.\ climate data, several assembled specifications yield statistically significant peak responses up to twice the response estimated using the fixed-weight benchmark measure. Component-space weights emphasize high-wind variables, while rank-space weights concentrate on distributional tails. Estimated macroeconomic responses can therefore vary materially across alternative climate measures.} These results show that measurement, the mapping from disaggregates to the VAR variable, is a quantitatively important margin for structural inference, with the choice of measurement affecting the resulting empirical conclusions, including those relevant for damage assessment and policy design.

Several directions remain open for future work. The aggregation weights could be allowed to evolve over time, and the VAR could be made modestly nonlinear so that propagation, and not only the measurement of extremes, can respond nonlinearly to the shocks the rank-space index emphasizes. On the climate side, replacing the ACI's regional structure with high-resolution gridded data (such as ERA5 or PRISM) would allow the estimator to learn which geographic cells carry macro-relevant variation, while assembling across competing climate model outputs would replace \textit{ad hoc} model averaging with supervised combination. More broadly, the framework applies wherever an aggregate enters a VAR and where its construction from subcomponents is contested or conventional, a description that fits not only climate indices but also core inflation, financial conditions indices, uncertainty indices, and measures of global economic activity.

\newpage
\setlength{\bibsep}{4pt}
\setstretch{1.0}
\bibliographystyle{chicago}
\bibliography{ref}

\newpage
\setstretch{1.27}
\appendix
\renewcommand{\thesection}{\Alph{section}}
\def\@seccntformat#1{\csname the#1\endcsname:\hskip 0.5em}

\renewcommand{\thefigure}{\Alph{section}.\arabic{figure}}
\setcounter{figure}{0}

\section{Weight Uncertainty}\label{sec:weight_uncertainty}

\setcounter{figure}{0}
\setcounter{equation}{0}

We use two bootstrap procedures that address different questions. The fixed-weights residual bootstrap measures impulse-response uncertainty conditional on the estimated assemblage weights, whereas the weight-reestimating pairs bootstrap assesses weight stability and whether the assembled index and impulse responses remain robust when the weights are re-estimated.

The fixed-weights residual bootstrap, described in Appendix~\ref{sec:fixed_weights_boot}, follows the fixed-tuning convention of Section~\ref{sec:penalty_selection}. The fixed-weights residual bootstrap bands hold both the selected penalty $\widehat\lambda$ and the resulting weights $\widehat{\boldsymbol w}(\widehat\lambda)$ fixed, thereby excluding weight-estimation uncertainty. To assess that uncertainty, the weight-reestimating pairs bootstrap holds $\widehat\lambda$ fixed but re-estimates the weights and VAR parameters jointly in each of $B=1{,}000$ replications. Because each weight-reestimating pairs bootstrap replication re-solves a constrained optimization problem, it does not employ the nested studentization used for the main-text bands.

Appendix~\ref{sec:emp_climate_boot} reads off the resulting bootstrap distribution of the weights, the assembled index and the impulse responses in turn. Appendix~\ref{sec:rolling} complements it by re-estimating the weights on expanding sub-samples. The message is the same in both: the individual component loadings are noisy, since the panel holds many overlapping series, but every object the paper's conclusions rest on, namely the dominant weight allocations, the tail concentration, the assembled index, and its turning points, survives resampling and re-estimation.

\subsection{\textcolor{black}{Fixed-weights residual bootstrap}}
\label{sec:fixed_weights_boot}

The main-text bands are $68\%$ equal-tailed percentile-$t$ intervals obtained from a fixed-weights residual bootstrap \citep{Kunsch1989,Hall1992,BrueggemannJentschTrenkler2016}. Under the fixed-tuning convention of Section~\ref{sec:penalty_selection}, both the selected penalty $\widehat\lambda$ and the resulting weights $\widehat{\boldsymbol w}(\widehat\lambda)$ are held fixed.

We center the fitted VAR residual vectors and resample overlapping blocks of length $\ell=\lfloor n^{1/3}\rfloor=8$, where $n$ is the number of fitted residuals. Resampling the residual vectors jointly retains their contemporaneous cross-equation dependence, while the moving blocks retain temporal dependence within blocks. For each of $B_{\mathrm{outer}}=1{,}000$ draws, an artificial series is generated recursively from
\begin{equation}
    \boldsymbol y^*_t
 =\widehat{\boldsymbol\nu}
 +\sum_{j=1}^{p}\widehat{\mathbf A}_j\boldsymbol y^*_{t-j}
 +\boldsymbol u^*_t,
\end{equation}
initialized at the first $p=12$ observed rows. The VAR is then refitted with the same lag order and intercept, and the impulse responses are recomputed.

At horizon $h$, the interval is
\begin{equation}
   \left[
\widehat\theta_h-q_h(1-\alpha/2)\widehat\sigma_h,\;
\widehat\theta_h-q_h(\alpha/2)\widehat\sigma_h
\right],
\qquad \alpha=0.32, 
\end{equation}
where $q_h(\cdot)$ denotes a quantile of
$(\widehat\theta^*_{b,h}-\widehat\theta_h)/\widehat\sigma^*_{b,h}$ across the outer draws. Each $\widehat\sigma^*_{b,h}$ is computed from $B_{\mathrm{inner}}=100$ recursive draws from outer replication $b$, while $\widehat\sigma_h$ is computed from 200 additional refits from the fitted model.

A replication whose companion matrix has an eigenvalue with modulus at least one is redrawn up to 50 times and otherwise recorded as missing. No outer or inner replication required a redraw, so every quantile is based on all $1{,}000$ outer statistics. Computations use \texttt{set.seed(123)}.

\subsection{\textcolor{black}{Weight-reestimating pairs bootstrap}}\label{sec:emp_climate_boot}

{\color{black}For the weight-reestimating pairs bootstrap, at the observed cross-validated penalty \(\widehat\lambda\), we form the aligned current-outcome and lag-regressor pairs used to evaluate \(S_T(\boldsymbol w)\) and draw overlapping blocks of consecutive pair indices with replacement. The pairs are aligned before resampling, so every retained observation carries its true lag structure. The block length is $\ell=\max\{p+1,\lfloor T^{1/3}\rfloor\}=13$ in the final sample, where the floor at $p+1$ binds: blocks shorter than the VAR memory would sever the lag dependence the criterion exploits and maximize the fraction of pairs manufactured at block joins.

In each replication, holding \(\widehat\lambda\) fixed, the aggregation weights are re-estimated on the resampled pairs subject to the same nonnegativity and normalization constraints as the point estimate, and the VAR coefficients and covariance matrix are re-estimated from the same pairs. Impulse responses follow from the drawn $(\mathbf B^*,\boldsymbol\Sigma^*)$ by Cholesky factorization with the assembled index ordered first, cumulated exactly as in the point estimate. The optimizer is started at $\widehat{\boldsymbol w}$ in every replication. The procedure does not separately resample raw observations or residuals, generate an artificial series recursively, or use bootstrap initial conditions.

We attempted 1{,}000 replications for each of the six full-sample specifications. All 1{,}000 returned a weight solution and a usable VAR fit in every case. An optimization error would cause that draw to be recorded as failed and discarded, without retry or replacement, but no such error occurred in the final runs.

The weight bands are pointwise percentile intervals, which respect the simplex support of the constrained estimates. They are interpreted as resampling measures of weight stability rather than as boundary-corrected confidence intervals. The reconstructed-index envelopes are likewise pointwise percentile intervals of the indices obtained from the bootstrap weight estimates. The impulse-response bands from the weight-reestimating pairs bootstrap are pointwise $68\%$ basic (reverse-percentile) intervals, $[2\widehat\theta-q^*_{0.84},\,2\widehat\theta-q^*_{0.16}]$ where $q^*_{h}(\gamma)$ is the empirical $\gamma$-quantile of $\widehat\theta^*_{b,h}$. These bands are not studentized.
}

\begin{figure}[!t]
	\caption{{Bootstrap stability of the assemblage weights }}
	\label{fig:boot_weights}
	\begin{center}
		\begin{subfigure}[t]{0.33\textwidth}\centering
			\vspace*{-0.3cm}\captionsetup{margin={1cm,0cm}}\caption{\asn}\vspace*{-0.25cm}
			\hspace*{-.05cm}\includegraphics[width=1.04\linewidth]{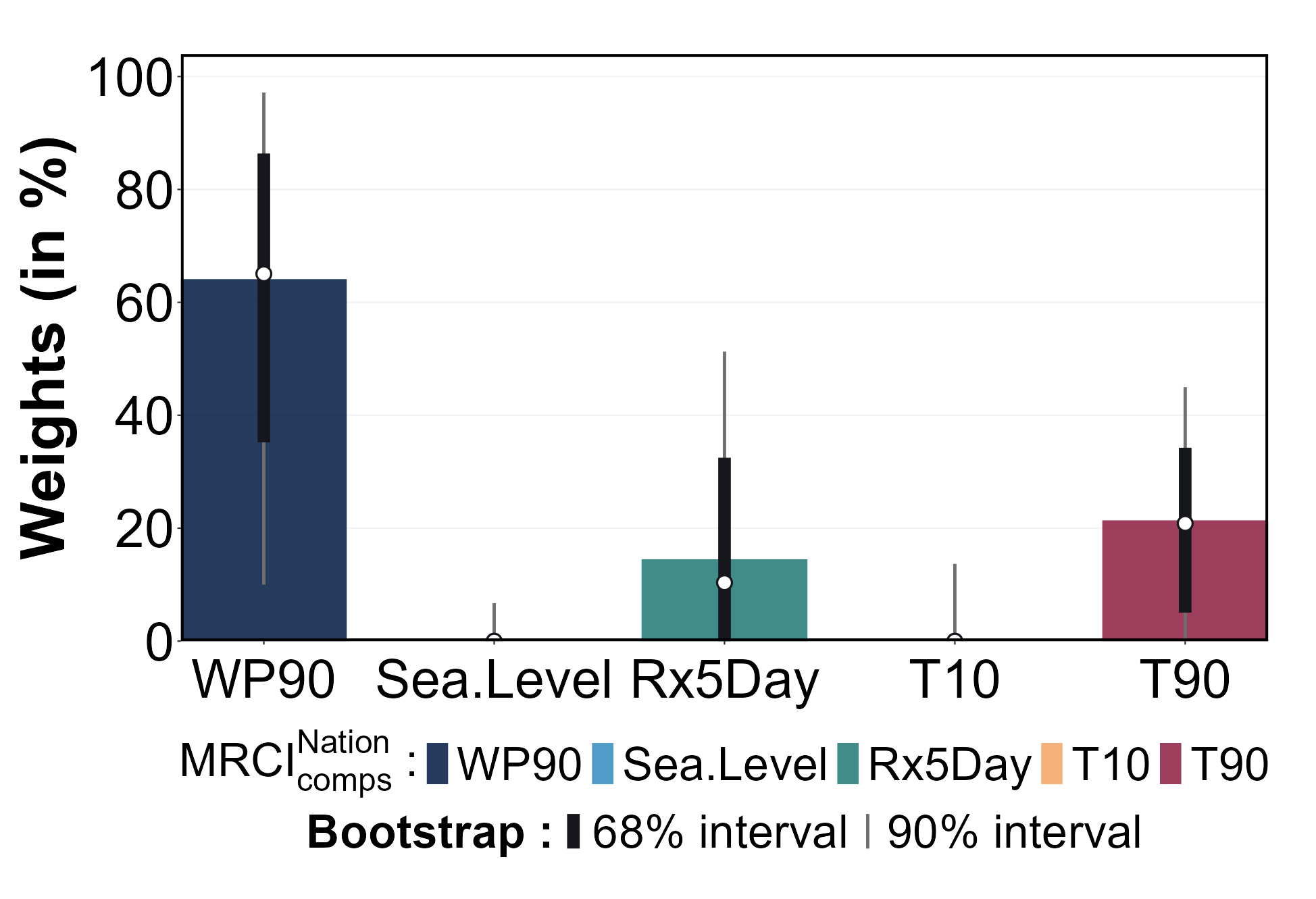}
		\end{subfigure}\hfill
		\begin{subfigure}[t]{0.33\textwidth}\centering
			\vspace*{-0.3cm}\captionsetup{margin={1cm,0cm}}\caption{\asr}\vspace*{-0.25cm}\hspace*{.025cm}
			\includegraphics[width=1.04\linewidth]{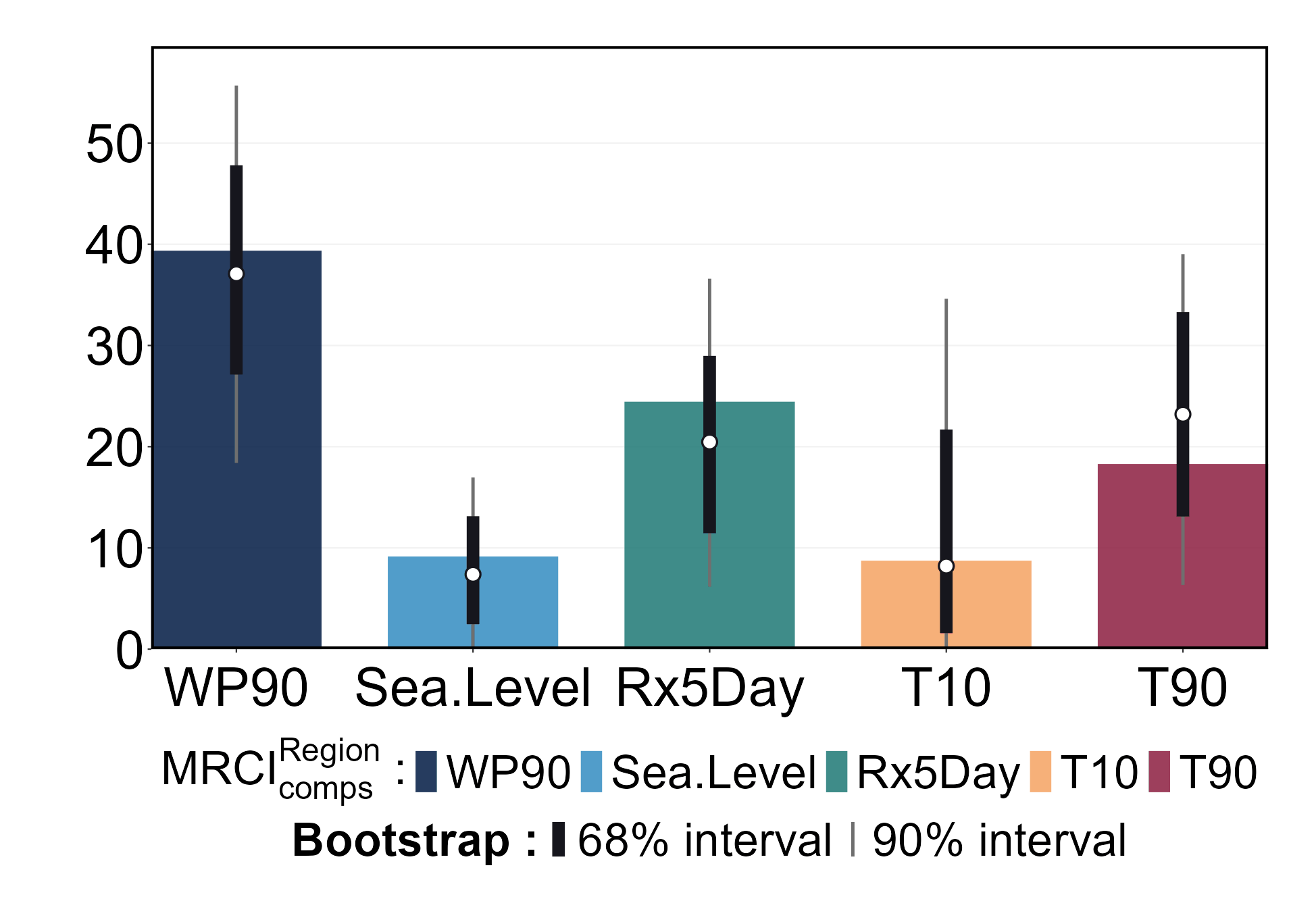}
		\end{subfigure}\hfill
		\begin{subfigure}[t]{0.33\textwidth}\centering
			\vspace*{-0.3cm}\captionsetup{margin={1cm,0cm}}\caption{\aso}\vspace*{-0.25cm}\hspace*{.05cm}
			\includegraphics[width=1.04\linewidth]{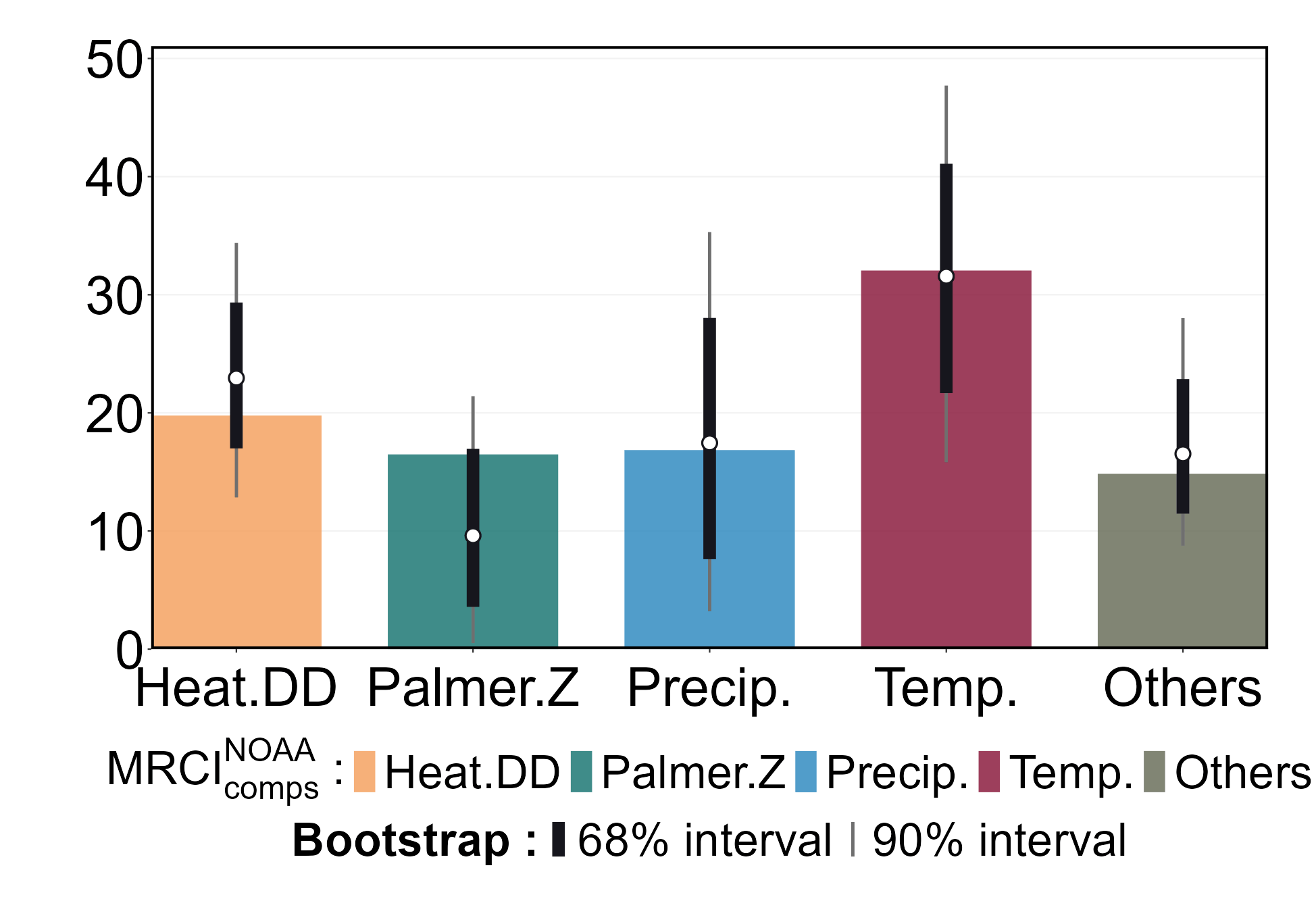}
		\end{subfigure}

		\vspace*{-0.1cm}
		\begin{subfigure}[t]{0.33\textwidth}\centering
			\hspace*{-.05cm}\includegraphics[width=1.04\linewidth]{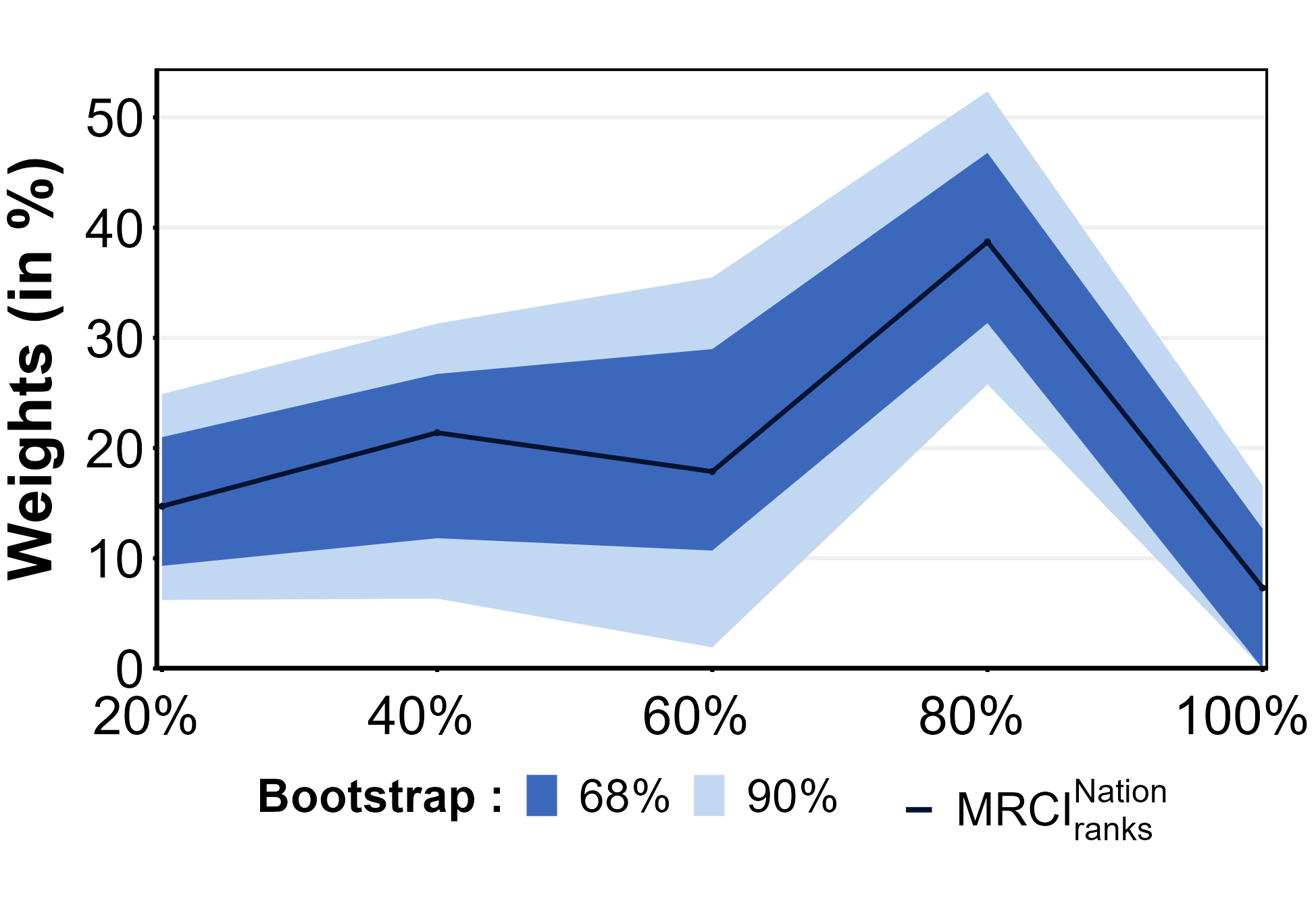}
		\end{subfigure}\hfill
		\begin{subfigure}[t]{0.33\textwidth}\centering
			\hspace*{.05cm}\includegraphics[width=1.04\linewidth]{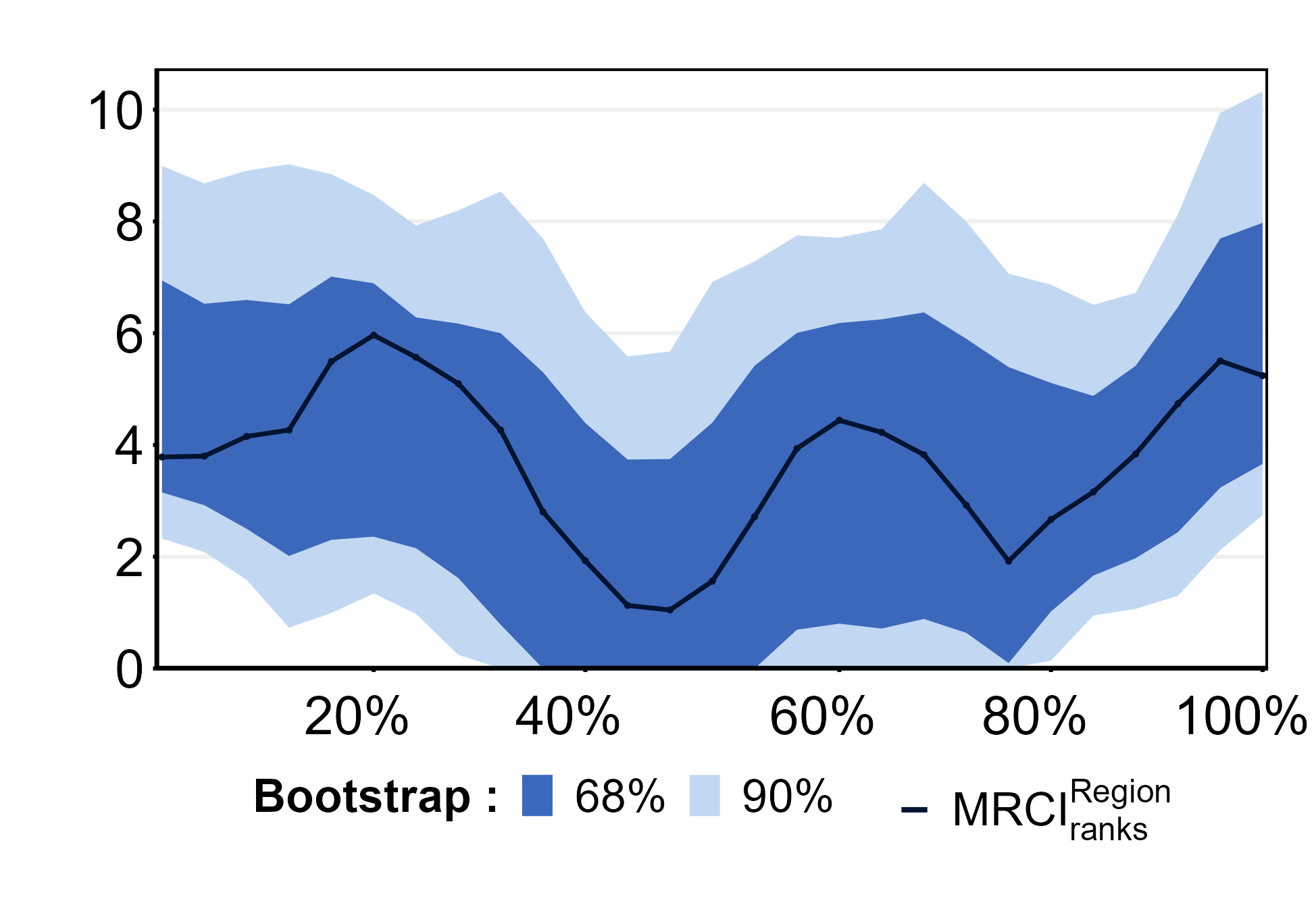}
		\end{subfigure}\hfill
		\begin{subfigure}[t]{0.33\textwidth}\centering
			\hspace*{.15cm}\includegraphics[width=1.04\linewidth]{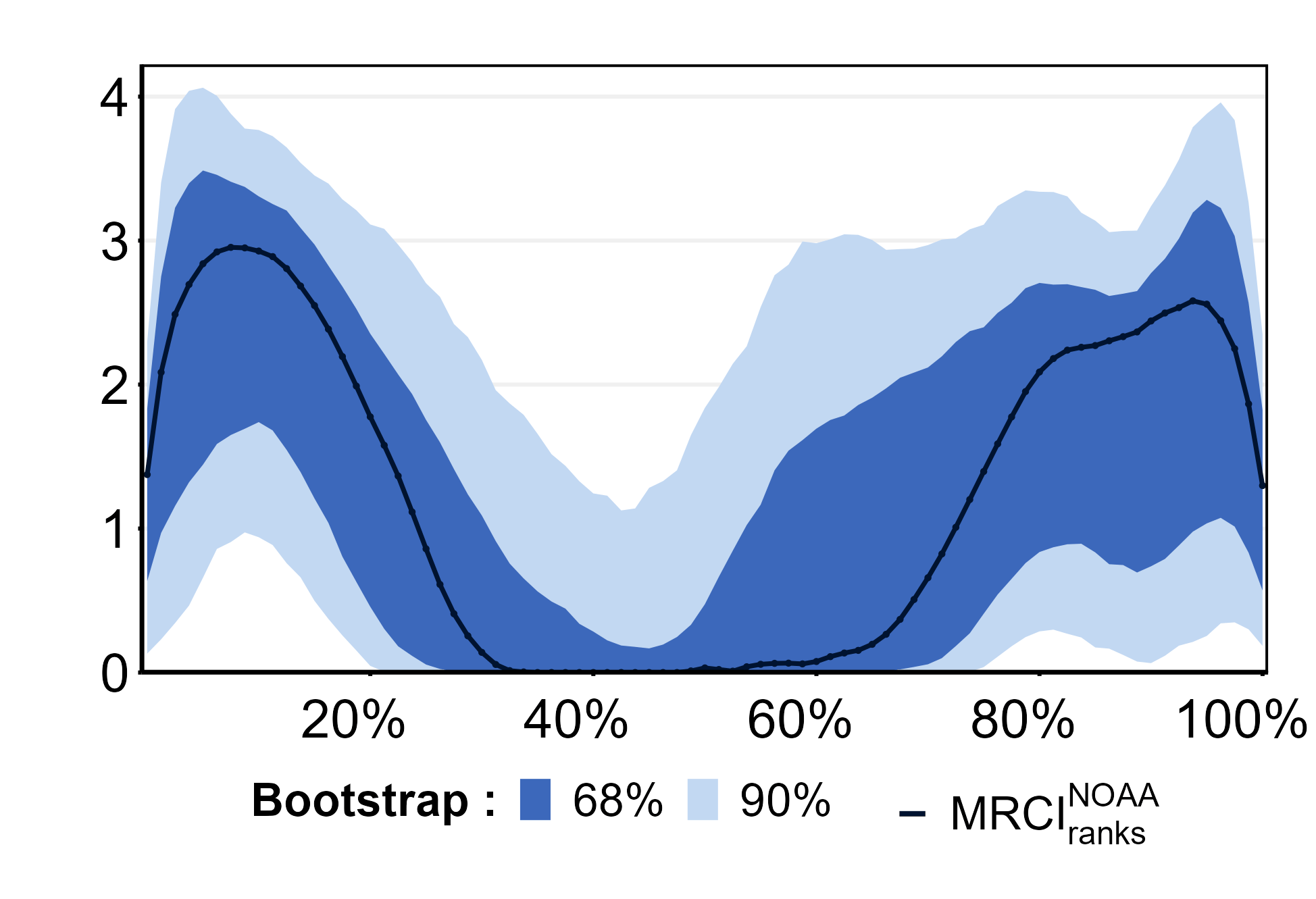}
		\end{subfigure}
	\end{center}
	\begin{threeparttable}
		\centering\vspace{-1cm}
		\begin{minipage}{\textwidth}
			\begin{tablenotes}[para,flushleft]
				\setlength{\lineskip}{0.2ex}
				\notsotiny
				{\textit{Notes}: Bootstrap distribution of the estimated aggregation weights $\widehat{\boldsymbol{w}}$ at the 2025 vintage, from $B=1{,}000$ \textcolor{black}{weight-reestimating pairs-bootstrap replications} \textcolor{black}{(all $1{,}000$ replications returned a weight solution and a usable VAR fit; none was discarded)}, with the cross-validated penalty $\widehat\lambda$ held fixed. \textbf{Top row}: component-space weights $\widehat{\boldsymbol{w}}_c(\widehat\lambda)$ aggregated to their variable families; colored bars give the point estimate, the thick dark bar the $68\%$ bootstrap interval, the thin bar the $90\%$ interval, and the white dot the bootstrap median. \textbf{Bottom row}: rank-space weights $\widehat{\boldsymbol{w}}_r(\widehat\lambda)$ shown as a fan of the smoothed rank profile (navy line) wrapped in its $68\%$ and $90\%$ bands. To ensure an adequate representation, the same smoothing is applied as in the main paper figures. Family totals are determined far more precisely than individual loadings.}
			\end{tablenotes}
		\end{minipage}
	\end{threeparttable}
\end{figure}

\paragraph{Bootstrap weights.}
Figure~\ref{fig:boot_weights} shows the bootstrap distribution of $\widehat{\boldsymbol{w}}$, aggregated to the named variable families. The family totals are well determined and the dominant weight allocations survive resampling, high winds at the national and regional levels and temperature at the NOAA level. What the resamples do not settle is which of several overlapping regions to credit: whether the Midwest or the Southern Plains carries the wind signal shifts from one draw to the next. That is what one expects when several regions proxy the same macro signal, and it washes out in the family totals we interpret. The rank-space profiles keep their tail-loaded shape across draws at the Regional and NOAA levels. The National profile is the exception, placing its highest mass near the 80th percentile, which is unsurprising with only five series to rank.

\begin{figure}[!t]
	\caption{{Bootstrap stability of the assembled index}}
	\label{fig:boot_index}
	\begin{center}
		\begin{subfigure}[t]{0.50\textwidth}\centering
        \caption{MRCI$_{\text{comps}}$}
			\hspace*{-.5cm}\includegraphics[width=1.03\textwidth]{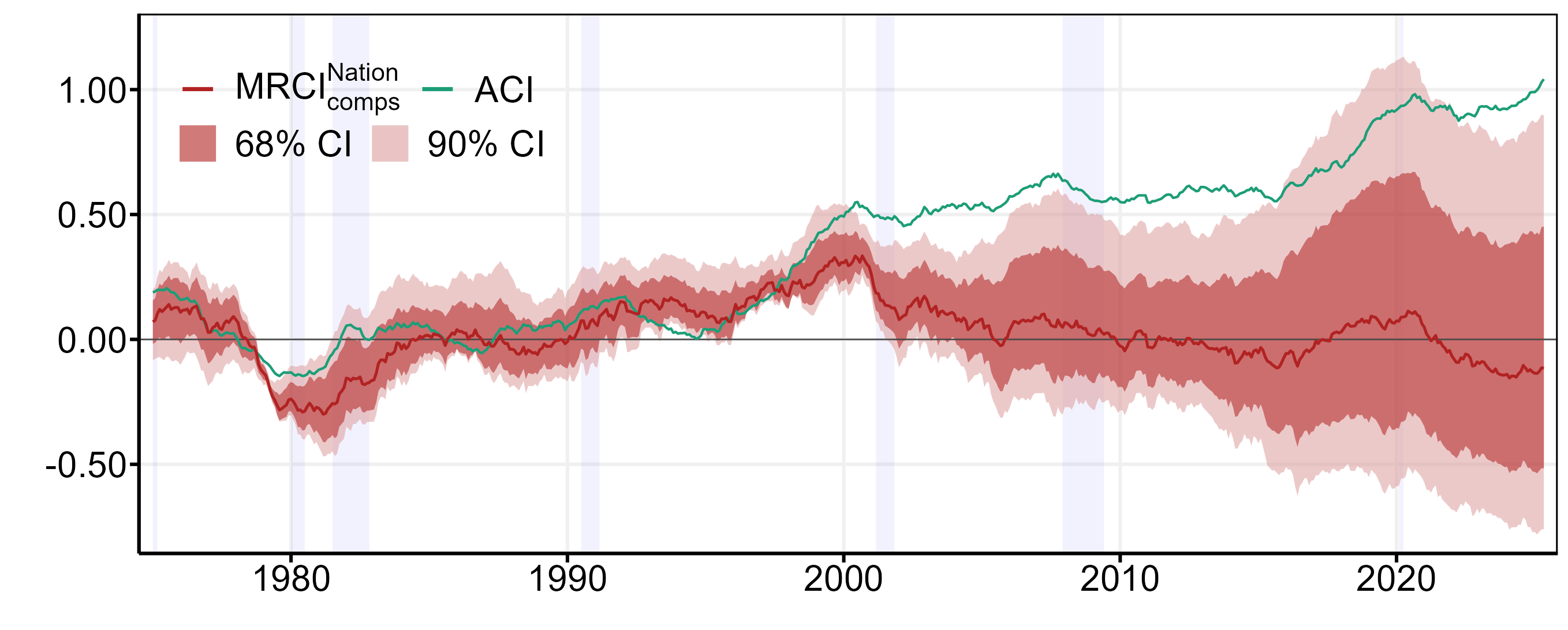}
		\end{subfigure}\hfill
		\begin{subfigure}[t]{0.5\textwidth}\centering
        \caption{MRCI$_{\text{ranks}}$}
			\includegraphics[width=1.03\textwidth]{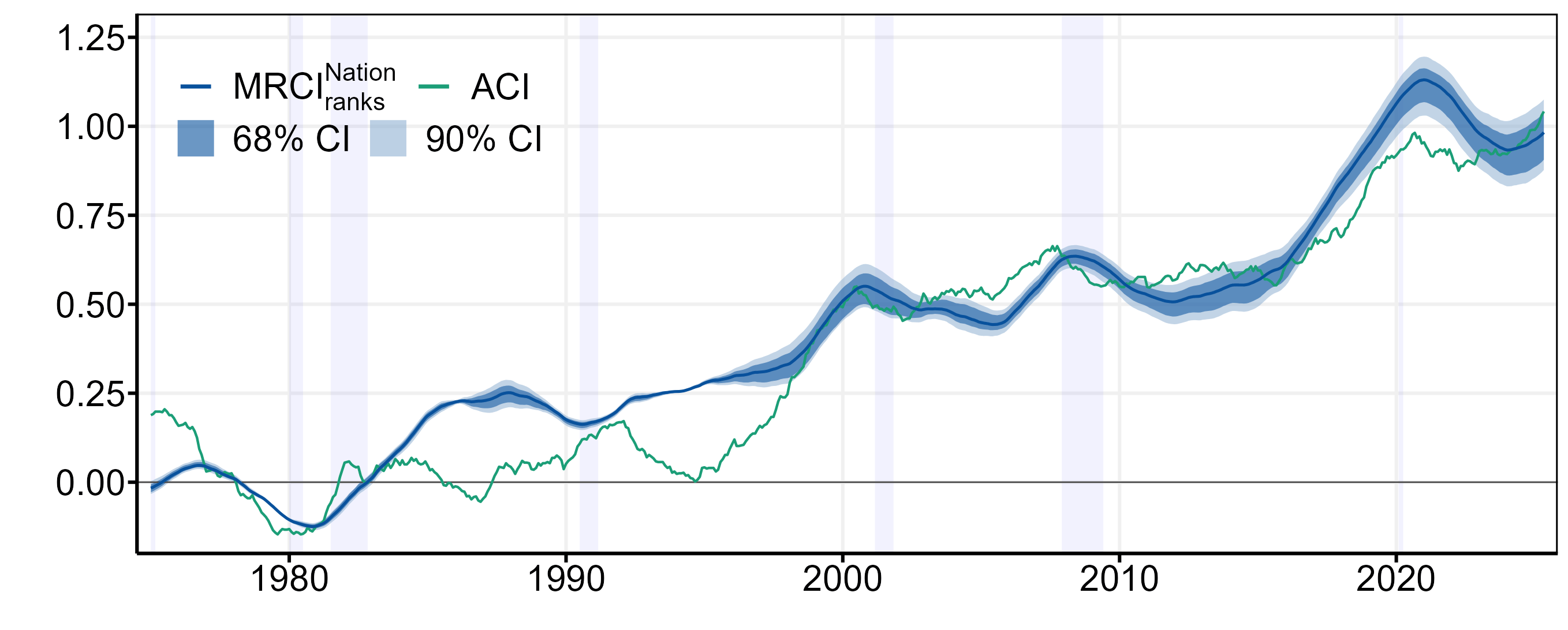}
		\end{subfigure}

		\vspace*{0.12cm}
		\begin{subfigure}[t]{0.50\textwidth}\centering
			\hspace*{-.5cm}\includegraphics[width=1.03\textwidth]{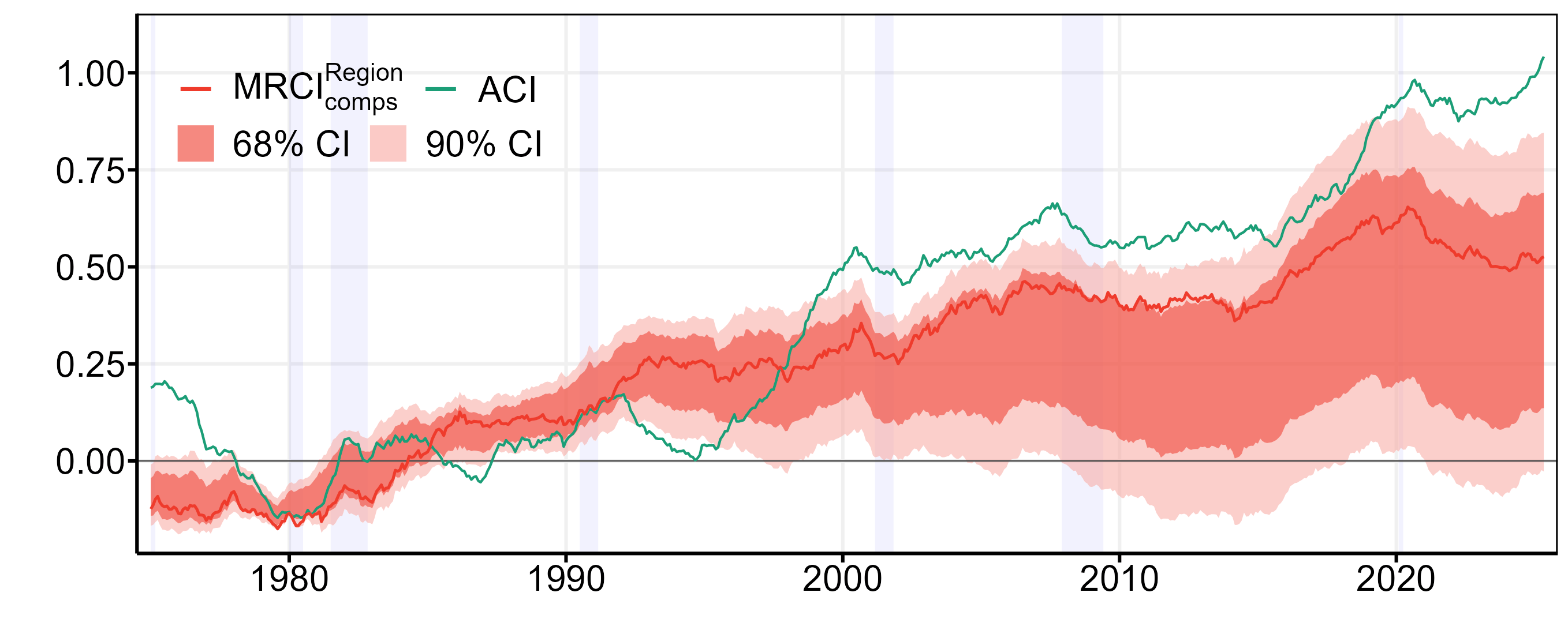}
		\end{subfigure}\hfill
		\begin{subfigure}[t]{0.50\textwidth}\centering
			\hspace*{0cm}\includegraphics[width=1.03\textwidth]{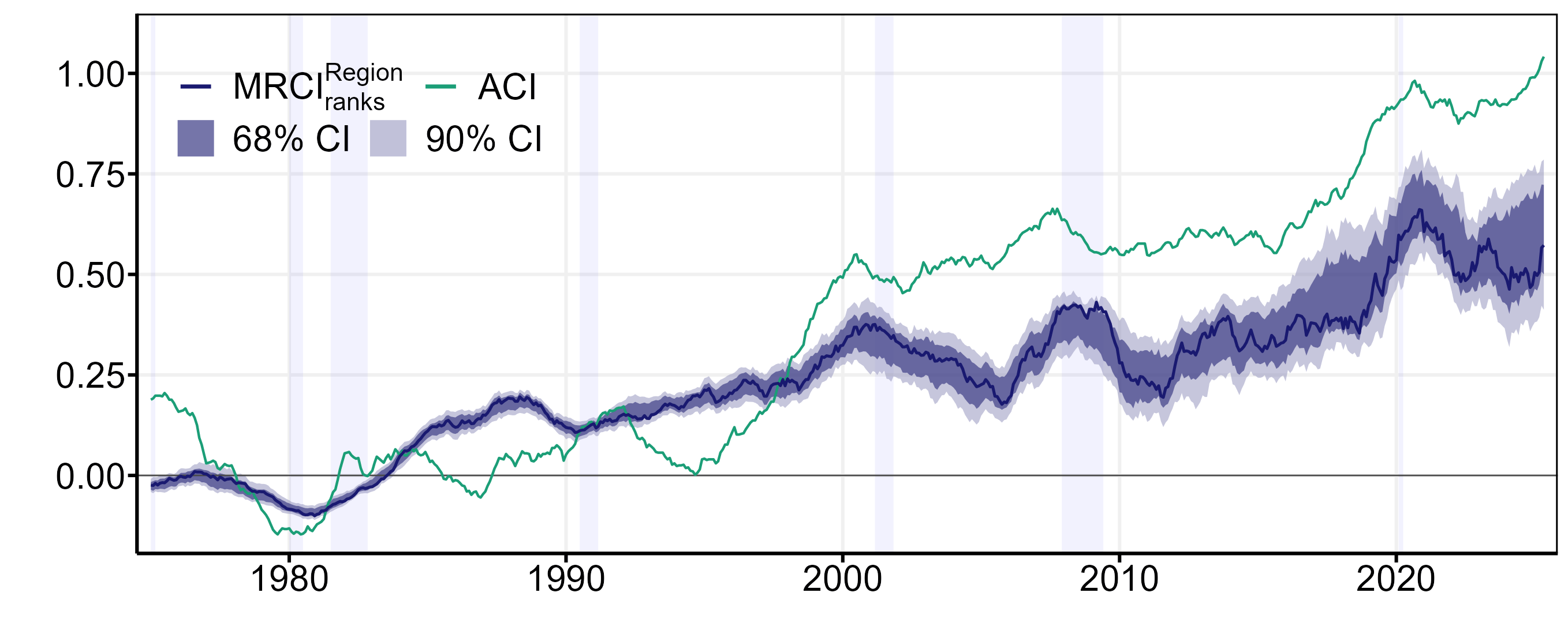}
		\end{subfigure}

		\vspace*{0.12cm}
		\begin{subfigure}[t]{0.50\textwidth}\centering
			\hspace*{-.5cm}\includegraphics[width=1.03\textwidth]{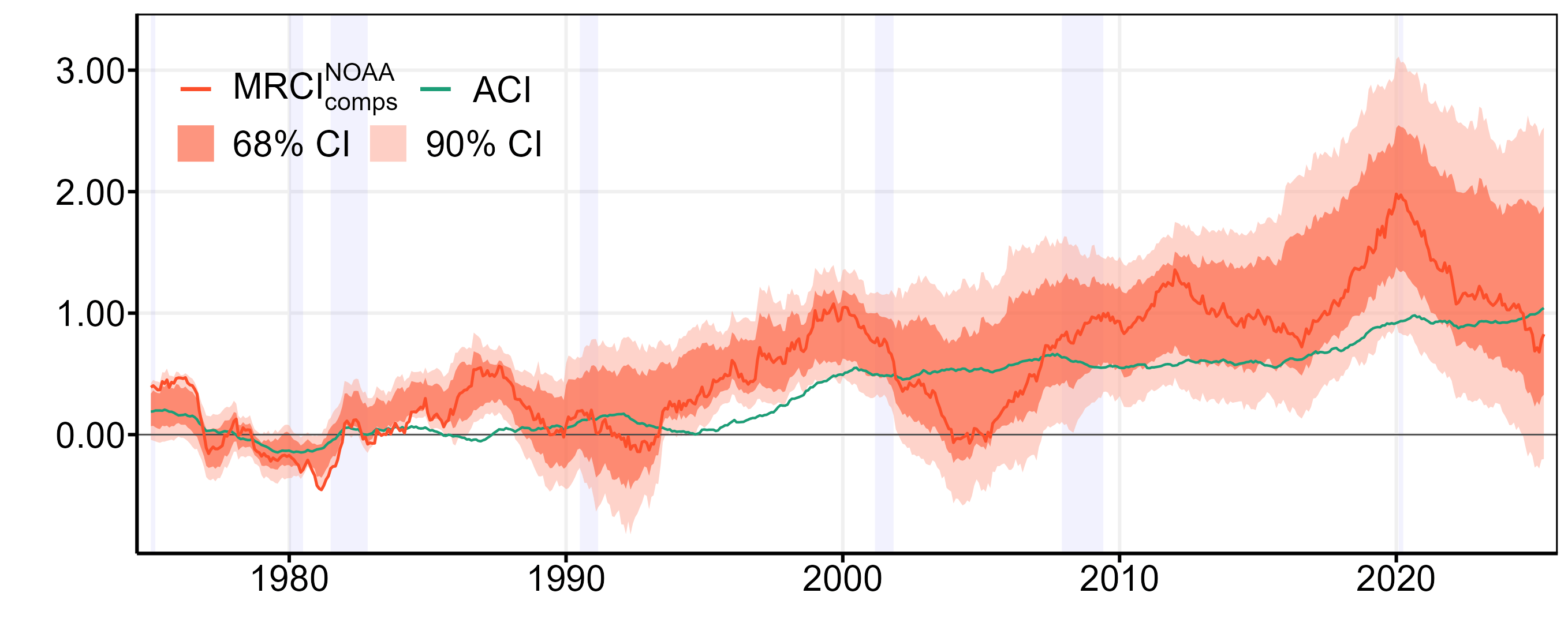}
		\end{subfigure}\hfill
		\begin{subfigure}[t]{0.50\textwidth}\centering
			\hspace*{0cm}\includegraphics[width=1.03\textwidth]{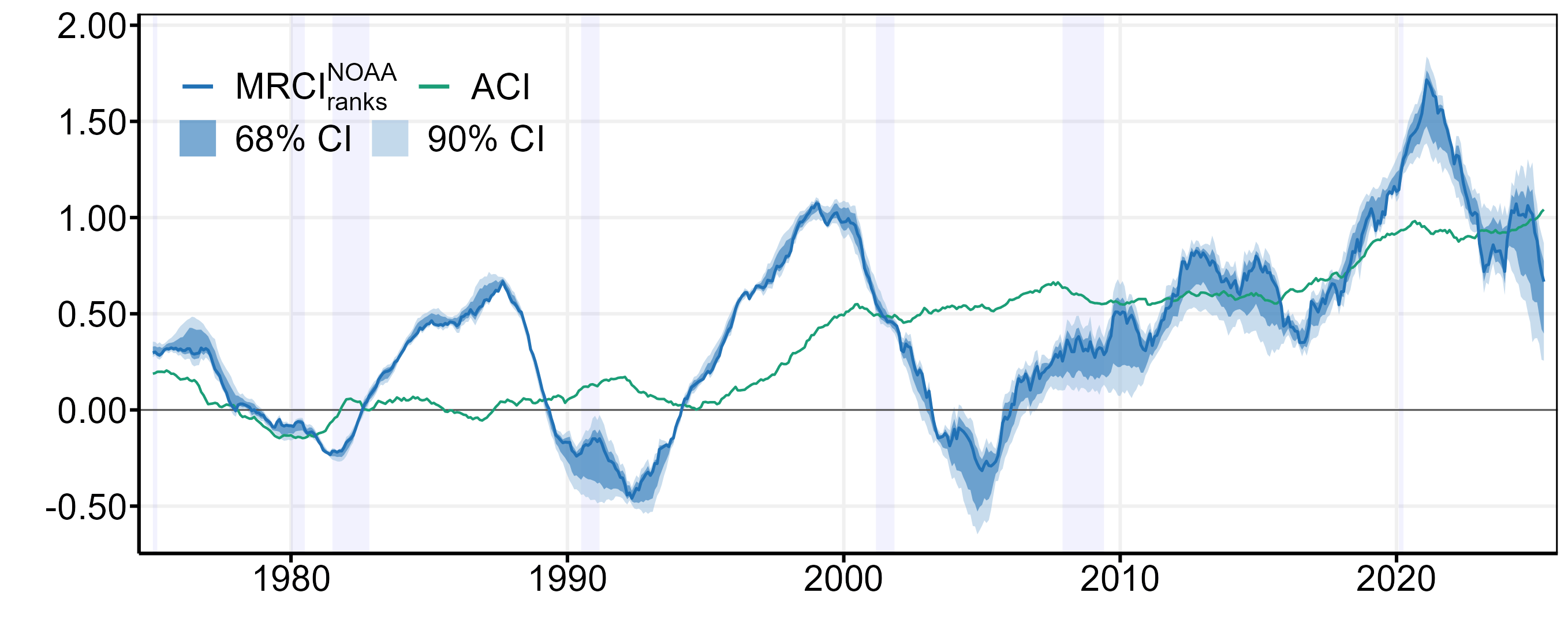}
		\end{subfigure}
	\end{center}
	\begin{threeparttable}
		\centering \vspace*{-1cm}
		\begin{minipage}{\textwidth}
			\begin{tablenotes}[para,flushleft]
				\setlength{\lineskip}{0.2ex}
				\notsotiny
				{\textit{Notes}: Each panel reconstructs the assembled index from every \textcolor{black}{estimated} bootstrap weight vector (all $B=1{,}000$ replications; none discarded) and reports the resulting
				$68\%$ and $90\%$ pointwise bands (shaded) around the point-estimate index
				(solid). The headline equal-weighted ACI is shown for reference (green line).
				The two columns are the component-space (MRCI$_\text{comps}$, \textbf{left column})
				and the rank-space (MRCI$_\text{ranks}$, \textbf{right column}); the three rows are
				the Nation (\textbf{top row}), Region (\textbf{middle row}), and NOAA (\textbf{bottom row}) aggregation levels. All
				series are five-year moving averages of the standardized components, with
				weights estimated over 1975--2025; the rank-space indices are additionally
				smoothed exactly as in the main paper.}\vspace{-.3cm}
			\end{tablenotes}
		\end{minipage}
	\end{threeparttable}
\end{figure}

\paragraph{The assembled index.}
What ultimately enters the VAR is the scalar index $x_t^*(\widehat{\boldsymbol{w}})=\widehat{\boldsymbol{w}}'\mathbf{X}_t$. It is the precision of this index, rather than that of the individual weights, which is important for inference.
 Reconstructing it from each bootstrap weight vector (Figure~\ref{fig:boot_index}) leaves the chronology emphasized in the main text intact: the point estimate shifts only a little, and its broad arc is preserved, the post-2000 flattening and, in some specifications, a mild downward drift after it. The point index also stays central to its own bootstrap distribution, falling within the central $68\%$ at about $88\%$ of dates across the three component-space specifications and about $99\%$ in rank-space, the latter tighter in part by construction. The reason the index is so much steadier than its ingredients is that several of the standardized series carry overlapping information, so shifting weight among them changes the weighted sum they produce far less than it changes the weights themselves.

\begin{figure}[t!]
	\caption{{\textcolor{black}{Weight-Reestimating Pairs Bootstrap of Component-Space Climate Measure Shocks}}} \label{fig:Caci_boot_irf}
	\begin{center}
		
		\begin{subfigure}[t]{.344\textwidth}
			\vspace*{-0.4cm}
			\captionsetup{margin={.40cm,0cm}} 
			\centering\caption{National}
			\vspace*{-0.16cm}
			\hspace*{-1mm}\includegraphics[width=1.01\textwidth,trim = 17mm 166mm 11.5mm 47mm, clip]{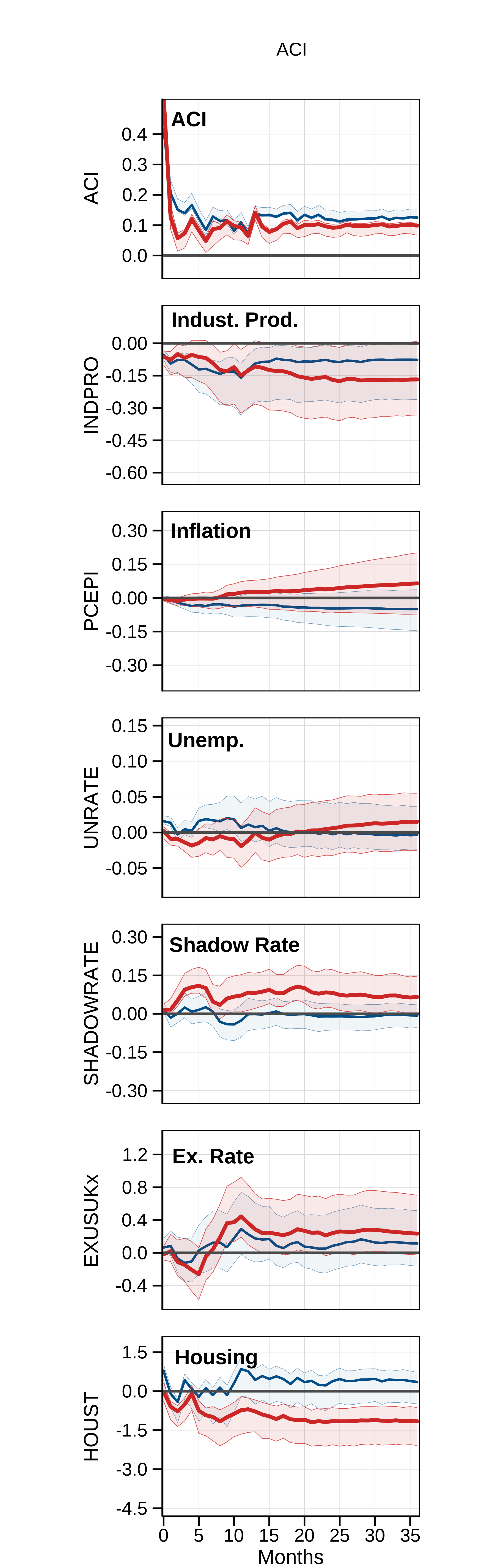}
		\end{subfigure}
		\hspace*{-4.25mm}
		\begin{subfigure}[t]{.344\textwidth}
			\vspace*{-0.4cm}
			\captionsetup{margin={-0.3cm,0cm}} 
			\centering\caption{Regional}
			\vspace*{-0.16cm}
			\hspace*{2.8mm}\includegraphics[width=.89\textwidth, trim = 24.5mm 166mm 9.5mm 47mm, clip]{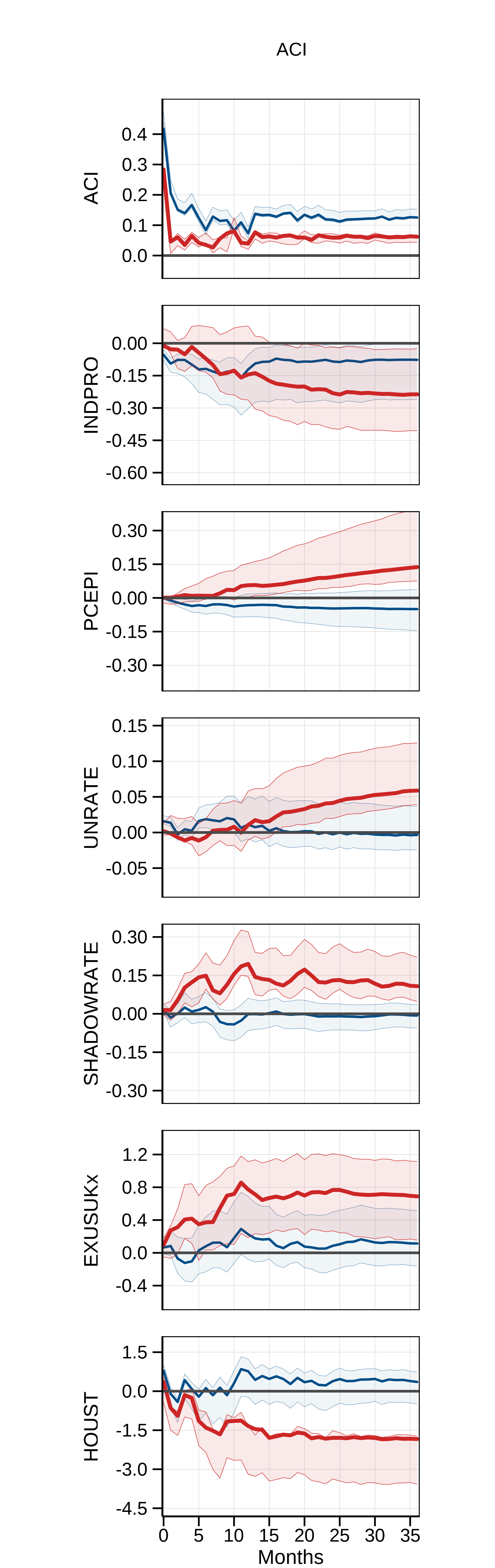}
		\end{subfigure}%
		\hspace*{-7mm}
		\begin{subfigure}[t]{.344\textwidth}
			\vspace*{-0.4cm}
			\captionsetup{margin={-0.3cm,0cm}} 
			\centering\caption{NOAA}
			\vspace*{-0.16cm}
			\hspace*{4.25mm}\includegraphics[width=.89\textwidth, trim = 24.5mm 166mm 9.5mm 47mm, clip]{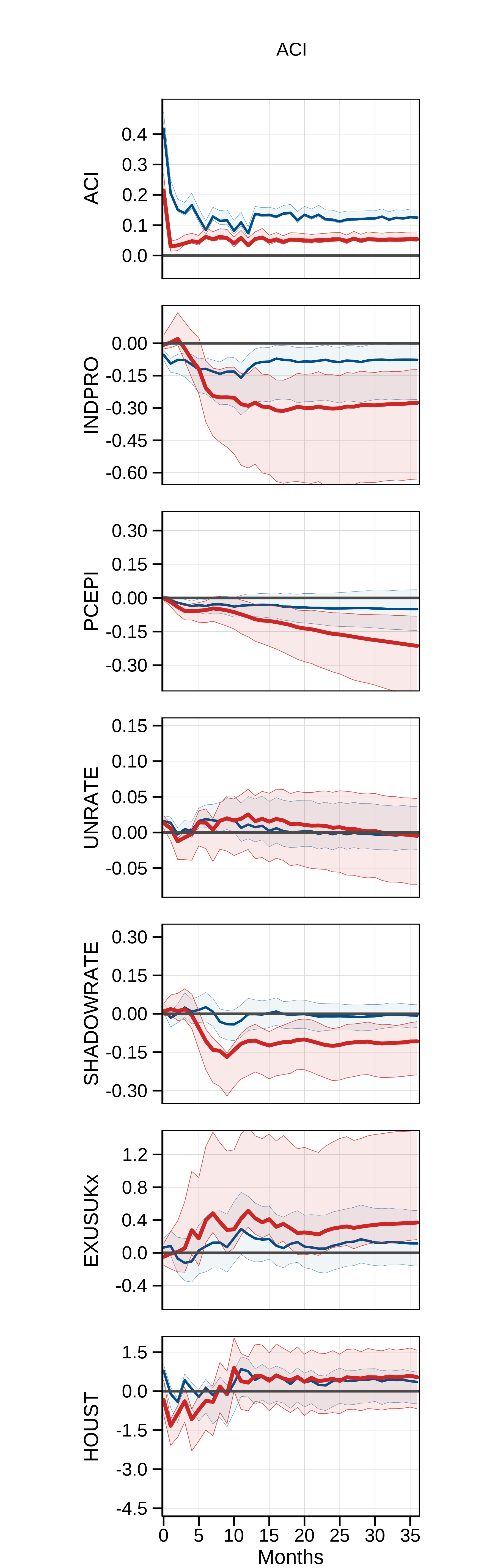}
		\end{subfigure}%
		
		\begin{subfigure}[t]{.344\textwidth}
			\vspace*{0.18cm}
			\hspace*{-1.5mm}\includegraphics[width=1.02\textwidth,trim = 16.5mm 102.9mm 11.5mm 110mm, clip]{figure/IRFBOOT_ACI_Comp_lvl_1_shockACI_Sfull_RUN0814.1049_v2026-08-19.png}
		\end{subfigure}
		\begin{subfigure}[t]{.344\textwidth}
			\vspace*{.16cm}
			\hspace*{0.6mm}\includegraphics[width=.89\textwidth, trim = 24.5mm 103mm 9.5mm 110mm, clip]{figure/IRFBOOT_ACI_Comp_lvl_2_shockACI_Sfull_RUN0814.0906_v2026-08-19.png}
		\end{subfigure}%
		\hspace*{-10.5mm}
		\begin{subfigure}[t]{.344\textwidth}
			\vspace*{.16cm}
			\hspace*{5.1mm}\includegraphics[width=.89\textwidth, trim = 24.5mm 103mm 9.5mm 110mm, clip]{figure/IRFBOOT_ACI_Comp_lvl_3_shockACI_Sfull_RUN0814.0903_v2026-08-19.png}
		\end{subfigure}%
		
		\begin{subfigure}[t]{.344\textwidth}
			\vspace*{-0.075cm}
			\hspace*{-.8mm}\includegraphics[width=1.01\textwidth,trim = 17mm 3mm 11.5mm 202.3mm, clip]{figure/IRFBOOT_ACI_Comp_lvl_1_shockACI_Sfull_RUN0814.1049_v2026-08-19.png}
		\end{subfigure}
		\begin{subfigure}[t]{.344\textwidth}
			\vspace*{-.09cm}
			\hspace*{.6mm}\includegraphics[width=.89\textwidth, trim = 24.5mm 3mm 9.5mm 202mm, clip]{figure/IRFBOOT_ACI_Comp_lvl_2_shockACI_Sfull_RUN0814.0906_v2026-08-19.png}
		\end{subfigure}%
		\hspace*{-10.5mm}
		\begin{subfigure}[t]{.344\textwidth}
			\vspace*{-.09cm}
			\hspace*{5.1mm}\includegraphics[width=.89\textwidth, trim = 24.5mm 3mm 9.5mm 202mm, clip]{figure/IRFBOOT_ACI_Comp_lvl_3_shockACI_Sfull_RUN0814.0903_v2026-08-19.png}
		\end{subfigure}%
		
		\begin{subfigure}[t]{.38\textwidth}
			\vspace*{-0.08cm}
			\hspace*{-1mm}\includegraphics[width=1.05\textwidth, trim = 13mm 3.8mm 5.5mm 235mm, clip]{figure/Naming_ACI.png}
		\end{subfigure}%
		
		\begin{subfigure}[t]{.38\textwidth}
			\vspace*{-0.07cm}
			\hspace*{-4mm}\includegraphics[width=1.05\textwidth, trim = 13mm 0mm 5.5mm 237.5mm, clip]{figure/Naming_ACI.png}
		\end{subfigure}%
	\end{center}
	
	\begin{threeparttable}
		\centering \vspace*{-0.6cm}
		\begin{minipage}{\textwidth}
			\begin{tablenotes}[para,flushleft]
				\setlength{\lineskip}{0.2ex}
				\notsotiny 
				{\textit{Notes}: Component-space impulse responses to a one-standard-deviation innovation in the assembled climate index $x_t^*(\widehat{\boldsymbol{w}})$, with \textcolor{black}{weight-reestimating pairs bootstrap bands}: each of $B=1{,}000$ replications re-solves for $\widehat{\boldsymbol{w}}$ and then re-estimates $(\mathbf{B},\boldsymbol{\Sigma})$ on the resulting index, with the penalty fixed at its cross-validated value. VAR models are estimated on data from 1975:M1 to 2025:M4, excluding the COVID-19 period (2020:M1--2021:M6), with 12 lags and the ordering ACI/MRCI, \mn{INDPRO}, \mn{PCEPI}, \mn{UNRATE}, \mn{SHADOWRATE}, \mn{EXUSUKx}, \mn{HOUST}. Bands are at the $68\%$ level. Selected variables are shown; complete results are in Appendix~\ref{annexe_climate} (Figure~\ref{fig:FULL_Caci_boot_irf}). This is the \textcolor{black}{weight-reestimating pairs bootstrap counterpart to} Figure~\ref{fig:Caci_irf}, with the aggregation weights re-estimated in every replication.}\vspace{-.3cm}
			\end{tablenotes}
		\end{minipage}
	\end{threeparttable}
\end{figure}

\begin{figure}[t!]
	\caption{{\textcolor{black}{Weight-Reestimating Pairs Bootstrap of Rank-Space Climate Measure Shocks}}} \label{fig:Raci_boot_irf}
	\begin{center}
		
		\begin{subfigure}[t]{.344\textwidth}
			\vspace*{-0.4cm}
			\captionsetup{margin={.40cm,0cm}} 
			\centering\caption{National}
			\vspace*{-0.16cm}
			\hspace*{-1mm}\includegraphics[width=1.01\textwidth,trim = 17mm 166mm 11.5mm 47mm, clip]{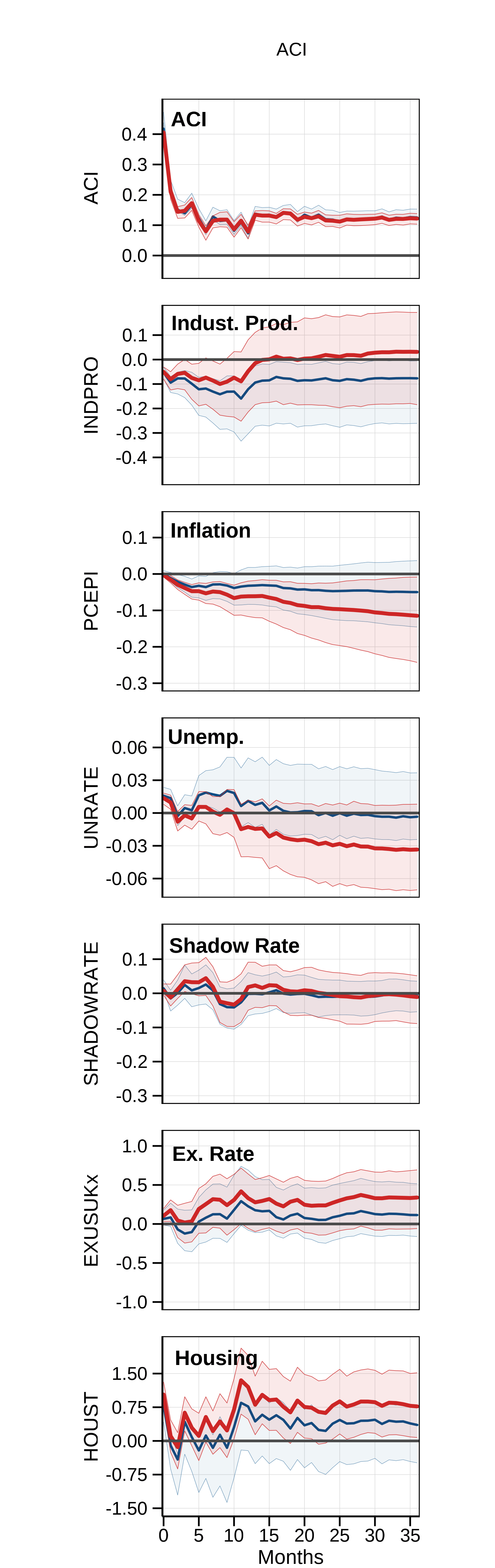}
		\end{subfigure}
		\hspace*{-4.25mm}
		\begin{subfigure}[t]{.344\textwidth}
			\vspace*{-0.4cm}
			\captionsetup{margin={-0.3cm,0cm}} 
			\centering\caption{Regional}
			\vspace*{-0.16cm}
			\hspace*{2.8mm}\includegraphics[width=.89\textwidth, trim = 24.5mm 166mm 9.5mm 47mm, clip]{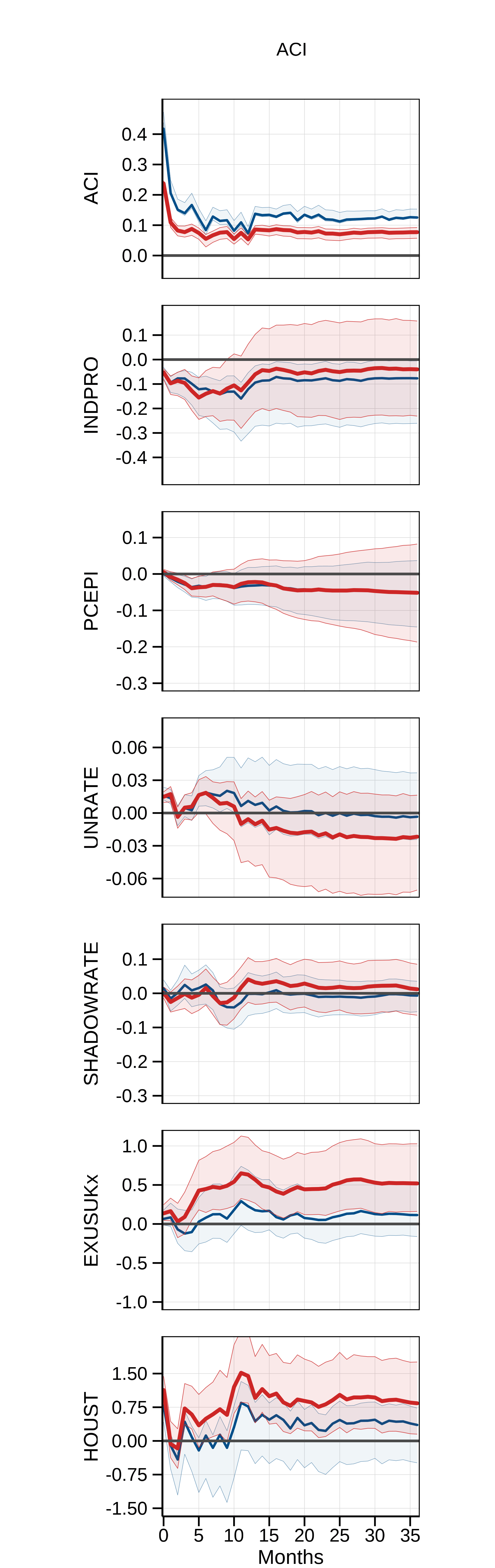}
		\end{subfigure}%
		\hspace*{-7mm}
		\begin{subfigure}[t]{.344\textwidth}
			\vspace*{-0.4cm}
			\captionsetup{margin={-0.3cm,0cm}} 
			\centering\caption{NOAA}
			\vspace*{-0.16cm}
			\hspace*{4.25mm}\includegraphics[width=.89\textwidth, trim = 24.5mm 166mm 9.5mm 47mm, clip]{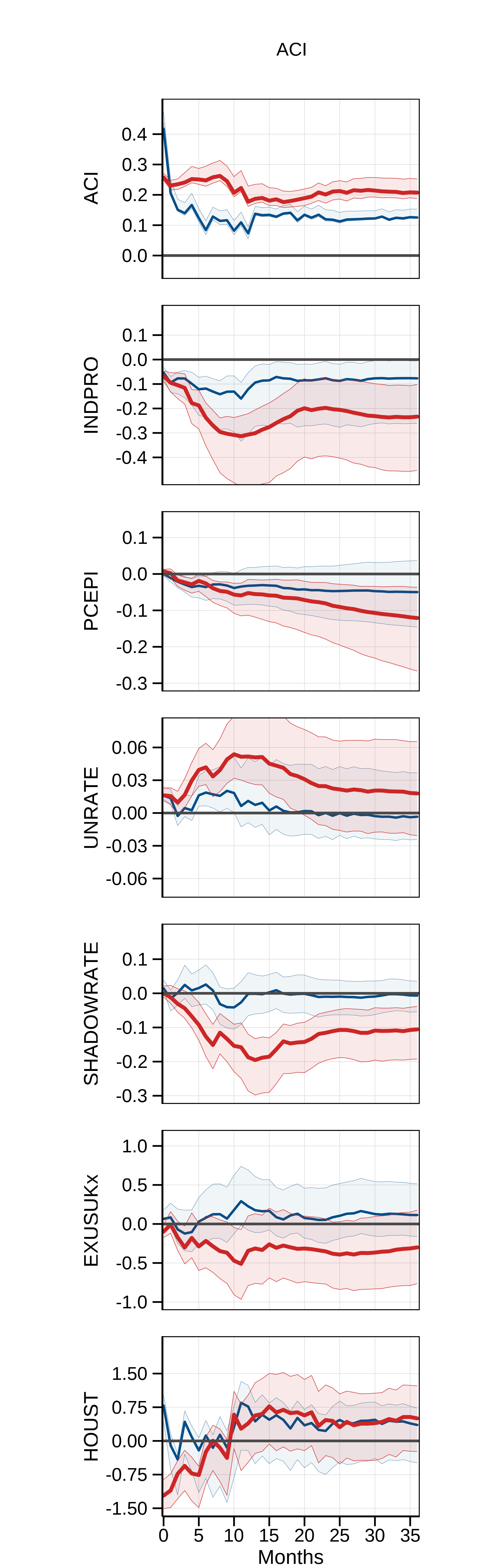}
		\end{subfigure}%
		
		\begin{subfigure}[t]{.344\textwidth}
			\vspace*{0.18cm}
			\hspace*{-1.5mm}\includegraphics[width=1.02\textwidth,trim = 16.5mm 102.9mm 11.5mm 110mm, clip]{figure/IRFBOOT_ACI_Rank_lvl_1_shockACI_Sfull_RUN0814.1043_v2026-08-20.png}
		\end{subfigure}
		\begin{subfigure}[t]{.344\textwidth}
			\vspace*{.16cm}
			\hspace*{0.6mm}\includegraphics[width=.89\textwidth, trim = 24.5mm 103mm 9.5mm 110mm, clip]{figure/IRFBOOT_ACI_Rank_lvl_2_shockACI_Sfull_RUN0814.0959_v2026-08-20.png}
		\end{subfigure}%
		\hspace*{-10.5mm}
		\begin{subfigure}[t]{.344\textwidth}
			\vspace*{.16cm}
			\hspace*{5.1mm}\includegraphics[width=.89\textwidth, trim = 24.5mm 103mm 9.5mm 110mm, clip]{figure/IRFBOOT_ACI_Rank_lvl_3_shockACI_Sfull_RUN0814.1010_v2026-08-20.png}
		\end{subfigure}%
		
		\begin{subfigure}[t]{.344\textwidth}
			\vspace*{-0.075cm}
			\hspace*{-.8mm}\includegraphics[width=1.01\textwidth,trim = 17mm 3mm 11.5mm 202.3mm, clip]{figure/IRFBOOT_ACI_Rank_lvl_1_shockACI_Sfull_RUN0814.1043_v2026-08-20.png}
		\end{subfigure}
		\begin{subfigure}[t]{.344\textwidth}
			\vspace*{-.09cm}
			\hspace*{.6mm}\includegraphics[width=.89\textwidth, trim = 24.5mm 3mm 9.5mm 202mm, clip]{figure/IRFBOOT_ACI_Rank_lvl_2_shockACI_Sfull_RUN0814.0959_v2026-08-20.png}
		\end{subfigure}%
		\hspace*{-10.5mm}
		\begin{subfigure}[t]{.344\textwidth}
			\vspace*{-.09cm}
			\hspace*{5.1mm}\includegraphics[width=.89\textwidth, trim = 24.5mm 3mm 9.5mm 202mm, clip]{figure/IRFBOOT_ACI_Rank_lvl_3_shockACI_Sfull_RUN0814.1010_v2026-08-20.png}
		\end{subfigure}%
		
		\begin{subfigure}[t]{.38\textwidth}
			\vspace*{-0.08cm}
			\hspace*{-1mm}\includegraphics[width=1.05\textwidth, trim = 13mm 3.8mm 5.5mm 235mm, clip]{figure/Naming_ACI.png}
		\end{subfigure}%
		
		\begin{subfigure}[t]{.38\textwidth}
			\vspace*{-0.07cm}
			\hspace*{-4mm}\includegraphics[width=1.05\textwidth, trim = 13mm 0mm 5.5mm 237.5mm, clip]{figure/Naming_ACI.png}
		\end{subfigure}%
	\end{center}
	
	\begin{threeparttable}
		\centering \vspace*{-0.6cm}
		\begin{minipage}{\textwidth}
			\begin{tablenotes}[para,flushleft]
				\setlength{\lineskip}{0.2ex}
				\notsotiny 
				{\textit{Notes}: Rank-space impulse responses to a one-standard-deviation innovation in the assembled climate index $x_t^*(\widehat{\boldsymbol{w}})$, with \textcolor{black}{weight-reestimating pairs bootstrap bands} constructed as in Figure~\ref{fig:Caci_boot_irf}. VAR models are estimated on data from 1975:M1 to 2025:M4, excluding the COVID-19 period (2020:M1--2021:M6), with 12 lags and the ordering ACI/MRCI, \mn{INDPRO}, \mn{PCEPI}, \mn{UNRATE}, \mn{SHADOWRATE}, \mn{EXUSUKx}, \mn{HOUST}. Bands are at the $68\%$ level. Selected variables are shown; complete results are in Appendix~\ref{annexe_climate} (Figure~\ref{fig:FULL_Raci_boot_irf}). This is the \textcolor{black}{weight-reestimating pairs bootstrap counterpart to} Figure~\ref{fig:Raci_irf}, with the aggregation weights re-estimated in every replication.}\vspace{-.3cm}
			\end{tablenotes}
		\end{minipage}
	\end{threeparttable}
\end{figure}

\paragraph{Impulse responses under the \textcolor{black}{weight-reestimating pairs bootstrap}.} Figures~\ref{fig:Caci_boot_irf} and~\ref{fig:Raci_boot_irf} show that the \textcolor{black}{weight-reestimating pairs bootstrap responses} in component-space and rank-space remain close to their \textcolor{black}{fixed-weights residual bootstrap counterparts} in Figures~\ref{fig:Caci_irf} and~\ref{fig:Raci_irf}. The contraction in industrial production remains the dominant feature, and \textcolor{black}{the difference between the assembled-specification responses and the equal-weight benchmark response is preserved}.

\subsection[Expanding-window stability of the weights]{Expanding-window stability of the weights}\label{sec:rolling}
\begin{figure}[!t]
	\caption{Expanding-window stability of the weights}
	\label{fig:rolling_weights}
	\begin{center}
		\begin{subfigure}[t]{0.33\textwidth}\centering
			\vspace*{-0.3cm}\captionsetup{margin={1cm,0cm}}\caption{\asn}\vspace*{-0.25cm}
			\hspace*{-.05cm}\includegraphics[width=1.04\linewidth]{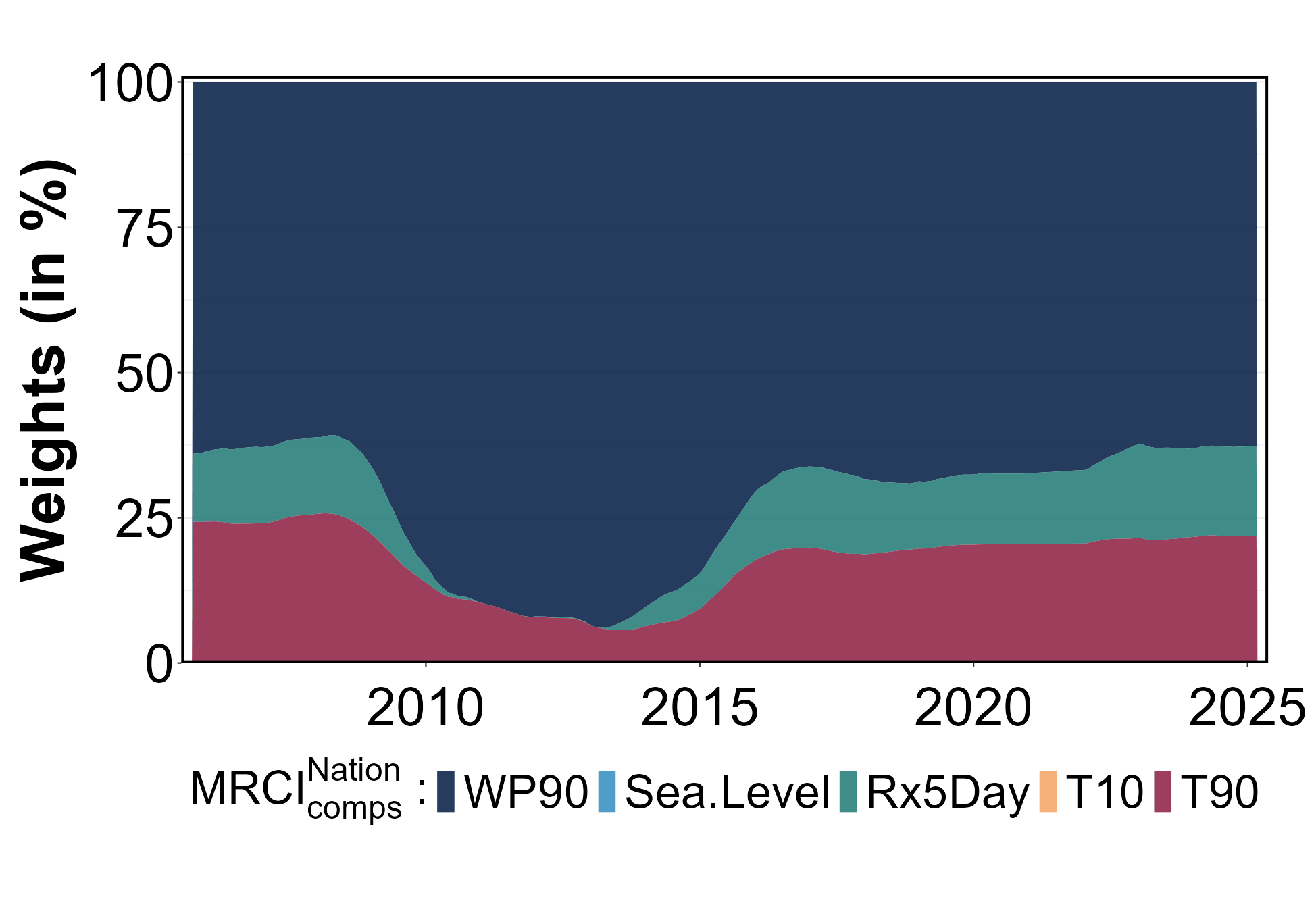}
		\end{subfigure}\hfill
		\begin{subfigure}[t]{0.33\textwidth}\centering
			\vspace*{-0.3cm}\captionsetup{margin={1cm,0cm}}\caption{\asr}\vspace*{-0.25cm}\hspace*{-.045cm}
			\includegraphics[width=1.04\linewidth]{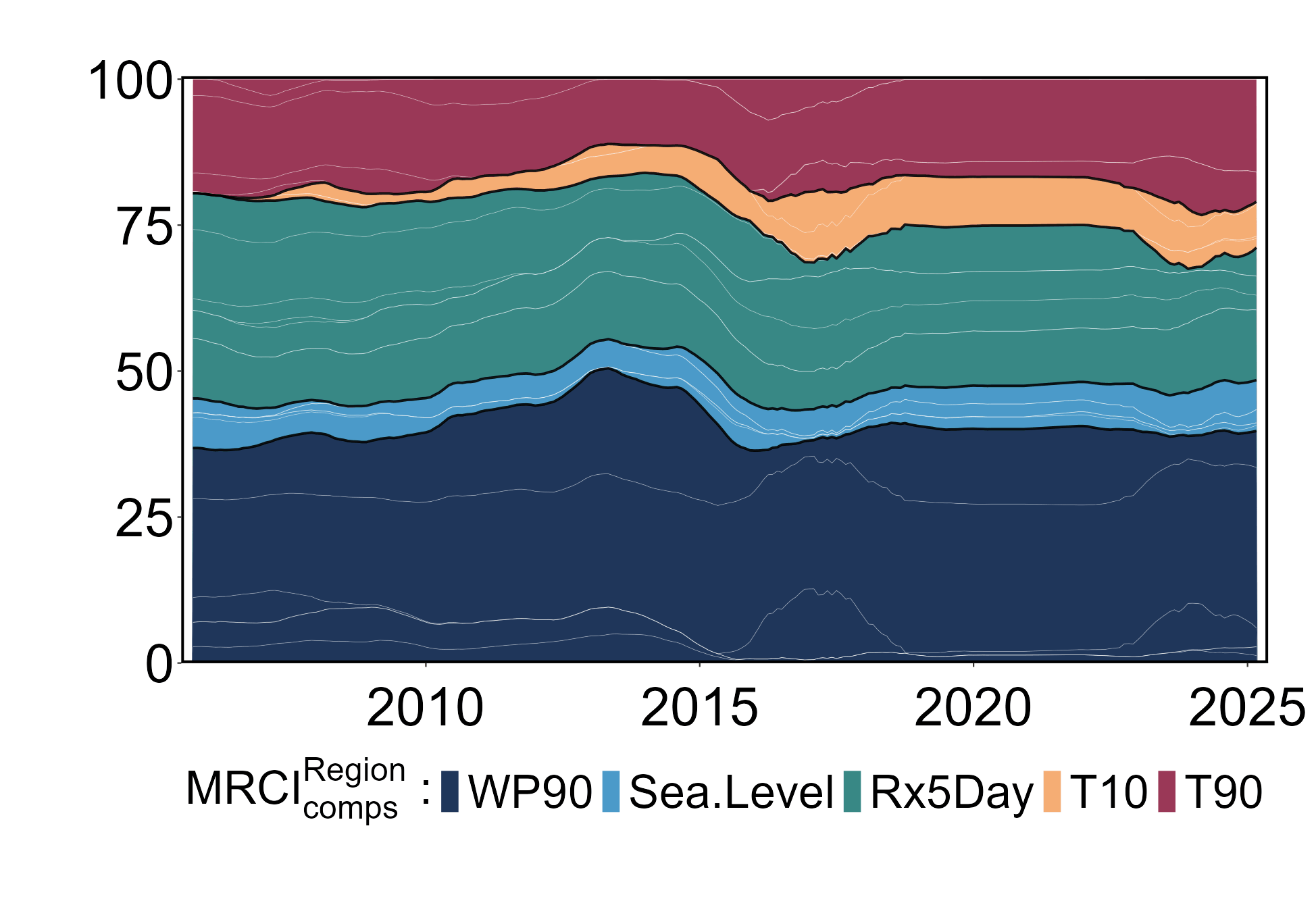}
		\end{subfigure}\hfill
		\begin{subfigure}[t]{0.33\textwidth}\centering
			\vspace*{-0.3cm}\captionsetup{margin={1cm,0cm}}\caption{\aso}\vspace*{-0.25cm}\hspace*{.05cm}
			\includegraphics[width=1.04\linewidth]{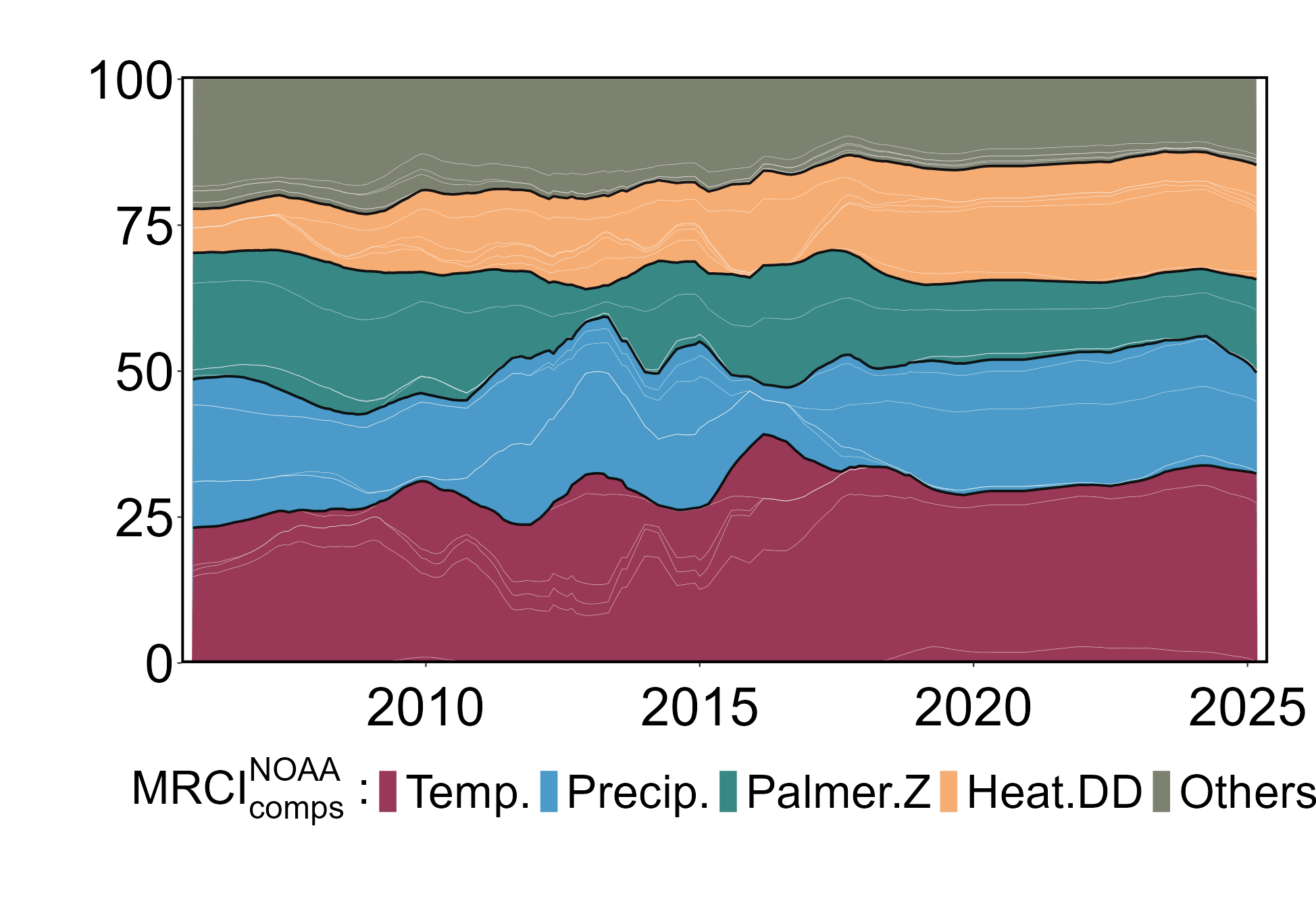}
		\end{subfigure}

		\vspace*{-0.2cm}
		\begin{subfigure}[t]{0.33\textwidth}\centering
			\hspace*{-.05cm}\includegraphics[width=1.04\linewidth]{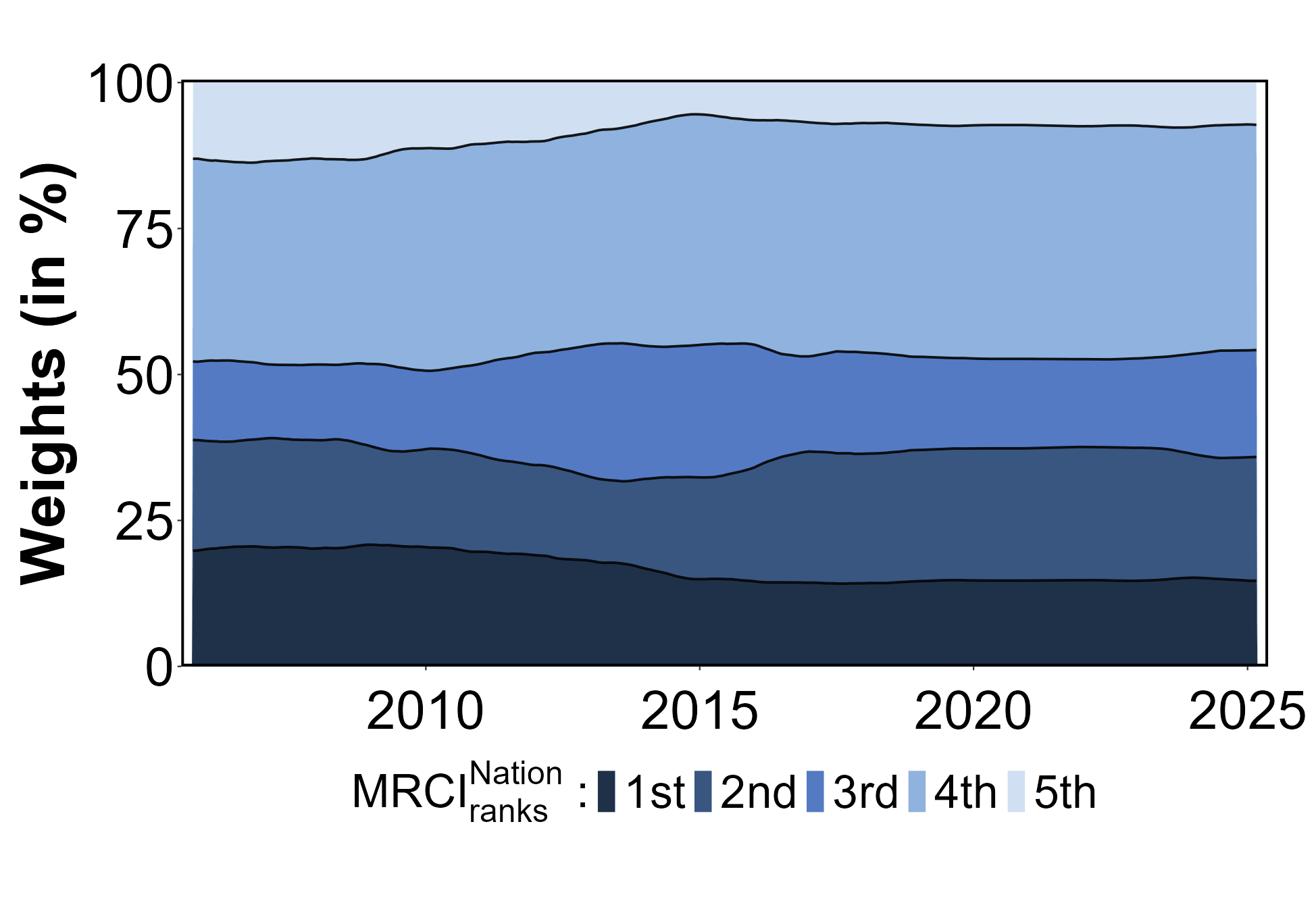}
		\end{subfigure}\hfill
		\begin{subfigure}[t]{0.33\textwidth}\centering
			\hspace*{.05cm}\includegraphics[width=1.04\linewidth]{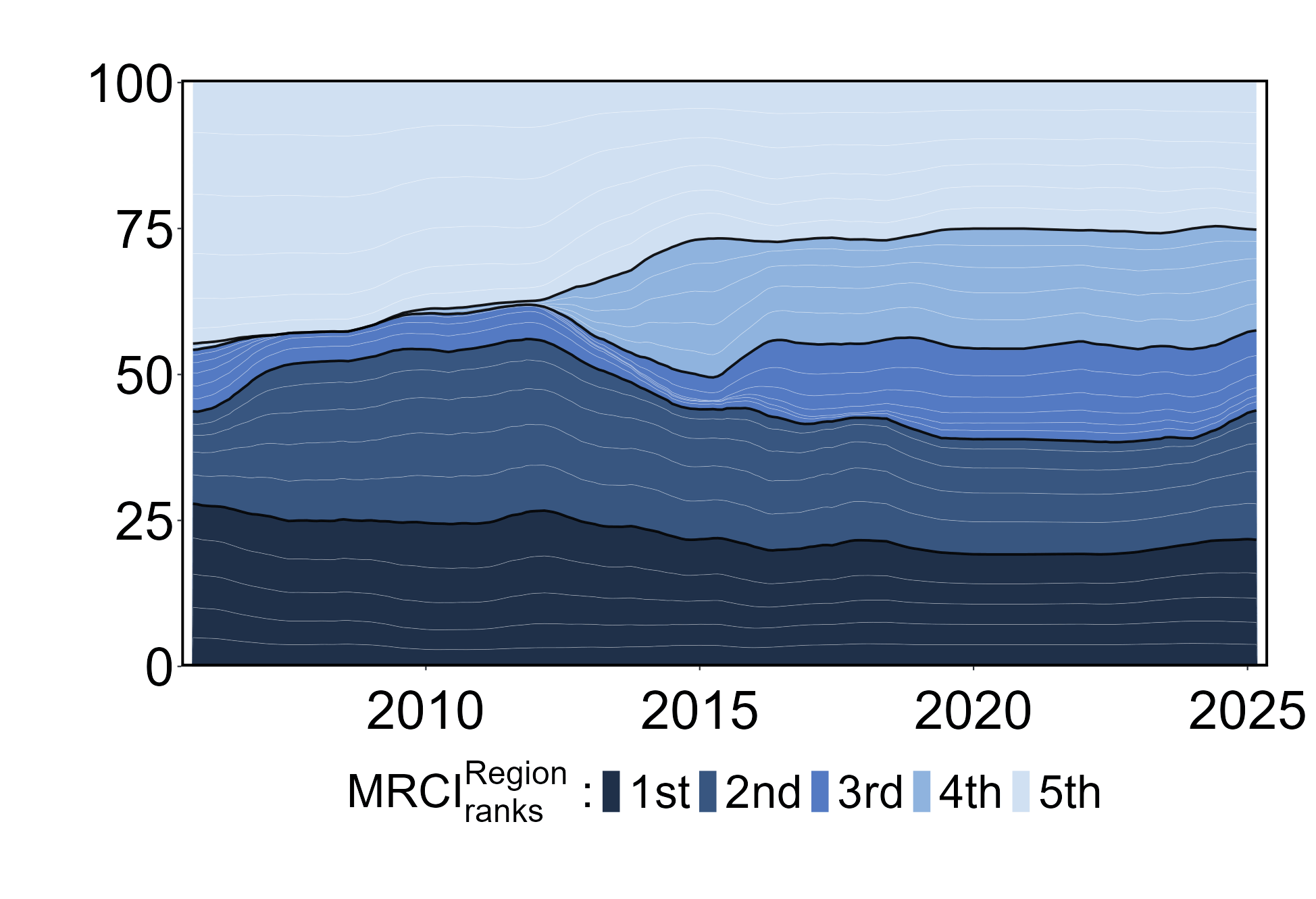}
		\end{subfigure}\hfill
		\begin{subfigure}[t]{0.33\textwidth}\centering
			\hspace*{.15cm}\includegraphics[width=1.04\linewidth]{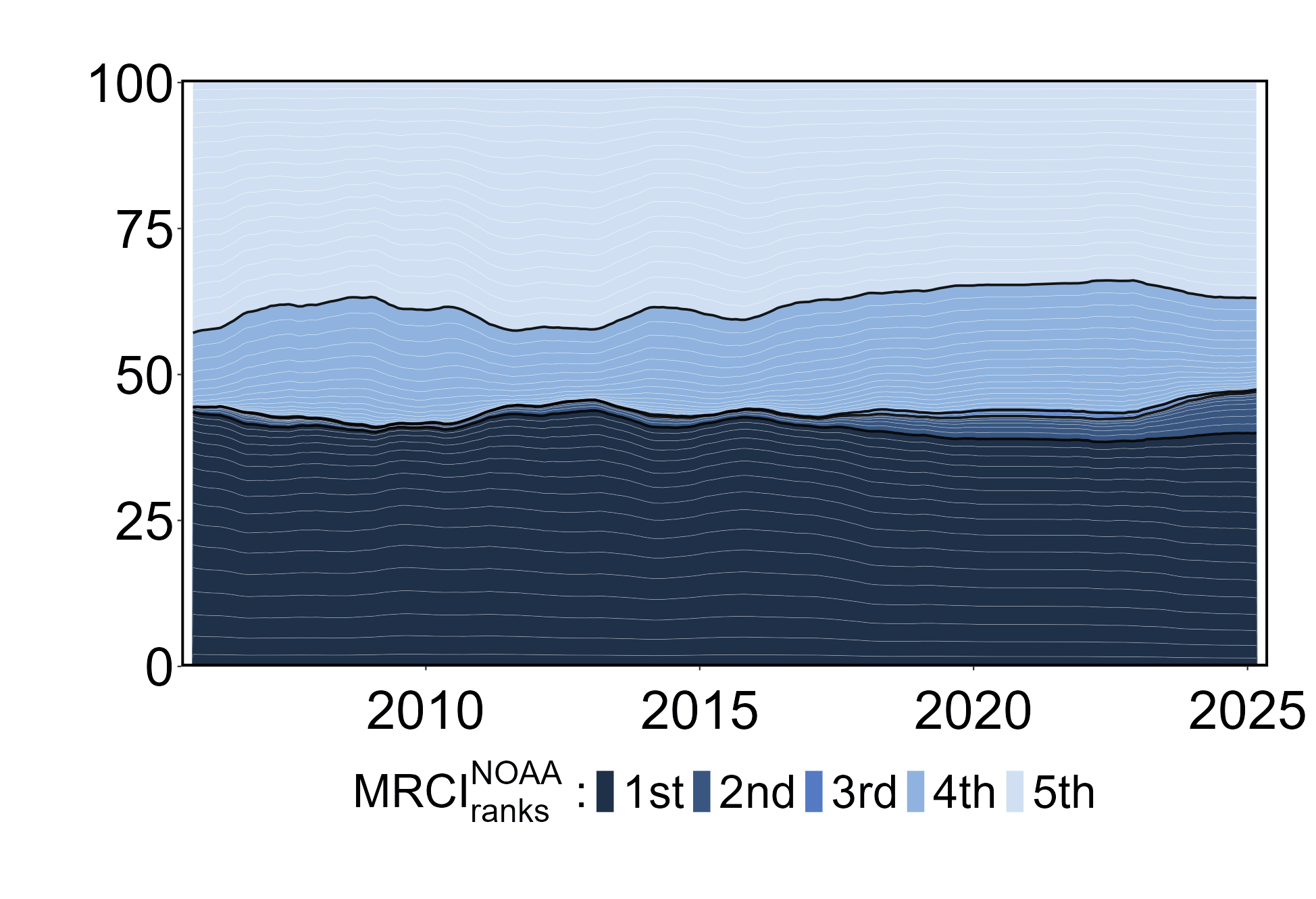}
		\end{subfigure}
	\end{center}
	\begin{threeparttable}
		\centering\vspace{-1cm}
		\begin{minipage}{\textwidth}
			\begin{tablenotes}[para,flushleft]
				\setlength{\lineskip}{0.2ex}
				\notsotiny
				{\textit{Notes}: Weights re-estimated on an expanding window and plotted vintage by vintage over the $2005$--$2025$ expanding-window period (the COVID window, 2020:M1--2021:M6, is held fixed at its December-2019 value). Each band is the share of the corresponding variable family (component-space, \textbf{top row}) or rank quintile (rank-space, \textbf{bottom row}) in the assembled index; thin white lines mark individual components, thick black lines mark family/quintile boundaries. To ensure an adequate representation, the series are smoothed over a year.}
			\end{tablenotes}
		\end{minipage}
	\end{threeparttable}
\end{figure}

The bootstrap asks whether the allocation survives resampling of a fixed sample. The natural complement asks whether the same allocation is recovered as the estimation window grows.
Figure~\ref{fig:rolling_weights} re-estimates the weights on an expanding window and plots their composition as the window grows over $2005$--$2025$. In the National component-space specification, \texttt{WP90} remains the dominant component throughout, accounting on average for roughly \textcolor{black}{70\%} of the total weight and never falls below \textcolor{black}{63\%}. \texttt{T90} declines slightly from \textcolor{black}{24\% to 22\%}, while the proportion of \texttt{Rx5Day} increased from \textcolor{black}{12\% to 15\%}. However, it should be noted that there was a significant collapse between 2010 and 2015 to about \textcolor{black}{2\%}. Sea level and \texttt{T10} remain essentially inactive over the full sample. 

The Regional specification evolves more smoothly: \texttt{WP90} increases from \textcolor{black}{37\% to 40\%}, \texttt{Rx5Day} declines from \textcolor{black}{35\% to 23\%}, and \texttt{T10} becomes more important towards the end of the sample and increases from \textcolor{black}{1\% to 8\%}. 

For NOAA, temperature-related indicators gain weight, with temperature rising from about \textcolor{black}{23\% to 32\%} and heating degree days more than doubling from about \textcolor{black}{8\% to 19\%}, while precipitation (\textcolor{black}{26\% to 18\%}) and the Palmer Z index (\textcolor{black}{21\% to 16\%}) both lose importance. These are slow reallocations, not abrupt reversals. 

rank-space remains stable in all three panels. The only visible movement is a modest reallocation within the upper tail in the Regional specification, with some mass shifting from the most extreme rank quintile to the adjacent one before stabilizing again after 2017.

\section{From Ranks to Named Components}\label{sec:rankstocomps}

\setcounter{figure}{0}
\setcounter{equation}{0}

\textcolor{black}{This appendix provides a descriptive mapping from rank-space weights to named inputs. Because a rank weight attaches to a position in the cross-sectional distribution rather than to a fixed variable, the named input occupying that position can change from one month to the next. We use the period-by-period sort to recover these time-varying named loadings and to summarize their average usage. This mapping describes the composition of the index, not the physical composition of its recursively identified innovation. It neither identifies a unique wind, flood, temperature, or other physical shock nor makes innovations in different VAR specifications a common shock. Their impulse responses remain responses to innovations in alternative climate measures.}

\paragraph{Mapping rank weights to named inputs.}
At date $t$ let $X_{(1),t}\le\cdots\le X_{(d),t}$ be the sorted component values and $\pi_t$ the permutation linking rank slots to names, $X_{(j),t}=\widetilde X_{\pi_t(j),t}$, where $\widetilde X_{m,t}$ is named component $m$. The estimated weight $\widehat w_j$ loads on rank slot $j$; because the component occupying a slot changes from month to month, the effective loading on a name is time varying,
\begin{equation}
  \omega_{m,t} \;=\; \widehat w_{\,\pi_t^{-1}(m)},
  \qquad\text{so that}\qquad
  x^*_t \;=\; \sum_{m=1}^{d}\omega_{m,t}\,\widetilde X_{m,t}.
  \label{eq:tvloadings}
\end{equation}
The rank index is thus an ordinary weighted average of the named components, only with weights that reshuffle as the sort does.

\paragraph{A size-neutral measure of named usage.}
Because every component is standardized, the order-statistic basis describes positions \emph{within} each observation rather than persistent identities: when the standardized components are close to exchangeable, each should visit every rank about equally, and the average loading $\bar\omega_m=T^{-1}\sum_t\omega_{m,t}$ collapses towards the equal weight $1/d$, so a raw average hides the signal. We read the departure from equal weighting through the rank-occupancy probabilities $p_{mr}$ (the fraction of periods in which component $m$ occupies rank $r$) which combine with the rank weights into the size-neutral usage
\begin{equation}
  U_m \;=\; \frac{\bar\omega_m}{1/d}-1 \;=\; d\sum_{r=1}^{d}\widehat w_r\,p_{mr}-1,
  \label{eq:rankusage}
\end{equation}
the proportional deviation of component $m$'s implied loading from equal weight: $U_m>0$ marks an over-weighted name, $U_m<0$ an under-weighted one. For the NOAA panel we average $U_m$ within a variable family, so a pooled residual is not inflated by the number of cells it absorbs.

\paragraph{ACI panel.}
\textcolor{black}{For the national and regional ACI panels, the implied named usage is diffuse and close to equal weight. Their rank weights are less concentrated in the tails than those estimated from the NOAA panel (Figures~\ref{fig:assemblage_aci} and~\ref{fig:assemblage_noaa}); similarly, the expanding-window National quintile shares are nearly flat and the Regional profile is only moderately tilted towards the extremes (Appendix~\ref{sec:rolling}, Figure~\ref{fig:rolling_weights}). Because the standardized ACI inputs move across ranks over time, the implied loadings within each variable family tend to fall on both sides of equal weight and largely offset one another. We therefore report this diffuse average-usage pattern in words rather than as a separate figure. It provides no basis for assigning either rank-space index to a single named climate family.}

\begin{figure}[t!]
   \caption{{Deviation from equal weight}}
  \label{fig:noaa_ranktocomp}
  \centering
  \includegraphics[width=0.72\textwidth]{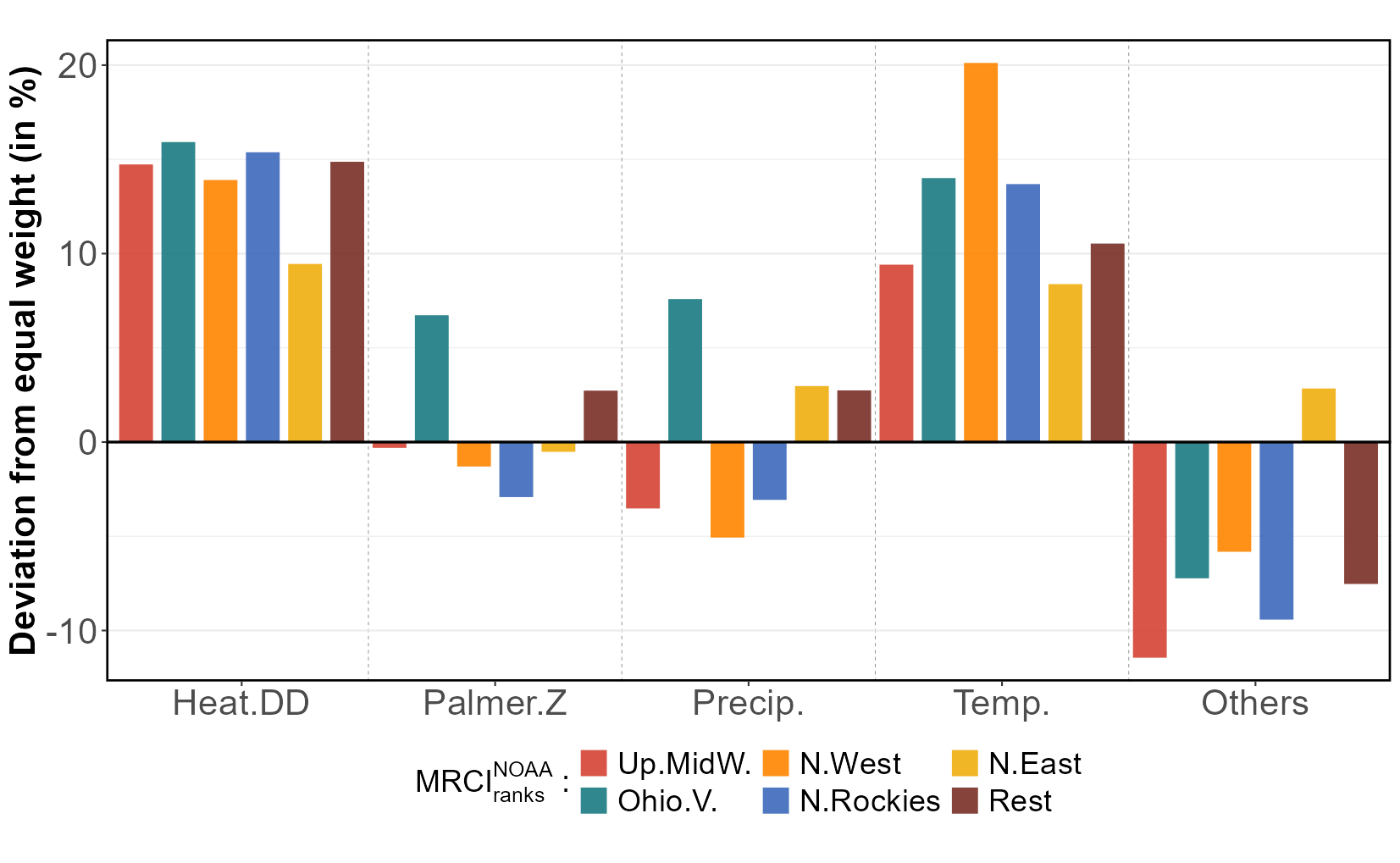}
  \vspace{1mm}
    \begin{minipage}{0.92\textwidth}
     \notsotiny \textit{Notes:} The NOAA order-statistic index MRCI$_{\text{ranks}}^{\text{NOAA}}$, over the full estimation sample 1975--2025. Each bar is the rank-implied usage $U_m$ (Equation~\ref{eq:rankusage}), in percent, averaged across the cells sharing a region (color, see legend) and a variable family (horizontal group), with $U_m=0$ the equal-weight baseline; ``Others'' collects the variables outside the four named families and ``Rest'' the regions outside the five named ones. Family-level usage and its rank correlation with the component-space weights (MRCI$_{\text{comps}}^{\text{NOAA}}$) are reported in the text.
  \end{minipage}
\end{figure}

\paragraph{NOAA panel.}
\textcolor{black}{For the NOAA panel, the estimated rank weights place about \textcolor{black}{three-quarters} of their mass in the distributional tails and about \textcolor{black}{three} percent on the $30$th--$55$th-percentile interior, with no single rank receiving more than $3\%$. Mapping these rank weights to names yields an average usage pattern tilted towards temperature-related families. Heating-degree days (\textcolor{black}{$U=+14\%$}) and temperature (\textcolor{black}{$+13\%$}) are the two most over-weighted families; precipitation and the Palmer drought index are within one percentage point of equal weight, and the pooled residual is the most under-weighted (\textcolor{black}{$-7\%$}). Average usage has a Spearman rank correlation of $\rho=0.80$ with the component-space weights (Figure~\ref{fig:noaa_ranktocomp}).}

\section{Aggregation of the NOAA Panel}\label{sec:aggregation}
\setcounter{topnumber}{1} 
\setcounter{figure}{0}
\setcounter{equation}{0}

The NOAA specification is estimated throughout using the full set of $82$ raw series. Each raw series is indexed along two dimensions: a geographic dimension, covering the U.S. climate regions, the contiguous-U.S. aggregate, and the Mauna Loa CO$_2$ records, and a climate-variable dimension, covering temperature, precipitation, degree-day measures, and the four Palmer drought indices. The consolidation described here is used only to display and interpret the component-space weights in Figure~\ref{fig:assemblage_noaa}; it is not used in estimation and does not alter the construction of the NOAA index. For this graphical representation, we group the raw series along both dimensions. Table~\ref{tab:region-aggregation} maps the raw geographic categories to consolidated regions, retaining the five U.S. climate regions that carry the bulk of the macro-relevant variation and collecting the remaining broad aggregates and non-U.S. CO$_2$ series into a single \emph{Rest} group. Table~\ref{tab:variable-aggregation} maps the raw climate variables to consolidated variable groups, keeping temperature, precipitation, heating-degree days, and the Palmer Z-Index as named families and pooling the remaining low-variance or closely related measures into \emph{Others}.  Applying the two mappings jointly partitions the $82$ raw inputs into region$\times$variable cells. For Figure~\ref{fig:assemblage_noaa} only, we report the sum of the estimated raw-series weights within each cell. Thus the displayed NOAA component-space weights are aggregated loadings on named region--hazard blocks, while the underlying index remains $x_t^* (\boldsymbol{w})=\boldsymbol{w}'\mathbf{X}_t$, where $\mathbf{X}_t$ contains the original unaggregated $82$ NOAA series.

\begin{table}[t]
\centering
\small
 \caption{Region aggregation for the NOAA specification}
\label{tab:region-aggregation}
\begin{threeparttable}
\begin{tabular}{p{0.62\linewidth} p{0.25\linewidth}}
\toprule
\textbf{Raw region} 
& \textbf{Aggregated region} \\
\midrule
Upper Midwest 
& Up.MidW. \\

Ohio Valley 
& Ohio.V. \\

Northwest 
& N.West \\

Northern Rockies \& Plains 
& N.Rockies \\

Northeast 
& N.East \\

\addlinespace

Conterminous (Contiguous) United States (Lower 48); 
South; West; Southeast; Southwest; World emissions series
& Rest \\
\bottomrule
\end{tabular}

\vspace{0.4em}
\notsotiny \textit{Notes:}  The mapping keeps five U.S. climate regions as separate groups and assigns every other geographic category (the contiguous-U.S. aggregate, the broad South/West/Southeast/Southwest groupings, and the global-emissions series) to \emph{Rest}. The two carbon-dioxide series measured at Mauna Loa, Hawaii, do not correspond to a U.S. climate region and are likewise assigned to \emph{Rest}. Combined with the variable mapping of Table~\ref{tab:variable-aggregation}, this partitions the raw panel into the region$\times$variable cells that are used to display summed weights.
\end{threeparttable}
\end{table}

\begin{table}[t]
\centering
\small
\caption{Variable aggregation for the NOAA specification}
\label{tab:variable-aggregation}
\begin{threeparttable}
\begin{tabular}{p{0.62\linewidth} p{0.25\linewidth}}
\toprule
\textbf{Raw variable} 
& \textbf{Aggregated variable} \\
\midrule
Temperature; Temperature, Average
& Temp. \\

Precipitation
& Precip. \\

Heating Degree Days
& Heat.DD \\

Palmer Z-Index
& Palmer.Z \\

\addlinespace

Carbon Dioxide, Mauna Loa, Hawaii, Average of Period;
Carbon Dioxide, Mauna Loa, Hawaii, Trend, Average of Period;
Palmer Drought Severity Index;
Cooling Degree Days;
Palmer Hydrological Drought Index;
Palmer Modified Drought Index
& Others \\
\bottomrule
\end{tabular}

\vspace{0.4em}
\notsotiny \textit{Notes:}  The mapping keeps temperature, precipitation, heating-degree days, and the Palmer Z-Index as separate variable families and pools all remaining measures (the CO$_2$ level and trend, cooling-degree days, and the three other Palmer drought indices) into \emph{Others}. Combined with the region mapping of Table~\ref{tab:region-aggregation}, this yields the consolidated region$\times$variable cells whose summed weights are displayed in Figure~\ref{fig:assemblage_noaa}.
\end{threeparttable}
\end{table}

\section{Full Impulse Response Functions}\label{annexe_climate}
\setcounter{topnumber}{2}
\setcounter{figure}{0}
\setcounter{equation}{0}
\setlength{\fboxrule}{0.0pt} 

\begin{figure}[t!]
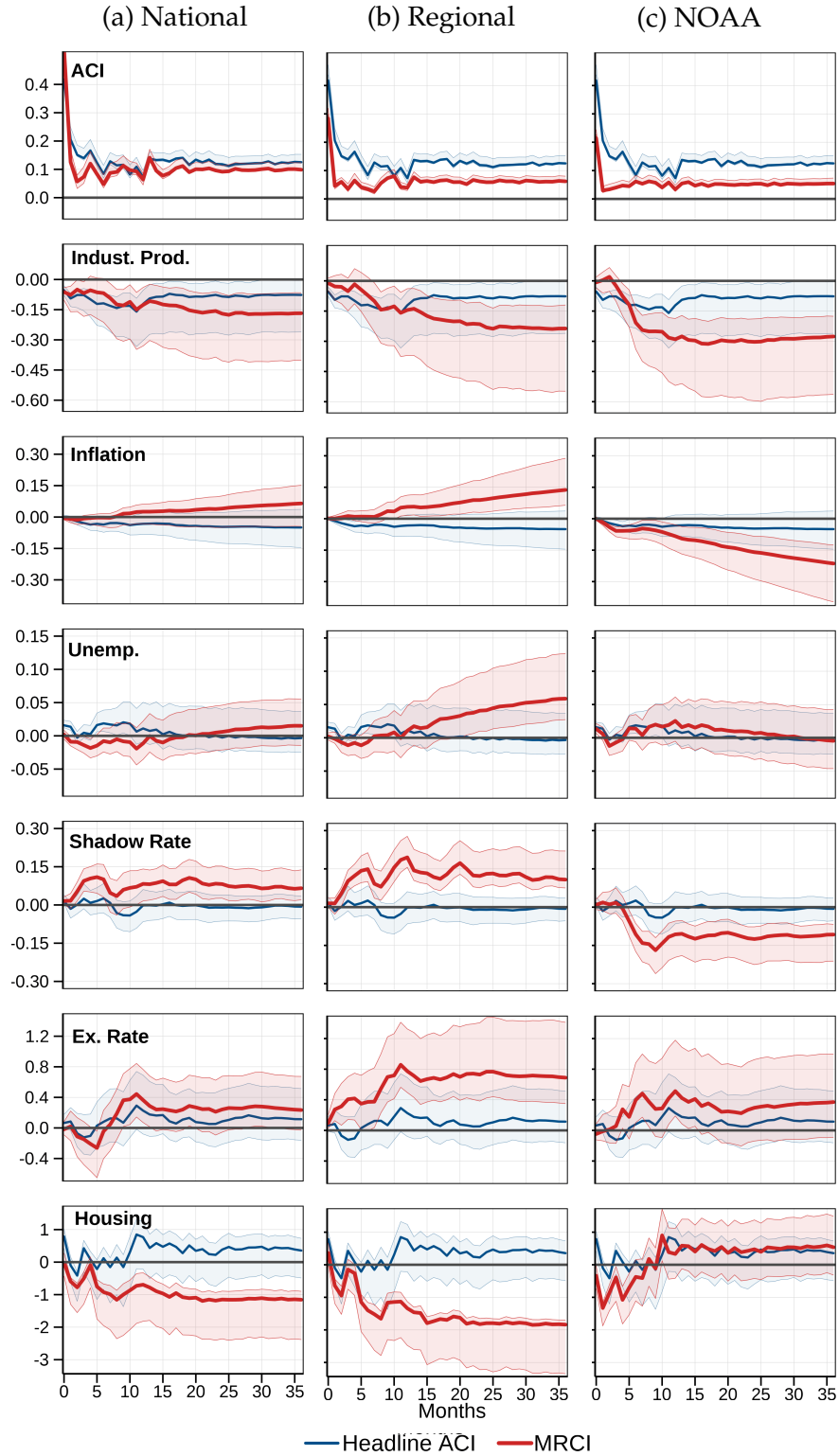

	\caption{{\textcolor{black}{Component-Space Climate Measure Shocks (All Variables)}}} \label{fig:FULL_Caci_irf}\vspace{.2cm}
    		    \centering\fbox{%
    \begin{minipage}{0.83\linewidth}
	\begin{center}
		\begin{subfigure}[t]{.274\textwidth}
			\vspace*{-0.3cm}
			\captionsetup{margin={.40cm,0cm}} 
			\centering\caption{National}
			\vspace*{-0.25cm}
			\hspace*{-1mm}\includegraphics[width=1.02\textwidth,trim = 16.5mm 4mm 11.5mm 12mm, clip]{figure/ACI_Comp_lvl_1_shockACI_Sfull_RUN0814.1049_v2026-08-14.png}
		\end{subfigure}
		\hspace*{-2.9mm}
		\begin{subfigure}[t]{.274\textwidth}
			\vspace*{-0.3cm}
			\captionsetup{margin={-0.0cm,0cm}} 
			\centering\caption{Regional}
			\vspace*{-0.235cm}
			\hspace*{2.9mm}\includegraphics[width=.894\textwidth, trim = 24.5mm 4mm 9.5mm 12mm, clip]{figure/ACI_Comp_lvl_2_shockACI_Sfull_RUN0814.0906_v2026-08-14.png}
		\end{subfigure}%
		\hspace*{-7mm}
		\begin{subfigure}[t]{.274\textwidth}
			\vspace*{-0.285cm}
			\captionsetup{margin={0.4cm,0cm}} 
			\centering\caption{NOAA}
			\vspace*{-0.25cm}
			\hspace*{7mm}\includegraphics[width=.894\textwidth, trim = 24.5mm 4mm 9.5mm 12mm, clip]{figure/ACI_Comp_lvl_3_shockACI_Sfull_RUN0814.0903_v2026-08-14.png}
		\end{subfigure}%
		
		\begin{subfigure}[t]{.35\textwidth}
			\vspace*{-0.02cm}
			\hspace*{0mm}\includegraphics[width=1.05\textwidth, trim = 13mm 3.8mm 5.5mm 235mm, clip]{figure/Naming_ACI.png}
		\end{subfigure}%
		
		\begin{subfigure}[t]{.35\textwidth}
			\vspace*{-0.07cm}
			\hspace*{-3mm}\includegraphics[width=1.05\textwidth, trim = 13mm 0mm 5.5mm 237.5mm, clip]{figure/Naming_ACI.png}
		\end{subfigure}%
	\end{center}
	
	\begin{threeparttable}
		\centering \vspace*{-0.6cm}
		\begin{minipage}{\textwidth}
			\begin{tablenotes}[para,flushleft]
				\setlength{\lineskip}{0.2ex}
				\notsotiny 
				 {\textit{Notes}: Component-space impulse responses of all seven VAR variables to a one-standard-deviation innovation in the assembled climate index $x_t^*(\widehat{\boldsymbol{w}})$, for the National (MRCI$_{\text{comps}}^{\text{Nation}}$), Regional (MRCI$_{\text{comps}}^{\text{Region}}$), and NOAA (MRCI$_{\text{comps}}^{\text{NOAA}}$) specifications. These are the full responses underlying the selected-variable Figure~\ref{fig:Caci_irf}. VAR models are estimated on data from 1975:M1 to 2025:M4, excluding the COVID-19 period (2020:M1--2021:M6), with 12 lags and the ordering ACI/MRCI, \mn{INDPRO}, \mn{PCEPI}, \mn{UNRATE}, \mn{SHADOWRATE}, \mn{EXUSUKx}, \mn{HOUST}. \textcolor{black}{Bands are at the $68\%$ level, conditional on the selected penalty $\widehat\lambda$ and the fitted weights $\widehat{\boldsymbol w}(\widehat\lambda)$.}}
			\end{tablenotes}
		\end{minipage}
	\end{threeparttable}
                \end{minipage}%
  }
\end{figure}

\begin{figure}[t!]
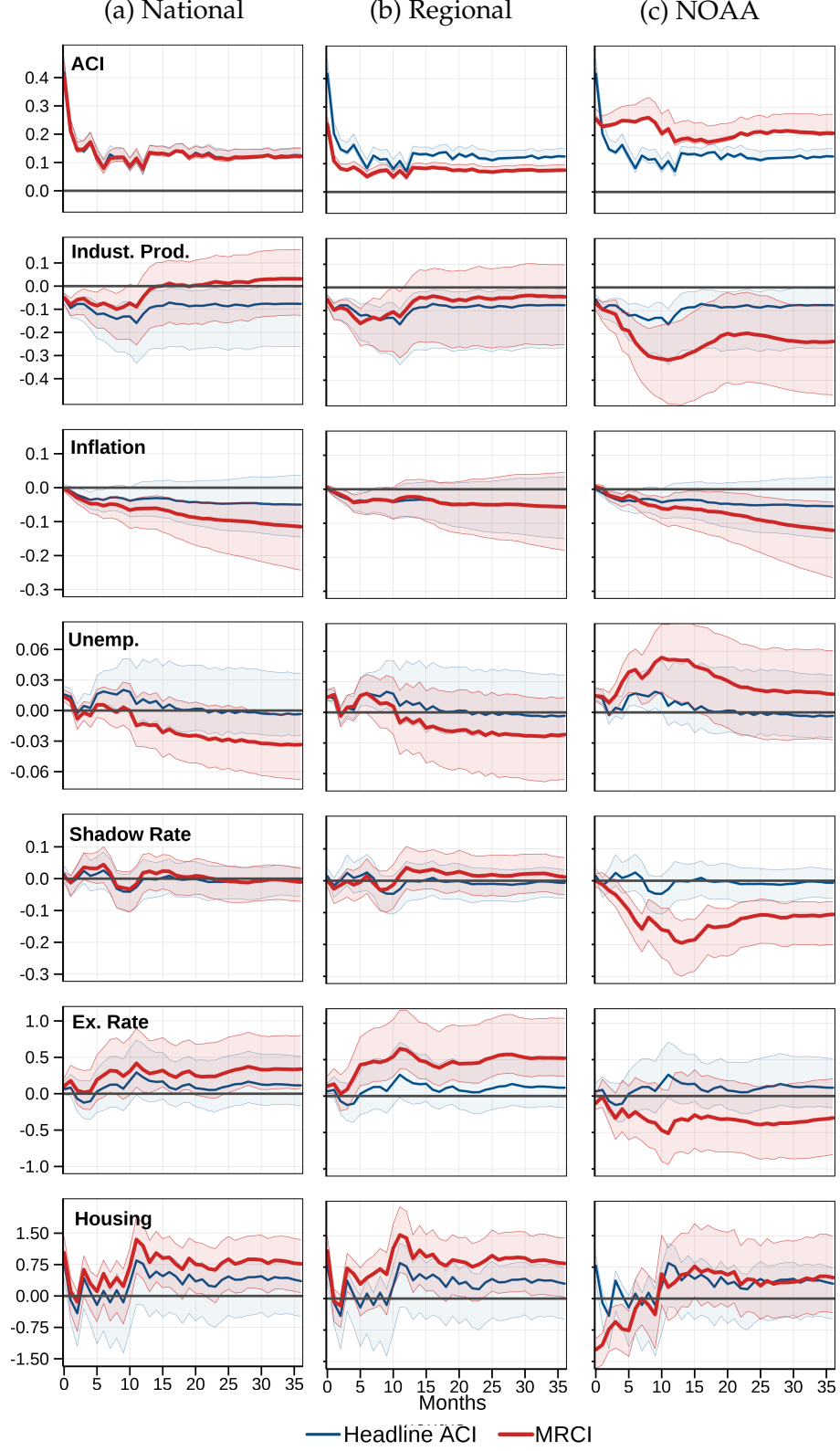

	\caption{{\textcolor{black}{Rank-Space Climate Measure Shocks (All Variables)}}} \label{fig:FULL_Raci_irf}
    \vspace{.2cm}
    		    \centering\fbox{%
    \begin{minipage}{0.86\linewidth}
	\begin{center}
		
		\begin{subfigure}[t]{.274\textwidth}
			\vspace*{-0.3cm}
			\captionsetup{margin={.40cm,0cm}} 
			\centering\caption{National}
			\vspace*{-0.25cm}
			\hspace*{-1mm}\includegraphics[width=1.02\textwidth,trim = 16.5mm 4mm 11.5mm 12mm, clip]{figure/ACI_Rank_lvl_1_shockACI_Sfull_RUN0814.1043_v2026-08-19.png}
		\end{subfigure}
		\hspace*{-2.9mm}
		\begin{subfigure}[t]{.274\textwidth}
			\vspace*{-0.3cm}
			\captionsetup{margin={-0.0cm,0cm}} 
			\centering\caption{Regional}
			\vspace*{-0.235cm}
			\hspace*{2.9mm}\includegraphics[width=.894\textwidth, trim = 24.5mm 4mm 9.5mm 12mm, clip]{figure/ACI_Rank_lvl_2_shockACI_Sfull_RUN0814.0959_v2026-08-19.png}
		\end{subfigure}%
		\hspace*{-7mm}
		\begin{subfigure}[t]{.274\textwidth}
			\vspace*{-0.285cm}
			\captionsetup{margin={0.4cm,0cm}} 
			\centering\caption{NOAA}
			\vspace*{-0.25cm}
			\hspace*{7mm}\includegraphics[width=.894\textwidth, trim = 24.5mm 4mm 9.5mm 12mm, clip]{figure/ACI_Rank_lvl_3_shockACI_Sfull_RUN0814.1010_v2026-08-19.png}
		\end{subfigure}%

		\begin{subfigure}[t]{.35\textwidth}
			\vspace*{-0.02cm}
			\hspace*{0mm}\includegraphics[width=1.05\textwidth, trim = 13mm 3.8mm 5.5mm 235mm, clip]{figure/Naming_ACI.png}
		\end{subfigure}%
		
		\begin{subfigure}[t]{.35\textwidth}
			\vspace*{-0.07cm}
			\hspace*{-3mm}\includegraphics[width=1.05\textwidth, trim = 13mm 0mm 5.5mm 237.5mm, clip]{figure/Naming_ACI.png}
		\end{subfigure}%
	\end{center}
	
	\begin{threeparttable}
		\centering \vspace*{-0.6cm}
		\begin{minipage}{\textwidth}
			\begin{tablenotes}[para,flushleft]
				\setlength{\lineskip}{0.2ex}
				\notsotiny 
				 {\textit{Notes}: Rank-space impulse responses of all seven VAR variables to a one-standard-deviation innovation in the assembled climate index $x_t^*(\widehat{\boldsymbol{w}})$, for the National (MRCI$_{\text{ranks}}^{\text{Nation}}$), Regional (MRCI$_{\text{ranks}}^{\text{Region}}$), and NOAA (MRCI$_{\text{ranks}}^{\text{NOAA}}$) specifications. These are the full responses underlying the selected-variable Figure~\ref{fig:Raci_irf}. Estimation sample, lag length, ordering, and band level are as in Figure~\ref{fig:FULL_Caci_irf}; \textcolor{black}{bands are conditional on the selected penalty $\widehat\lambda$ and the fitted weights $\widehat{\boldsymbol w}(\widehat\lambda)$.}}
			\end{tablenotes}
		\end{minipage}
	\end{threeparttable}
    \end{minipage}
    }
\end{figure}

\begin{figure}[t!]
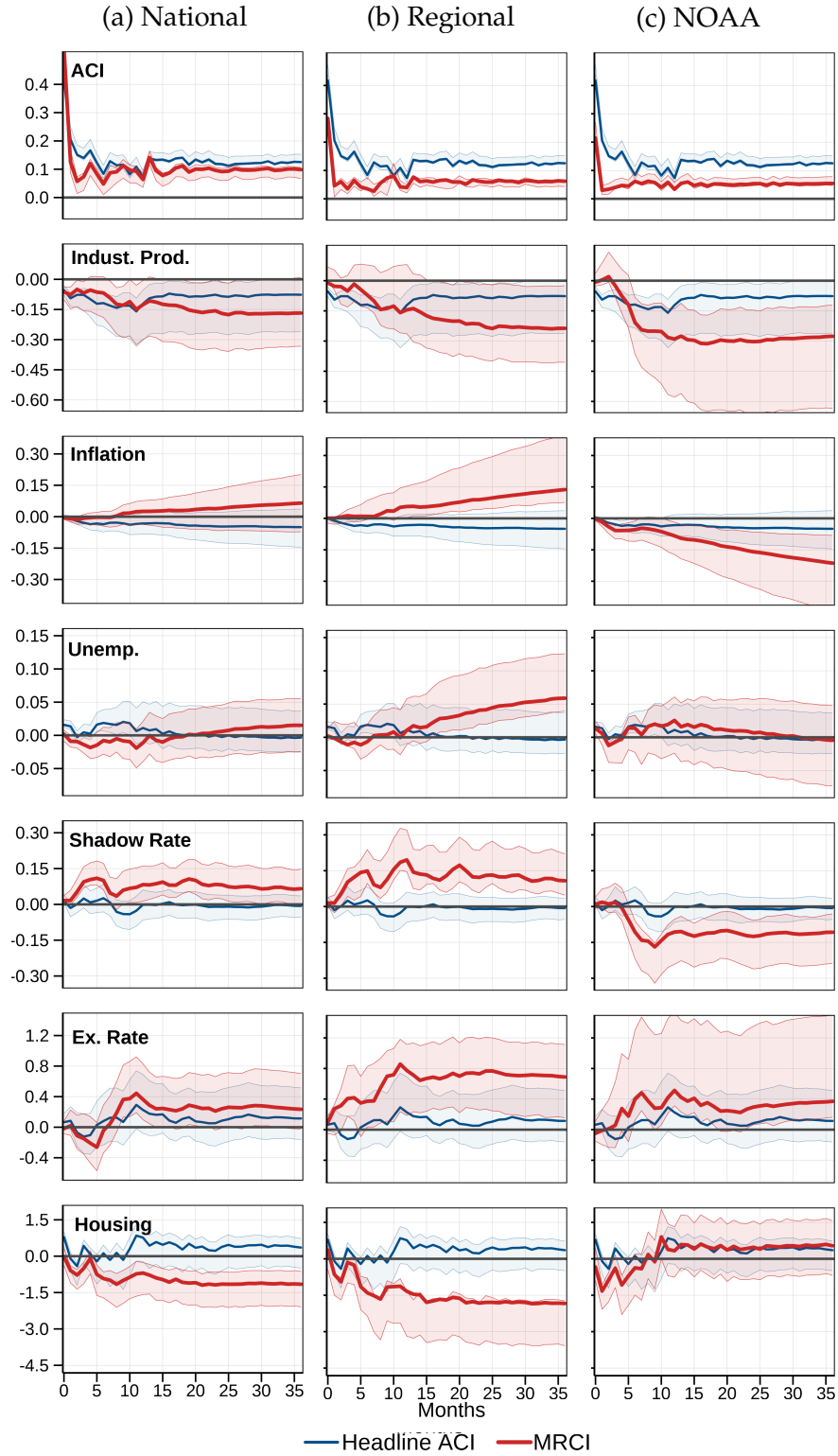

	\caption{{\textcolor{black}{Component-Space Climate Measure Shocks: Weight-Reestimating Pairs Bootstrap (All Variables)}}} \label{fig:FULL_Caci_boot_irf}\vspace{.2cm}
    		    \centering\fbox{%
    \begin{minipage}{0.83\linewidth}
	\begin{center}
		\begin{subfigure}[t]{.274\textwidth}
			\vspace*{-0.3cm}
			\captionsetup{margin={.40cm,0cm}} 
			\centering\caption{National}
			\vspace*{-0.25cm}
			\hspace*{-1mm}\includegraphics[width=1.02\textwidth,trim = 16.5mm 4mm 11.5mm 12mm, clip]{figure/IRFBOOT_ACI_Comp_lvl_1_shockACI_Sfull_RUN0814.1049_v2026-08-19.png}
		\end{subfigure}
		\hspace*{-2.9mm}
		\begin{subfigure}[t]{.274\textwidth}
			\vspace*{-0.3cm}
			\captionsetup{margin={-0.0cm,0cm}} 
			\centering\caption{Regional}
			\vspace*{-0.235cm}
			\hspace*{2.9mm}\includegraphics[width=.894\textwidth, trim = 24.5mm 4mm 9.5mm 12mm, clip]{figure/IRFBOOT_ACI_Comp_lvl_2_shockACI_Sfull_RUN0814.0906_v2026-08-19.png}
		\end{subfigure}%
		\hspace*{-7mm}
		\begin{subfigure}[t]{.274\textwidth}
			\vspace*{-0.285cm}
			\captionsetup{margin={0.4cm,0cm}} 
			\centering\caption{NOAA}
			\vspace*{-0.25cm}
			\hspace*{7mm}\includegraphics[width=.894\textwidth, trim = 24.5mm 4mm 9.5mm 12mm, clip]{figure/IRFBOOT_ACI_Comp_lvl_3_shockACI_Sfull_RUN0814.0903_v2026-08-19.png}
		\end{subfigure}%

		\begin{subfigure}[t]{.35\textwidth}
			\vspace*{-0.02cm}
			\hspace*{0mm}\includegraphics[width=1.05\textwidth, trim = 13mm 3.8mm 5.5mm 235mm, clip]{figure/Naming_ACI.png}
		\end{subfigure}%
		
		\begin{subfigure}[t]{.35\textwidth}
			\vspace*{-0.07cm}
			\hspace*{-3mm}\includegraphics[width=1.05\textwidth, trim = 13mm 0mm 5.5mm 237.5mm, clip]{figure/Naming_ACI.png}
		\end{subfigure}%
	\end{center}
	
	\begin{threeparttable}
		\centering \vspace*{-0.6cm}
		\begin{minipage}{\textwidth}
			\begin{tablenotes}[para,flushleft]
				\setlength{\lineskip}{0.2ex}
				\notsotiny 
				 {\textit{Notes}: Component-space impulse responses of all seven VAR variables under the \textcolor{black}{weight-reestimating pairs bootstrap}, in which each of $B=1{,}000$ replications re-solves for $\widehat{\boldsymbol{w}}$ and re-estimates $(\mathbf{B},\boldsymbol{\Sigma})$ at the fixed cross-validated penalty. These are the full responses underlying the selected-variable Figure~\ref{fig:Caci_boot_irf}. Estimation sample, lag length, ordering, and band level are as in Figure~\ref{fig:FULL_Caci_irf}.}
			\end{tablenotes}
		\end{minipage}
	\end{threeparttable}
                \end{minipage}%
  }
\end{figure}

The main text reports impulse responses for the variables of central interest: industrial production, unemployment, and housing starts. This appendix collects the responses of all seven variables in the VAR, for all specifications, so that the selected-variable figures of Section~\ref{sec:climate_results} (Figures~\ref{fig:Caci_irf}--\ref{fig:Raci_irf}) and the \textcolor{black}{weight-reestimating pairs bootstrap figures} of Appendix~\ref{sec:emp_climate_boot} (Figures~\ref{fig:Caci_boot_irf}--\ref{fig:Raci_boot_irf}) can be located within the full system response. Figures~\ref{fig:FULL_Caci_irf} and~\ref{fig:FULL_Raci_irf} give the \textcolor{black}{fixed-weights residual bootstrap responses} in component and rank-space; Figures~\ref{fig:FULL_Caci_boot_irf} and~\ref{fig:FULL_Raci_boot_irf} give the corresponding \textcolor{black}{weight-reestimating pairs bootstrap responses}. Three features are useful for reading the full-system figures. \textcolor{black}{First, each specification reports responses to a one-standard-deviation innovation in its own climate measure, so the figures show how full-system responses vary across alternative climate measures rather than responses to a common shock.} Second, the variables not discussed in the main text (PCE inflation, the shadow rate, and the exchange rate) display responses that are less robust across specifications than the real-activity variables. In several cases the effects are economically muted or imprecisely estimated. Accordingly, the main empirical action is most consistently concentrated in the real-activity block, rather than exclusively so.
Third, every VAR is estimated on 1975:M1--2025:M4 with the COVID-19 window (2020:M1--2021:M6) excluded, 12 lags, and $68\%$ bands, so the figures here are directly comparable to those in the main text and differ only in the set of response variables shown.

\newpage

\newpage
\begin{figure}[t!]
	\caption{{\textcolor{black}{Rank-Space Climate Measure Shocks: Weight-Reestimating Pairs Bootstrap \\(All Variables)}}} \label{fig:FULL_Raci_boot_irf}
    \vspace{.2cm}
    		    \centering\fbox{%
    \begin{minipage}{0.86\linewidth}
	\begin{center}
		
		\begin{subfigure}[t]{.274\textwidth}
			\vspace*{-0.3cm}
			\captionsetup{margin={.40cm,0cm}} 
			\centering\caption{National}
			\vspace*{-0.25cm}
			\hspace*{-1mm}\includegraphics[width=1.02\textwidth,trim = 16.5mm 4mm 11.5mm 12mm, clip]{figure/IRFBOOT_ACI_Rank_lvl_1_shockACI_Sfull_RUN0814.1043_v2026-08-20.png}
		\end{subfigure}
		\hspace*{-2.9mm}
		\begin{subfigure}[t]{.274\textwidth}
			\vspace*{-0.3cm}
			\captionsetup{margin={-0.0cm,0cm}} 
			\centering\caption{Regional}
			\vspace*{-0.235cm}
			\hspace*{2.9mm}\includegraphics[width=.894\textwidth, trim = 24.5mm 4mm 9.5mm 12mm, clip]{figure/IRFBOOT_ACI_Rank_lvl_2_shockACI_Sfull_RUN0814.0959_v2026-08-20.png}
		\end{subfigure}%
		\hspace*{-7mm}
		\begin{subfigure}[t]{.274\textwidth}
			\vspace*{-0.285cm}
			\captionsetup{margin={0.4cm,0cm}} 
			\centering\caption{NOAA}
			\vspace*{-0.25cm}
			\hspace*{7mm}\includegraphics[width=.894\textwidth, trim = 24.5mm 4mm 9.5mm 12mm, clip]{figure/IRFBOOT_ACI_Rank_lvl_3_shockACI_Sfull_RUN0814.1010_v2026-08-20.png}
		\end{subfigure}%

		\begin{subfigure}[t]{.35\textwidth}
			\vspace*{-0.02cm}
			\hspace*{0mm}\includegraphics[width=1.05\textwidth, trim = 13mm 3.8mm 5.5mm 235mm, clip]{figure/Naming_ACI.png}
		\end{subfigure}%
		
		\begin{subfigure}[t]{.35\textwidth}
			\vspace*{-0.07cm}
			\hspace*{-3mm}\includegraphics[width=1.05\textwidth, trim = 13mm 0mm 5.5mm 237.5mm, clip]{figure/Naming_ACI.png}
		\end{subfigure}%
	\end{center}
	
	\begin{threeparttable}
		\centering \vspace*{-0.6cm}
		\begin{minipage}{\textwidth}
			\begin{tablenotes}[para,flushleft]
				\setlength{\lineskip}{0.2ex}
				\notsotiny 
				 {\textit{Notes}: Rank-space impulse responses of all seven VAR variables under the \textcolor{black}{weight-reestimating pairs bootstrap}, constructed as in Figure~\ref{fig:FULL_Caci_boot_irf}. These are the full responses underlying the selected-variable Figure~\ref{fig:Raci_boot_irf}. Estimation sample, lag length, ordering, and band level are as in Figure~\ref{fig:FULL_Caci_irf}.}

			\end{tablenotes}
		\end{minipage}
	\end{threeparttable}
    \end{minipage}
    }
\end{figure}

\end{document}